%% file: tech_report_main.tex
\documentclass[manuscript]{acmart}
\usepackage{geometry}
\renewcommand\footnotetextcopyrightpermission[1]{} %
\usepackage{xcolor}
\usepackage{caption}
\usepackage{subcaption}
\usepackage[linesnumbered,ruled]{algorithm2e}
\usepackage[noend]{algpseudocode}
\usepackage{color}
\usepackage{enumitem}

\SetKwFor{Iter}{iter}{do}{end}
\SetKwFor{Concurrent}{concurrently foreach}{do}{end}
\SetKwProg{Pn}{Procedure}{}{end}
\SetKwProg{Event}{Event}{}{end}
\algnewcommand\KwAnd{\textbf{\upshape and} }
\algnewcommand\KwReturn{\textbf{\upshape return} }
\algnewcommand{\LeftComment}[1]{\(\triangleright\) #1}
\usepackage{soul}
\usepackage{xcolor}
\def\HiLi{\leavevmode\rlap{\hbox to \hsize{\color{yellow!50}\leaders\hrule height .8\baselineskip depth .5ex\hfill}}}
\newcommand{\mathcolorbox}[2]{\colorbox{#1}{$\displaystyle #2$}}
\newcommand{\stitle}[1]{\noindent \textbf{#1} }
\newcommand{\sstitle}[1]{\noindent $\bullet\,$ \textsl{\underline{#1}} }
\newcommand{\mybullet}[1]{\noindent $\bullet\,$ \textsl{#1} }

\newcommand{\tid}{\texttt{TID}}

\newcommand{\added}[1]{#1}    
\newcommand{\icde}[1]{}
\newcommand{\techrep}[1]{#1}
\newtheorem{claim}{claim}
\newtheorem{myrule}{rule}
\newcommand{\deleted}[1]{}

\newcommand{\RIFig}[1]{}
\newcommand{\RIIFig}[1]{}
\newcommand{\RIIIFig}[1]{}
\newcommand{\META}[1]{}
\newcommand{\RI}[1]{}
\newcommand{\RII}[1]{}
\newcommand{\RIII}[1]{}
\newcommand{\revision}[1]{#1}
\newcommand{\marginnote}[1]{}
\newcommand{\CR}[1]{#1}

\AtBeginDocument{%
  \providecommand\BibTeX{{%
    \normalfont B\kern-0.5em{\scshape i\kern-0.25em b}\kern-0.8em\TeX}}}

\setcopyright{acmcopyright}
\copyrightyear{2018}
\acmYear{2018}
\acmDOI{XXXXXXX.XXXXXXX}

\acmConference[Conference acronym 'XX]{Make sure to enter the correct
  conference title from your rights confirmation emai}{June 03--05,
  2018}{Woodstock, NY}
\acmPrice{15.00}
\acmISBN{978-1-4503-XXXX-X/18/06}

\begin{document}

\title{Knock Out 2PC with Practicality Intact: a High-performance and General Distributed Transaction Protocol \\ (Technical Report)}
\author{Ziliang Lai}
\affiliation{%
  \institution{The Chinese University of Hong Kong}\country{zllai@cse.cuhk.edu.hk}}
\author{Hua Fan}
\affiliation{%
  \institution{Alibaba Group}\country{guanming.fh@alibaba-inc.com}}
\author{Wenchao Zhou}
\affiliation{%
  \institution{Alibaba Group}\country{zwc231487@alibaba-inc.com}}
\author{Zhanfeng Ma}
\affiliation{%
  \institution{Alibaba Group}\country{zhanfeng.mzf@alibaba-inc.com}}
\author{Xiang Peng}
\affiliation{%
  \institution{Alibaba Group}\country{pengxiang.px@alibaba-inc.com}}
\author{Feifei Li}
\affiliation{%
  \institution{Alibaba Group}\country{lifeifei@alibaba-inc.com}}
\author{Eric Lo}
\affiliation{%
  \institution{The Chinese University of Hong Kong}\country{ericlo@cse.cuhk.edu.hk}}
\fancyhead{}

\input{abstract}

\maketitle

\input{introduction}

\input{background}
\input{methodology}

\input{experiments}
\input{related_work}
\input{conclusion}

\input{tech_report_main.bbl}

\techrep{
\input{appendix}

}
\end{document}

%% file: abstract.tex
\begin{abstract}
Two-phase-commit (2PC) has been widely adopted for distributed transaction processing, but it also jeopardizes throughput by introducing two rounds of network communications and two durable log writes to a transaction’s critical path. Despite the various proposals that eliminate 2PC such as deterministic database and access localization, 2PC remains the de facto standard since the alternatives often lack generality (e.g., requiring workloads without branches based on query results). In this paper, we present Primo, a distributed transaction protocol that supports a more general set of workloads  without 2PC. Primo features write-conflict-free concurrency control that guarantees once a transaction enters the commit phase, no concurrency conflict (e.g., deadlock) would occur when installing the write-set
--- hence the prepare phase is no longer needed to account for any potential conflict from any partition. 
In addition, Primo further optimizes the transaction path 
using asynchronous group commit. With that, the durability delay is also taken off the transaction’s critical path. Empirical results on Primo are encouraging -- in YCSB and TPC-C, Primo attains 1.42$\times$ to 8.25$\times$ higher throughput than state-of-the-art general protocols including Sundial and COCO, while having similar latency as COCO which also employs group commit.
\end{abstract}

%% file: introduction.tex
\section{Introduction} \label{sec:intro}

Recent years have witnessed the growing popularity of shared-nothing databases \cite{f1,spanner,cockroachdb,oceanbase}. 
They offer high scalability by horizontal partitioning and high availability by replication.
In these databases, a distributed transaction is typically processed by an (1) \emph{execution phase} where the transaction commands are executed to collect the transaction's write-set; and then a (2) \emph{commit phase} where two-phase-commit (2PC) ensures the write-set is \textbf{atomically} and \textbf{durably} installed to the involved partitions.
For serializability,
a concurrency control scheme (e.g., 2PL) is employed to guard the read operations in the execution phase and the write operations in the commit phase.
Unfortunately, since locks can only be released after commit,
2PC would prolong the lock duration (i.e., the contention footprint \cite{calvin}) by two network roundtrips and two durable log writes --- significantly degrading the system throughput  \cite{calvin,deter-eval,leap,2pc-overhead}.

Over the years, researchers have been seeking alternatives to replace 2PC.
However, it is challenging due to the \emph{non-deterministic aborts} (e.g., deadlocks) \cite{calvin} that {\bf may} occur in some partitions. 2PC is a pessimistic approach that pessimistically assumes aborts would happen and hence prepare a series of steps to handle that issue \textbf{if} it really happens.
Deterministic database (DDB) \cite{calvin,bohm,pwv,slog,ov2,deter-case,deter-eval,deter-overview,caracal,quecc}, in contrast, is a \emph{preventive} approach that leverages \textbf{full} transaction information (e.g., read-write sets) and carefully plans the execution schedule a priori to avoid any non-deterministic aborts to happen. 
Access localization (e.g., LEAP \cite{leap}) is another preventive approach 
that aggressively avoids any transaction to span across partitions by \emph{pre-localizing} the remote data at the same site (i.e., all ship to one site), which also requires knowing the transactions read-write sets beforehand.
However, inferring read-write sets before execution is not always possible for a general workload, which limits their practicality \cite{aria, sensitivity}. For example, a transaction may choose a record to update according to query results; unless exhausting all the query results, the exact read-write set cannot be obtained.

In face of that, we present Primo\footnote{Primo means the first part of a duet, which echos our protocol that only requires the first part (execution) in a distributed transaction, eliminating the need for 2PC.}, a distributed transaction protocol targeting high throughput and scalability without 2PC. In Primo, we observe that completely preventing non-deterministic aborts like DDB is an overkill. 
As shown in \cite{calvin}, non-deterministic aborts have two types: (i) \emph{conflict-induced} aborts (e.g., deadlock) and (ii) \emph{crash-induced} aborts. We found novel approaches to handle them in a more flexible manner such that Primo retains generality.

For conflict-induced aborts, 
we observe that eschewing them {\bf only} when installing the write-set (i.e., during the commit phase) is sufficient for eliminating 2PC (unlike deterministic databases who do that for the whole lifetime of a transaction and thus need strong assumptions).
With this observation, we derive a \emph{write-conflict-free} (WCF) scheme that acquires an \textbf{exclusive-lock} for each \textbf{read} record (instead of a shared-lock) in the execution phase.
At first glance, such a locking scheme seems a step backward in reducing concurrency.
It, however, eliminates the major bottleneck of 2PC.
Specifically, if a transaction's read-set covers its write-sets\footnote{Technically, read-set covering write-set is not even an assumption, because it can be enforced by adding a dummy read to avoid the blind write, which has insignificant overhead.} (e.g., 
commonly seen in TPC-C \cite{tpcc}, TATP \cite{tatp}, and Smallbank \cite{smallbank}), by acquiring exclusive-locks for reads, locks for write operations are already held by the transaction and thus no conflict can occur when installing the write-set during commit. 
Hence, if no crash (crash handling will be discussed soon), each partition is able to commit the transaction immediately after installing the write-set (without 2PC) since there is no chance for any partition to abort in the commit phase.
Many concurrency control algorithms have already leveraged that \cite{silo,bcr,blotter,update-replication}, 
but none of them have fully exploited it for distributed transactions and we are the first to leverage it to eliminate 2PC.

Primo deliberately does not prevent or handle any potential partition crash in the transaction execution path,
effectively removing the need for 2PC-like synchronization.
Instead, Primo 
handles crash-induced aborts in granules of batches.
It uses a new \emph{distributed group commit} to commit transactions
because existing ones do {\bf not} scale.
For example, COCO \cite{coco} processes transactions and commits them epoch-by-epoch. An epoch can only start after a coordinator has confirmed \emph{all the partitions} have finished the previous epoch, which limits its scalability.
Primo addresses that by making distributed group commit \emph{asynchronous} to the transaction execution.
During normal operation, partitions execute transactions \emph{continuously} (without epoch barrier).
Synchronizations are only needed at recovery time, 
where partitions agree on which part of the transaction history should
be recovered.

Overall, with 2PC roundtrips eliminated, and durable log writes taken out of the critical path,
Primo improves the throughput by shrinking the transactions' contention footprint to only span the duration of their reads and writes.
In summary, the key contributions of this paper are:
\begin{itemize}
    \item We show that to eliminate 2PC, non-deterministic aborts need \textbf{not} be completely eschewed like deterministic database. With that observation, practical protocols can be derived to obtain high throughput and better generality.
    \item A write-conflict-free distributed concurrency control scheme that eliminates 2PC by handling conflict-induced aborts.
    \item An asynchronous distributed group commit scheme that takes failure handling off the critical path and offers better scalability than existing ones.
    \item We demonstrate the effectiveness of Primo through a theoretical analysis and an empirical study. Empirical results on YCSB and TPC-C show that Primo achieves 1.42$\times$ to 8.25$\times$ improvements over state-of-the-art.
\end{itemize}

The remainder of this paper is organized as follows. Section \ref{sec:background} provides the background. We give an overview of Primo in Section \ref{sec:primo}. We introduce the write-conflict-free concurrency control scheme in Section \ref{sec:waf} and the asynchronous group commit in Section \ref{sec:watermark}. The evaluation of Primo is presented in Section \ref{sec:exp}. We discuss related works in Section \ref{sec:related_work} and conclude the paper in Section \ref{sec:conclusion}.

%% file: background.tex
\section{Background} \label{sec:background}
In this section, we provide essential background for understanding Primo. For discussions on the related work, we refer readers to Section \ref{sec:related_work}.

\subsection{Classic Distributed Transaction Protocol} \label{sec:2pc}
In conventional distributed databases, 2PC often works with a concurrency control scheme to provide atomicity, durability, and serializability. We take the classic combination of 2PL and 2PC (similar to the one in Spanner \cite{spanner}) as an example.

On receiving a distributed transaction $T$, the database engine forwards $T$ to the partition that owns the first record assessed by $T$, which we call the \emph{coordinator}.
We call other involved partitions as \emph{participants}.
$T$ is then processed in an execution phase and a commit phase. 

\begin{enumerate}[leftmargin=1.2em]
    \item \textbf{Execution phase}. In this phase, the coordinator executes the transaction in which a shared-lock is acquired for each (local) read; remote reads are sent to the participants who acquire the corresponding shared-locks and return the read results to the coordinator. Local computations are carried out and
    write operations are \emph{buffered} locally at the coordinator.
    \item \textbf{Commit phase}. After the write-set is buffered, the transaction enters the commit phase, where 2PC is invoked. The coordinator acquires exclusive-locks for local writes and installs them to its partition. It also sends remote writes to each corresponding participant in a batch with a \texttt{PREPARE} message.
    On receiving the message, each participant acquires exclusive-locks, installs the writes and persists $T$'s log records in a similar manner as the coordinator.
    If the aforementioned operations succeed, it responds a \texttt{YES} to the coordinator. Otherwise, a \texttt{NO} is responded.
    If all participants respond \texttt{YES}, the coordinator logs this commit decision for $T$ and sends \texttt{COMMIT} messages to the participants.
    On receiving that, the participants also log the commit decision and commit the transaction.
    If one of the participants replies \texttt{NO}, the coordinator sends \texttt{ABORT} to the participants, and all partitions would abort $T$. 
    Only after the commit or abort, the involved partitions would release the acquired locks.
\end{enumerate}

\added{
The protocol discussed above is incorporated with common optimizations including Presumed-Abort (i.e., do not log the abort decision and presume abort if no commit decision is found during recovery), and Unsolicited-Vote \cite{ingress} (i.e., the PREPARE message is combined with the write requests to save the communication overhead).
However, optimizations of 2PC \cite{presumed-opt-1,NPrC,ingress,aria,lotus} cannot completely remove the network roundtrips in the transactions' contention footprint and thus still
hurt the throughput significantly \cite{calvin}.}

\subsection{Deterministic Database} \label{sec:ddb}
\added{
Most DDBs eliminate 2PC by preventing any non-deterministic aborts.
However, strong assumptions are often required to do so.
In particular, 
For conflict-induced aborts, DDB assumes (I) the transactions' read-write sets can be inferred before execution. With that, a \emph{conflict-free} schedule can be devised a priori for the subsequent execution to follow.
For crash-induced aborts, DDB assumes (II) the transactions' logic is deterministic. Hence, by having a deterministic transaction executor, a server can always reproduce the same state as before crash by replaying the input, such that no transaction is aborted due to crash.
}

Unfortunately, assumption (I) does not always hold for general transactions. 
For example,
it is hard to extract the read-write sets of interactive transactions or stored procedures that branch based on query results.
Furthermore, assumption (II) may also be violated due to the use of operators like \texttt{Rand()} and \texttt{Date()}. In addition, the query engine itself contains many non-deterministic code which is shown to be hard to remove completely \cite{fabric}.
\CR{Aria \cite{aria} as the state-of-the-art DDB
however assumes (II) only.
With (II), it can handle crash-induced aborts by lightweight input logging like other DDBs. However, it does not eliminate conflict-induced aborts as in the other DDBs.
Aria still guarantees 
the resulting states would be identical if the same inputs were given.
However,
without seeing the transactions' read-write sets a priori
and the fact that different partitions are having different sets of data (i.e., the input states are different), Aria has to retain 2PC-like roundtrips for finalizing the abort-vs-commit decisions. 
Hence, the efficiency of Aria is actually attributed more to the elimination of logging during 2PC.
}

\subsection{Epoch-based Transaction Processing with Distributed Group Commit} \label{sec:dgc}
Distributed group commit takes the durability concern off the critical path of distributed transaction processing.
To explain, we introduce the scheme in COCO \cite{coco} as an example.
COCO processes transactions in epochs 
and \emph{synchronously} coordinates each epoch to commit.
Specifically, an epoch in COCO is processed as follows:

\begin{enumerate}[leftmargin=1.2em]
    \item \textbf{Epoch execution}. Within an epoch, transactions are executed using a standard distributed transaction protocol (such as 2PL + 2PC). However, it is not necessary to synchronously flush log records for every individual transaction. Because durability is only confirmed at the echo boundary, the outcome of a transaction (commit or abort), is not immediately returned to the clients in this step.
    \item \textbf{Epoch group commit}. This step ensures log of the current epoch is durable on all partitions. After a designated time period (e.g.,10ms), one of the partitions takes the role of coordinator and issues a \texttt{GROUP-PREPARE} message to all the partitions. Upon receiving the message, each partition ensures that all log records from the current epoch are persisted and responds with a \texttt{GROUP-READY} message. Once the coordinator receives \texttt{GROUP-READY} messages from all partitions, it issues a \texttt{GROUP-COMMIT} message. 
    If a partition crashes and fails to send \texttt{GROUP-READY}, the coordinator would send out a \texttt{GROUP-ABORT} message on timeout.
    On receiving \texttt{GROUP-COMMIT}, each partition returns the results (commit or abort) of all transactions in the epoch and advances to the next epoch. However, the whole epoch would be aborted if \texttt{GROUP-ABORT} is received. In this case, the cluster waits for the crashed partition to recover or to be replaced by a replica.
\end{enumerate}

Distributed group commit has two advantages: (1) it reduces the cost of durable log writes by batching; and (2) it reduces the contention footprint by moving logging out of the critical path.
It also has two tradeoffs:
(1) the latency is typically increased from several micro-seconds to several milli-seconds in a main-memory database. 
However, that latency difference is practically unnoticeable 
from the user's perspective;
(2) if a partition crashes before the current epoch is group-committed, the whole epoch has to be aborted.
However, given the low crash rate in modern hardware, 
such a drawback is acceptable \cite{coco,impementation-tech}.
Nonetheless, the global synchronization step in distributed group commit limits the system scalability.

\added{
Aria \cite{aria} and Lotus \cite{lotus} also commit transactions in the granule of an epoch, but they expand the contention footprint of each distributed transaction to the duration of the whole epoch (e.g., Lotus requires a distributed transaction to hold locks until the end of the epoch), which hurts the throughput especially under high contention.
}

%% file: methodology.tex
\section{Primo Overview} \label{sec:primo}
Primo is designed for main-memory shared-nothing databases in which records are assigned to partitions and those partitions can be replicated to provide high availability. Our goal is to optimize the throughput of short read-write transactions, such as those typically found in TPC-C \cite{tpcc}. %

\icde{
\begin{figure}
    \centering
    \includegraphics[width=0.9\linewidth]{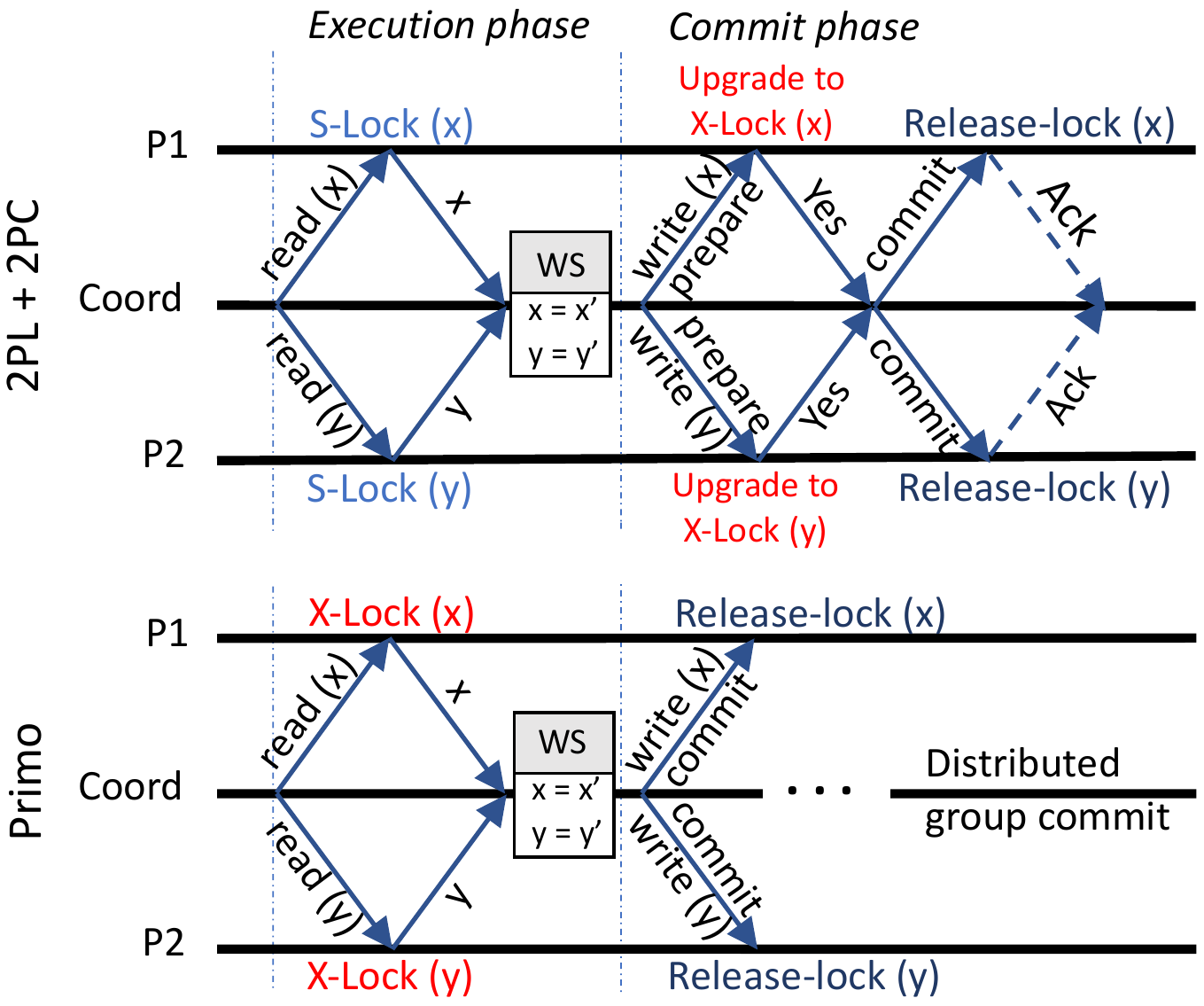}
    \caption{\revision{Primo compared with a 2PC-based scheme}}
    \vspace{-1em}
    \label{fig:compare}
\end{figure}
}

\techrep{
\begin{figure}
    \centering
    \includegraphics[width=0.6\linewidth]{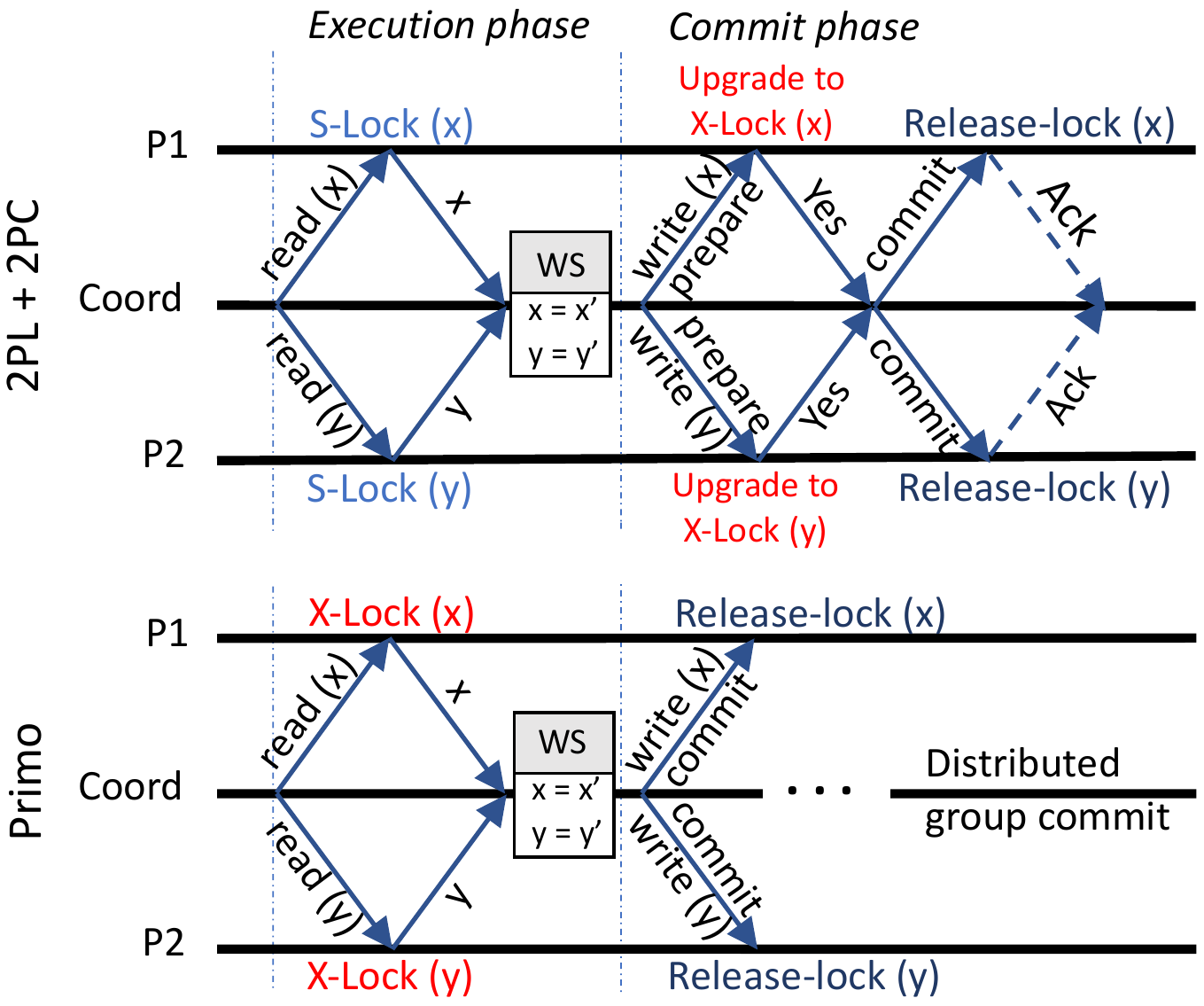}
    \caption{Primo compared with 2PC-based scheme}
    \label{fig:compare}
\end{figure}
}

Recall the classic 2PL+2PC protocol (Section \ref{sec:2pc}) executes a transaction and computes its write-set during the execution phase, and then installs the writes using 2PC during the commit phase (top part of Figure \ref{fig:compare}). 
2PC in the commit phase ensures atomicity in case some partitions cannot acquire the exclusive-locks to install the writes (i.e., conflict-induced abort) or some partitions crash (i.e., crash-induced abort).
Primo's approach is two-fold: (I) it acquires exclusive locks earlier during the execution phase to avoid conflict-induced aborts during the commit phase, and (II) it handles crash-induced aborts using distributed group commit. With these two strategies, 2PC becomes unnecessary.

Notice that (I) 
is different from upgrading shared locks to exclusive locks
because the lock upgrade approach would degenerate back to 2PL+2PC (that needs three roundtrips),
where the first roundtrip is to get the shared locks and values from the remote,
and the second roundtrip (the prepare phase of 2PC) is to  upgrade the lock and write the values to the remote,
and the last roundtrip is to propagate the commit/abort decision.
Instead, Primo acquires \textbf{exclusive-locks} for \textbf{read} records regardless of whether the read records would be updated later (bottom part of Figure \ref{fig:compare}). Therefore, given a transaction's read-set covering the write-set, 
(I) is achieved.
As a tradeoff, aggressively acquiring exclusive-locks for read operations could harm concurrency when most read records are not in the write-set. Nonetheless, we are able to mitigate this problem by employing TicToc \cite{tictoc}, an optimistic concurrency control (OCC) scheme that is more immune to the extra exclusive-locks for local transactions (more details in Section \ref{sec:waf}).

For (II) failure handling, Primo confirms durability in batches like COCO (Section \ref{sec:dgc}) for higher throughput at the cost of the increased latency and more aborts due to failure.
However, the typical added latency is only 10$ms$, which is negligible for end-users. For applications running on unstable machines that are more susceptible to failures, Primo can be configured to use smaller batch sizes for lower abort rates.
Unlike COCO \cite{coco} which synchronizes the entire cluster to process epochs, our approach enables partitions to make progress independently and only exchange a \emph{watermark} asynchronously.
This watermark is a threshold of the transaction timestamp (see Section \ref{sec:complete_waf}), which represents the progress of durable log writes, and helps partitions to abort the same set of  transactions in case of failure.

\section{Write-Conflict-Free Distributed Concurrency Control} \label{sec:waf}
In this section, we provide details of the write-conflict-free concurrency control (WCF) scheme.
We first introduce the basic form of WCF based on 2PL, and discuss how we combine OCC with it to mitigate the impact of the extra exclusive-locks. 
\added{
For lucidity, we assume a single server for each partition in this section and discuss replication in Section \ref{sec:watermark}.
} We also first assume each transaction satisfies $\texttt{write-set} \subseteq \texttt{read-set}$ and discuss how the violations can be handled seamlessly in Section \ref{sec:complete_waf}.

\subsection{The Basic Form: a 2PL Variant} \label{sec:waf_basic}
In Primo, a transaction $T$ is forwarded to the first partition that $T$ accesses, who serves as the coordinator and assigns $T$ a  \tid, which is made globally unique by combining the coordinator's server ID and its local transaction counter (incremented by one upon receiving a new transaction). 
The basic form of WCF processes $T$ in the following phases. 
\begin{enumerate}[leftmargin=1.2em]
    \item \textbf{Execution phase}. Like classic distributed transaction protocols, the coordinator executes the transaction commands, with only reads actually performed while writes are only buffered at the coordinator. Unlike 2PL, an \emph{exclusive-lock} is acquired for each read.
    \item \textbf{Commit phase}. Once all commands of $T$ are executed (all writes are also buffered), $T$ enters the commit phase. At this point, the coordinator installs the local writes and sends remote writes in batches to the corresponding participants. Each partition then installs the writes (the required exclusive-locks are already held by the transaction since $\texttt{write-set} \subseteq \texttt{read-set}$) and immediately releases locks without requiring any further communication. 
\end{enumerate}
By acquiring exclusive-locks for the read operations, Primo ensures that a distributed transaction does not encounter conflicts during the commit phase. 
In case $T$ fails to acquire locks in the execution phase (e.g., due to deadlock), Primo simply aborts $T$ because it does not make any effects yet.
This simple protocol readily yields better performance than 2PC-based schemes, especially in write-heavy workloads, by eliminating the major overhead of 2PC.
However, WCF may acquire extra exclusive-locks which could block innocent local transactions that contribute to the major portion of transactions in typical workloads (e.g., TPC-C \cite{tpcc}).
To address this issue, we combine WCF with an OCC scheme for local transactions, as described in the next section.

\subsection{The Complete Scheme: Combined with OCC} \label{sec:complete_waf}
Primo adopts an OCC scheme for processing local transactions.
In OCC, read operations of $T$ do not need any locks during execution, and $T$ goes through a validation phase that checks serializability and determines whether $T$ can commit. 
Therefore, local transactions using OCC are not blocked by exclusive locks during execution. Although exclusive locks may still affect the validation phase, we found that TicToc \cite{tictoc}, a recent OCC scheme, can be utilized to mitigate this issue.
It is worth noting that multi-version concurrency control (MVCC) schemes \cite{mvcc-1,ssi-1,ssi-2,cicada} could also address this problem, but maintaining multiple versions is an overhead.
We first provide a brief introduction of TicToc and discuss in detail why it is a perfect fit for working with WCF.

\subsubsection{Analysis of TicToc} \label{sec:tictoc}
The idea of TicToc is to determine whether a transaction can commit based on the timestamps attached to its accessed records. In TicToc, a record is associated with two \emph{logical timestamps}: a write timestamp (\texttt{wts}) and a read timestamp (\texttt{rts}), meaning that the record is valid to be read within the (logical) time interval of $[\texttt{wts}, \texttt{rts}]$ (i.e., the valid interval, and $\texttt{rts} \ge \texttt{wts}$).
As long as a transaction $T$ has a logical timestamp $ts$ within the valid intervals of all the records read, \emph{$T$ can be committed even if the records read are exclusively locked by other transactions}.
Notice that the logical timestamp is independent of the \tid~and irrelevant to the wall clock.
Specifically, $T$'s logical timestamp $ts$ is assigned based on the following constraints:
\begin{itemize}[leftmargin=2.5em]
    \item For each record $x$ read by $T$, $T.ts \ge x.\texttt{wts}$. Intuitively, that's because $T$ reads $x$ after $x$ is written.
    \item For each record $y$ written by $T$, $T.ts > y.\texttt{rts}$. Intuitively, that's because $T$ overwrites $y$ after the last time when $y$ is read.
\end{itemize}
$T$'s logical timestamp $ts$ is then assigned to be the minimal number that satisfies the above constraints.
TicToc validates the transaction $T$ by (1) locking all the records in $T$'s write-set, (2) computing $T$'s logical timestamp $ts$ based on the above constraints, and (3) checking whether $ts$ is within the valid interval of all records read by $T$.
In the case that a read record $x$ has $x.\texttt{rts} < T.ts$, $T$ tries to update $x.\texttt{rts} = T.ts$ such that $T.ts \in [\texttt{wts}, \texttt{rts}]$.
\ul{Only when $T$ \textbf{updates the \texttt{rts}} of a read record, an exclusive-lock on the record held by other transactions could cause $T$ to abort.} 
With that, we observe in the experiments that WCF only cause less than 2\% of extra local transactions to abort even in a read-heavy workload. 
Intuitively, that's because once $T$ has increased the \texttt{rts} of $x$, the prolonged valid interval of $x$ allows more transactions to commit.
Therefore, with local transactions employing TicToc, the impact of the extra exclusive-locks on local transactions can largely be mitigated.
Although write operations could still conflict with exclusive-locks, that's not the problem of WCF because write operations intrinsically conflict with all the operations of other transactions.
Notice that TicToc alone still requires 2PC to process distributed transactions, which prolongs the contention footprint.
Thus, combining TicToc with WCF obtains the best of both worlds.

\subsubsection{Put together}

We are now ready to introduce the complete concurrency control scheme in Primo. Primo distinguishes between local transactions and distributed transactions, with local transactions following TicToc, and distributed transactions employing the WCF. In fact, for general transactions, it is not possible to determine whether a transaction $T$ is distributed before execution. Therefore, Primo first executes $T$ in local mode and switches to the distributed mode when $T$ performs its first remote access.

\begin{algorithm}
\footnotesize
\SetAlgoLined

\Pn{Execution phase}{
\LeftComment{Execute the transaction commands, which triggers the events:}\\
\Event{$read (\texttt{key}$)}{
    \uIf{\texttt{key} is local}{
        $\texttt{record} = find\_record(\texttt{key})$ \\
        $X\_Lock(record)$ \Comment{Abort if fail} \\
    }\uElse{
        $\texttt{record} = remote\_read(\texttt{key})$ \Comment{Abort if fail} \\
    }
    $\texttt{read\_set}.append(\texttt{record})$ \\
    $\KwReturn~ \texttt{record}$
}
\Event{write (\texttt{key}, \texttt{value})\Comment{Writes are only buffered }}{ 
    $\texttt{write\_set}.append(\text{\textless}\texttt{key}, \texttt{value}\text{\textgreater})$
}
}
\Pn{Commit phase}{
\HiLi$ts \gets compute\_ts()$ \Comment{Same as TicToc} \\
\HiLi\ForEach{local read r $\in$ \texttt{read\_set}}{
\HiLi\If{$r.\texttt{rts} < ts$}{
         \HiLi$r.\texttt{rts} = ts$
}
}

\ForEach{local write \text{\textless}\texttt{key}, \texttt{value}\text{\textgreater} $\;\in$ \texttt{write\_set}} {
    $\texttt{record} = find\_record(\texttt{key})$ \\
    $\texttt{record}.\texttt{value} = \texttt{value}$ \\
\HiLi$\texttt{record}.\text{\textless}\texttt{wts}, \texttt{rts}\text{\textgreater} = \text{\textless} ts, ts \text{\textgreater}$
}
$unlock(\{\text{all local records} \})$ \\
\ForEach{participant $P$}{
    $P_{write} = \texttt{write\_set}.filter(partition == P) $\\
    $send(P, \mathcolorbox{yellow!50}{ts}, P_{write})$
}
}
\caption{WCF at the coordinator (distributed mode)}
\label{alg:waf}
\end{algorithm}

Algorithm \ref{alg:waf} shows the pseudocode of processing a distributed transaction $T$ at the coordinator. The detail of switching to the distributed mode is omitted here, and we illustrate that with an example later.
The algorithm is largely identical to the basic form introduced in Section \ref{sec:waf_basic}, except for the highlighted part that mimics TicToc to maintain the logical timestamps such that TicToc and WCF can work in harmony.
Specifically, the execution phase executes the transaction commands with each read record exclusively locked (Line 6, and the same at Line 8 when the participant receives the remote read), and each write is buffered (Line 13).
In the commit phase, $T$ calculates its logical timestamp $ts$ in the same way as TicToc (Line 17). 
Unlike TicToc, $T$ needs no validation phase because it can always prolong the valid intervals of its read records (Lines 18 -- 20) since $T$ holds exclusive-locks on them.
After that, $T$ installs local writes (Lines 23 -- 26) and releases locks on the local records (Line 28).
For the remote writes, they are sent to the corresponding participants along with the logical timestamp $ts$ (Line 31).

The mechanism for each participant $P$ is similar. On receiving a remote read (sent from the coordinator in Line 8), $P$ acquires an exclusive-lock on the read record for $T$ and returns the record (including the \texttt{wts} and \texttt{rts}) to the coordinator. On receiving the write-set from the coordinator (Line 30), $P$ first prolongs the valid interval of $T$'s read records on $P$ and installs the writes. After that, $P$ releases $T$'s locks. We omit the pseudocode on the participants since it largely overlaps with Algorithm \ref{alg:waf}.

\vspace{1mm}
\stitle{Example.} Figure \ref{fig:example} shows an example. where a transaction $T$ reads $x$, $y$ and $z$, and then updates $y$ and $z$ ($x$, $y$, and $z$ are located at partitions 1, 2, and 3, respectively).
Partition 1 serves as the coordinator since it owns $T$'s first accessed record $x$.
On partition 1, $T$ is first executed in the local mode (TicToc), and it reads $x$ whose $\texttt{wts}=1$ and $\texttt{rts}=3$.
$T$ switches to the distributed mode (WCF) when it attempts to perform remote reads. Before that, it acquires the exclusive-lock on $x$ and checks whether $x$ has changed ($x$ maybe changed since it is first read under the local mode, which does not hold a lock). If $x$ is changed, $T$ aborts and retries directly in the distributed mode. Otherwise, $T$ successfully enters the distributed mode and reads $y$ and $z$ from remote partitions. The participants acquire the corresponding exclusive-locks, and return the results including the logical timestamps.
Following the transaction logic, $T$ computes its write-set (in this case, $\{y=10$, $z=20\}$), and then enters the commit phase.
In the commit phase, the coordinator computes $T$'s logical timestamp, which should be larger than the read record $x$'s \texttt{wts}, and the write records $y$ and $z$'s \texttt{rts}, resulting in $ts = 5$.
Since $ts$ does not fit in $x$'s valid interval, the coordinator updates $x.\texttt{rts}=ts=5$.
After that, the coordinator unlocks $x$ and sends the remote writes along with $ts=5$.
On receiving the remote write on $y$, partition 2 installs $y=10$, sets its $\texttt{rts} = \texttt{wts} = 5$ and unlocks $y$. Partition 3 performs similar operations.

\icde{
\begin{figure}
    \centering
    \includegraphics[width=\linewidth]{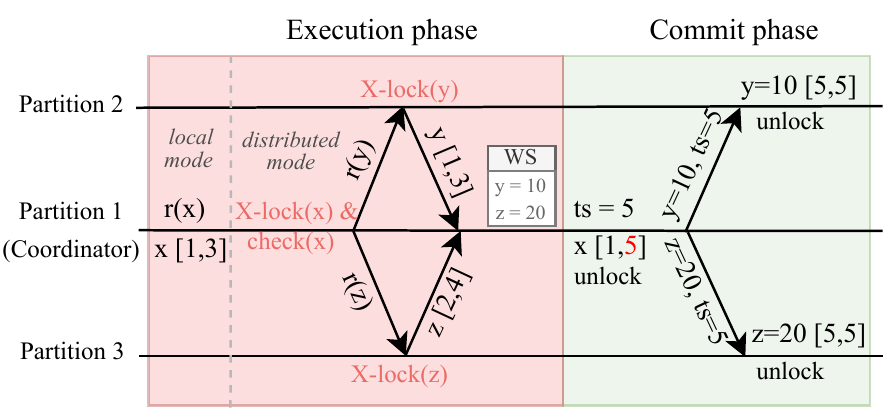}
    \caption{An example of WCF, where $T$ reads $x,y,z$ and updates $y,z$.} \vspace{-1em}
    \label{fig:example}
\end{figure}
}
\techrep{
\begin{figure}
    \centering
    \includegraphics[width=0.8\linewidth]{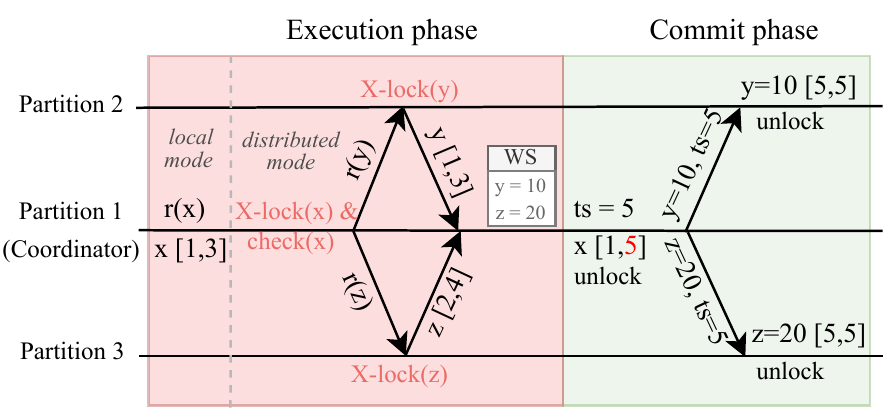}
    \caption{An example of WCF, where $T$ reads $x,y,z$ and updates $y,z$.}
    \label{fig:example}
\end{figure}
}

\vspace{1mm}
\stitle{Deadlock Prevention.}
To avoid deadlock, we adopt the common \texttt{WAIT\_DIE} policy \cite{2pl}.
Specifically,
when $T_i$ ($\tid=i$) acquires the lock that $T_j$ is holding,
$T_i$ waits for $T_j$ only if $i < j$. Otherwise, $T_i$ is aborted.
In this way, 
the wait-for relationship is always unidirectional
such that no deadlock would occur.
In Primo, if one of $T$'s read operations fails to acquire the exclusive-lock (to avoid deadlock), $T$ is aborted.
In this case, the coordinator sends an \texttt{ABORT} message to all the participants to abort $T$. Since WCF only acquires locks in the execution phase, no deadlock would occur in the commit phase.

\added{
\vspace{1mm}
\stitle{Blind-write Handling.}
Primo can handle blind-writes seamlessly at runtime by adding dummy reads.
Specifically, when a write is appended to the write-set (Line 13 of Algorithm \ref{alg:waf}), Primo checks whether it is in the read-set. If not, a dummy read is performed which acquires the exclusive-lock without actually reading the data. 
If the dummy read is a remote request, it can be batched with a normal remote read such that no extra roundtrip is required.
Nonetheless, when there is no normal remote read to batch with, an extra roundtrip is required. 
Notice that the dummy read would not add extra read-write dependencies to TicToc, since its purpose is only for acquiring an exclusive-lock for the blind-write.

\icde{
\vspace{1mm}
\stitle{Corner Cases.} We provide detailed descriptions on how to handle some corner cases including predicate read, constraint check, user-specified abort and read-only transactions in our technical report \cite{techreport}. Essentially, they all can be seamlessly handled without fundamental modifications to the protocol.
}
}
\techrep{
\stitle{Corner Cases.}
Some corner cases are not explicitly handled in the above description, and we discuss them individually here.

\mybullet{Predicate read.} Primo handles predicate read using predicate-locks like other protocols \cite{silo}, except that the predicate-lock in Primo is exclusive. In addition, predicate read may be caused by large scan operations. In this case, Primo uses normal shared predicate-locks and fallback to 2PC to avoid such operations from aggressively blocking other transactions.

\mybullet{Constraint checking.} Constraint checking (e.g., checking unique key when doing insertions) is in fact read operations following predicates on the read results (e.g., read and check whether a key exists). Therefore, constraints are checked in the execution phase similarly to the read operations.

\mybullet{User-specified abort.} The stored procedure may explicitly invoke \texttt{Rollback} command to abort the transaction (or the user calls \texttt{Rollback} in an interactive transaction). Since transaction commands are all executed in the execution phase, the coordinator can simply handle it by sending an \texttt{ABORT} message to all the participants to abort the transaction. 

\mybullet{Read-only transactions.} A transaction can be identified to be read-only if it only invokes a stored procedure that contains no \texttt{UPDATE} or \texttt{INSERT}. Primo can optimize read-only transactions with snapshots \cite{silo} without requiring any lock.

}

\subsection{Analysis} \label{sec:analysis}

\stitle{Performance.} 
To gain a better understanding of the tradeoff in WCF, we conducted a theoretical analysis to compare its performance with 2PC-based schemes. To avoid overwhelming the main text with mathematical equations, we provide the details in \icde{our technical report \cite{techreport}}\techrep{Appendix \ref{appendix}} and only present our conclusions here. Our theoretical analysis demonstrates that Primo can outperform 2PC-based schemes in most cases, with the improvement being amplified by the contention degree and the ratio of distributed transactions. Intuitively, this is because Primo eliminates the need for 2PC, and the impact of the extra exclusive locks on local transactions is mitigated by employing TicToc. However, there is one exception: the performance of Primo is not as good as 2PC-based schemes in a read-heavy and mostly-distributed workload (e.g., 80\% distributed with 90\% read) due to the large number of extra exclusive locks required in this setting. In fact, since read-heavy workloads have low contention, the contention footprint induced by 2PC is less of an issue. Therefore, Primo can fall back to 2PC in read-heavy workloads to avoid unnecessary overhead.

\vspace{1mm}
\stitle{Serializability.} 
\icde{Since WCF is based on 2PL and TicToc, it also offers serializability like its predecessors. We provide the proof in our technical report \cite{techreport}.}
\techrep{
We prove the following claim regarding the correctness of WCF:

\begin{claim}
WCF (combined with TicToc) ensures serializability.
\end{claim}

\begin{proof}
Given the fact that TicToc's validation phase ensures serializability, we prove the claim by showing all committed transactions in Primo effectively pass the same validation.
For local transactions, since they use TicToc directly, all committed local transactions pass the validation phase.
For distributed transactions, although they are not explicitly validated, Line 18 -- 20 of Algorithm \ref{alg:waf} ensures each distributed transaction has $ts$ within the valid intervals of all its read records, and thus they are guaranteed to pass TicToc's validation.
Overall, all committed transactions in Primo can pass TicToc's validation, and hence it provides serializability like TicToc does.
\end{proof}
}

\section{Watermark-based Distributed Group Commit} \label{sec:watermark}
While watermark has been widely used in stream processing \cite{flink-1, flink-2} and concurrency control \cite{ov1,ov2,centiman} to represent the system's progress, 
we found the concept useful for deriving an asynchronous distributed group commit scheme for handling the crash-induced aborts.
Our watermark-based distributed group commit (WM) is scalable in that it allows each partition to make progress \emph{independently}, with partitions communicating their progress \emph{asynchronously} via watermarks.
In Primo, %
the \emph{global-watermark} $W_g$ is a threshold of the logical timestamp that satisfies the following requirements:
\begin{itemize}[leftmargin=2em]
    \item \emph{Monotonicity}: each partition sees a monotonically increasing $W_g$.
    \item \emph{Durability}: every transaction with $ts < W_g$ is durable on all the involved partitions.
    \item \emph{Consistency}: the transactions with $ts < W_g$ form a prefix of the serial history of the database. %
\end{itemize}
The monotonicity requirement is essential to ensure that the system always makes progress, as is the case with other watermark-based protocols. The durability and consistency requirements are unique to distributed group commit.
With the durability guarantee, Primo can return the result of a transaction $T$ once the global watermark is raised beyond its $ts$ without the problem of cascading aborts.
The consistency requirement is also crucial for Primo to recover to a consistent state after a crash, as the recovered state must be produced by a serializable execution.
We first introduce the protocol in normal operation and then discuss how Primo recovers from failures.

\subsection{Normal Operation}
Like many shared-nothing databases \cite{f1,spanner,cockroachdb,oceanbase}, each partition in Primo has a leader that processes transactions and replicates transaction logs to replicas through a consensus protocol (e.g., Raft \cite{raft}). A transaction is persisted on the partition once its log records are replicated to a quorum of the partition replicas.
During normal operation, 
the partition leaders execute transactions as described in Section \ref{sec:complete_waf}, where locks are released immediately after installing the write-sets without waiting for their log records to be persisted.
Each partition leader persists the log records asynchronously, and it periodically generates a partition-watermark $W_p$ indicating that transactions with $ts < W_p$ are persisted on that partition. 
After learning the other partitions' watermarks, the minimum of them is regarded as the global-watermark $W_g$. This makes transactions with $ts < W_g$ durable because all the involved partitions have persisted them.
Different from COCO, the generation of watermarks requires no synchronization.

\vspace{1mm}
\stitle{Partition-Watermarks.}
In Primo, each partition leader $L$ \emph{independently} generates and persists a partition-watermark $W_p$ every $t_m$ interval.
There are two requirements regarding $W_p$. (R1) for durability, $W_p$
must be smaller than the minimum $ts$ of the active transactions on $L$; (R2) for monotonicity, once a new $W_p$ is generated, all transactions thereafter should have $ts > W_p$.
R1 is tricky because some active transactions may not be assigned a $ts$ yet. The violation of R2 can happen if a new transaction $T$ only accesses cold records that are rarely accessed and thus have small \texttt{wts} and \texttt{rts}. In this case, $T$ may be assigned with a small $ts < W_p$.

To ensure R1, for a transaction that has not determined $ts$, the partition leader $L$ assigns it a lower bound of the logical timestamp $l_{ts}$, which is set to the \texttt{wts} of its first accessed record on $L$.
The logical timestamp of $T$ is then guaranteed to be no smaller than $l_{ts}$ because the two constraints of computing $ts$ discussed in Section \ref{sec:tictoc} enforce $T.ts > \texttt{wts}$. With that, Primo determines $W_p$ by the following rule:

\begin{myrule}
For every $t_m$ interval, a partition leader $L$ sets $W_p$ to the minimum $\texttt{ts}$ (or $l_{ts}$ if $\texttt{ts}$ is not available) of the active transactions on that partition.
\end{myrule}

For efficiency, each worker-thread in Primo independently maintains a list of active transactions sorted by their $ts$ (or $l_{ts}$ if $ts$ is not available), and thus $W_p$ can be determined by examining the heads of the sorted-lists. 

To ensure R2, 
we consider two cases: (1) the partition leader $L$ is the coordinator of the transaction $T$ (including the case that $T$ is a local transaction on $L$); and (2) $L$ is a participant of $T$.
For (1), $L$ is the one who calculates $T$'s logical timestamp $ts$. Thus, it is able to add one more constraint (besides the constraints discussed in Section \ref{sec:tictoc}) when determining $ts$, i.e., $ts$ must be greater than the current $W_p$.
For (2), $L$ cannot directly determine $T$'s logical timestamp because it is assigned by the coordinator. Fortunately, $L$ is able to modify the \texttt{wts} and \texttt{rts} of the record that $T$ accesses on $L$, which affects $T$'s logical timestamp at the end. Specifically, when $T$ remote-reads $x$ on $L$, if $x.\texttt{wts} \le W_p$, $L$ sets $x.\texttt{wts}=x.\texttt{rts} = W_p+1$ before returning the results.
Since $T$ holds the exclusive-lock on $x$, updating the logical timestamps of $x$ can always succeed. In this way, the final $ts$ of $T$ would be greater than $W_p$.
Enforcing the logical timestamps to be greater than $W_p$ ensures $W_p$ grows monotonically (hence each partition always makes progress), but it
could cause some local transactions to abort because $W_p$ may exceed the valid interval of their read records. 
Nonetheless, $W_p$ is only upraised every $t_m$ interval, and thus it does not hurt much performance. Once $W_p$ is determined, $L$ persists a log record containing $W_p$ to notify its replicas and broadcasts $W_p$ to other partitions.

\added{
In case a partition is idle or less busy than the others, its partition watermarks may grow slower and could detain the progress of the global-watermark (see below). To address that, the partition leader enforces its transactions to have $ts > W_p + \Delta$ (like how we enforce the R2) if its partition-watermark $W_p$ is smaller than the average of the other partitions' partition-watermarks, where $\Delta$ is the difference between $W_p$ and the average value. If there is no transaction on the partition, the partition leader simply increases its $W_p$ by $\Delta$ every $t_m$ (if its $W_p$ is smaller than the average of the other partitions' $W_p$).
}

\icde{
\begin{figure}
    \centering
    \includegraphics[width=\linewidth]{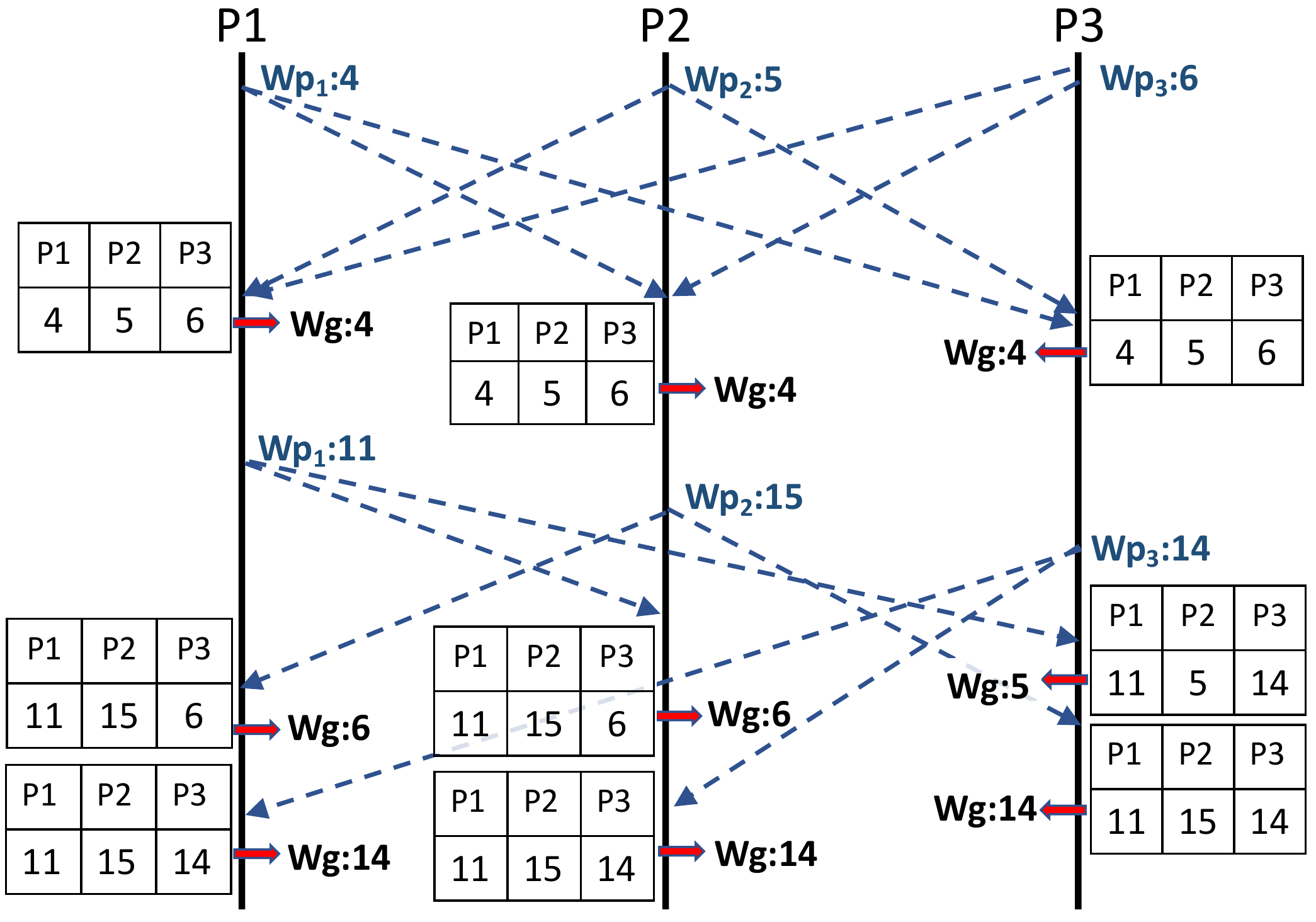}
    \caption{An example run of the watermark-based distributed group commit}
    \label{fig:watermark}
\end{figure}
}
\techrep{
\begin{figure}
    \centering
    \includegraphics[width=0.7\linewidth]{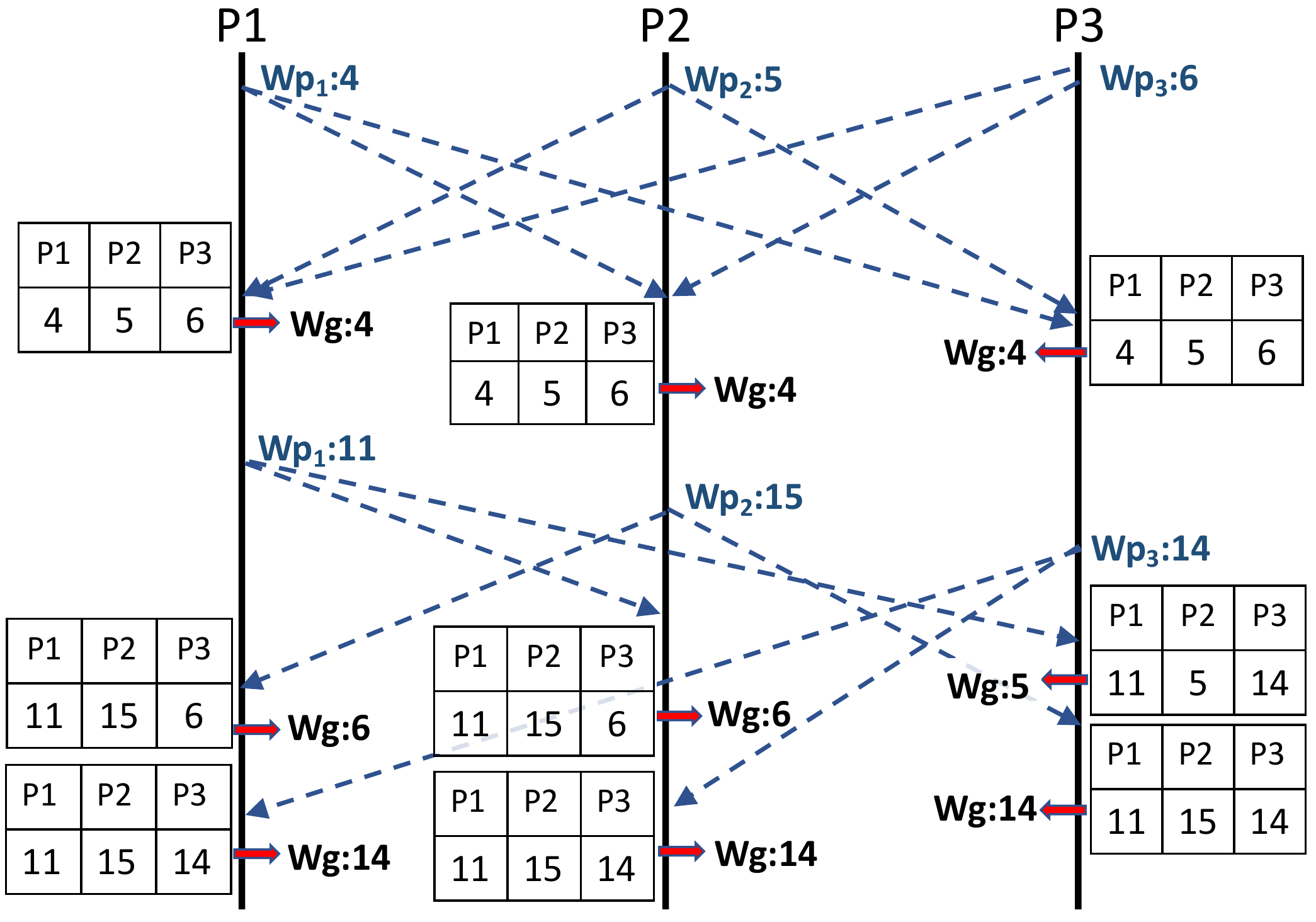}
    \caption{An example run of the watermark-based distributed group commit}
    \label{fig:watermark}
\end{figure}
}

\vspace{1mm}
\stitle{Global-Watermarks.}
Each partition leader $L$ learns the latest partition-watermarks of other partitions by receiving their messages.
With that, $L$ calculates a global-watermark $W_g$ which is the minimum of all the latest partition-watermarks.
After that, $L$ returns the results of the transactions with $ts < W_g$ to the clients.
Figure \ref{fig:watermark} shows an example. As shown in the figure, each partition leader maintains a table of the lastly received partition-watermarks of all the partitions. On receiving a new partition-watermark, the table is updated and the global-watermark is computed by taking the minimum number in the table. For example, after the leader of partition P1 receives $W_{p_2}=5$ and $W_{p_3}=6$, combined with its own partition-watermark $W_{p_1}=4$, it calculates $W_g=4$.
The partition-watermarks do not have to be broadcast synchronously (e.g., the leader of P3 may broadcast $W_{p_3}=14$ later than the other partitions), and the messages are also allowed to be delayed. Nonetheless, each partition would not block-waiting messages from other partitions because each of them processes and persists transactions independently and only communicate watermarks asynchronously. %

\techrep{
We next show this mechanism generates desired global-watermarks by proving the following claim:

\begin{claim}
    $W_g$ satisfies the monotonicity, durability, and consistency requirements.
\end{claim}

\begin{proof}
(1) Monotonicity: since the partition-watermarks grow monotonically, the minimum of them (i.e., $W_g$) also grows monotonically.
(2) Durability: given the fact that $W_g$ is no greater than the latest partition-watermarks of all the partitions,
each transaction $T$ with $ts < W_g$ is persisted on all the partitions.
(3) Consistency: we prove by contradiction. Suppose the transactions with $ts < W_g$ does not form a prefix of the serial history. This only happens when there exists $T_1$ whose $ts < W_g$ depends on another transaction $T_2$ whose $ts \ge W_g$ because Primo ensures serializability during normal operation. However, it is not possible by how logical timestamps are calculated. Specifically, if $T_1$ reads $T_2$'s write records, since these records have $\texttt{wts} = T_2.ts$, $T_1$ would be assigned a logical timestamp no smaller than the logical timestamp of $T_2$. A similar argument holds when $T_1$ writes $T_2$'s accessed records. Overall, the consistency requirement is satisfied.
\end{proof}
}

\stitle{Limitations and Discussions.}
Many epoch-based protocols \cite{silo,coco,calvin} suffer from stragglers and WM inherits similar problems. 
These problems may arise due to a long-running transaction or a partition leader that is slow to respond. However, long-running transactions in Primo only increase latency by detaining the growth of the watermarks, without hurting throughput because there is no global synchronization at the epoch boundary; Hence, the execution of the transactions is not detained even if the growth of $W_g$ is detained.
If a partition leader is slow to respond, leader re-election is triggered to allow a healthy replica to take over.
\revision{
Another limitation of Primo is disallowing unilateral aborts from the clients/coordinator in the commit phase to avoid 2PC. 
In fact, 
this aligns with recent 2PC-based protocols (e.g., MDCC \cite{mdcc}, TAPIR \cite{TAPIR}), and is acceptable in commercial products (e.g., CockraochDB \cite{cockroachdb}). \icde{For lucidity, we provide proof that $W_g$ satisfies the monotonicity, durability, and consistency requirements in our technical report \cite{techreport}.}
}

\subsection{Recovery}

Primo utilizes a membership service (e.g., Zookeeper \cite{zookeeper}) to detect server crashes and coordinate recovery. \added{
In practice, failure detection in Zookeeper can be augmented with server response time and hardware monitors to detect a wider range of failures (e.g., gray failures \cite{gray-failure}).}
In Primo, each server is a member of the Zookeeper cluster and once Zookeeper detects the failure of a partition leader, it triggers the recovery process.
Notice that Zookeeper would not affect the performance during normal operation because
the servers only need to send heartbeats to the Zookeeper leader regularly.
We discuss the mechanisms of the recovery process as follows.

\vspace{1mm}
\stitle{Recover Failed Partitions.}
To recover, the failed partition (whose leader crashes) triggers leader re-election in the replication group following the Raft protocol \cite{raft}. After that, the new leader retrieves the latest $W_p$ in its Raft log.
Notice that Raft consensus protocol ensures that the new leader has replicated all the transactions with $ts < W_p$ because the old leader has persisted them before persisting $W_p$ in the replication group.
The crashed server can recover from the log and rejoin the replication group.
To bound the time for recovering the crashed server, each server periodically persists a checkpoint of its state to a local disk such that it only needs to apply the redo log after the checkpoint to recover.

\stitle{Rollback.}
Only in this step, Primo synchronizes partitions for aborting transactions caused by failure.
To do so, partition leaders (1) agree on a global-watermark $\mathcal{W}_g$ and then (2) rollback the transactions with $ts \ge \mathcal{W}_g$.
The reason for (1) is that each partition maintains its own view of the global-watermark asynchronously during normal operation, and their views could diverge due to network delay.
To agree on a global-watermark, each partition leader publishes its latest $W_p$ on Zookeeper (attached with a \texttt{TERM-ID} to differentiate different runs of recovery). After that,
all the partitions adopt the maximum value as the agreed global-watermark $\mathcal{W}_g$.
The consensus mechanism in Zookeeper ensures all the partitions adopt the same $\mathcal{W}_g$.
The transactions with $ts \ge \mathcal{W}_g$ are then safe to rollback because their results are not returned to the clients yet. 
After rollback, Primo continues normal operations.
In case another failure occurs during recovery, the recovery process is restarted with a new \texttt{TERM-ID}.

\added{
\subsection{Correctness.}
\icde{
Combining WCF and WM, Primo guarantees agreement, validity, and termination properties like classic atomic commit protocols. 
We provided complete proof in our technical report \cite{techreport}.
}
}
\techrep{
Combining WCF and WM, Primo guarantees agreement, validity, and termination properties like classic atomic commit protocols. 
\begin{itemize}[leftmargin=1.2em]
    \item Agreement: no two servers decide on different values (i.e., commit or abort).
    \item Validity: the value decided upon must be proposed by some server in the cluster.
    \item Termination: all non-failed servers must eventually decide.
\end{itemize}

\begin{claim}
    Primo satisfies the agreement property.
\end{claim}

\begin{proof}
Since the replicas in each partition follow the decision of the partition leader by a consensus protocol, we only need to prove agreement among partition leaders. We show that by contradiction. Assume the partition leader $L_1$ decides to commit $T$ while the leader of another partition $L_2$ aborts $T$. Since $L_1$ commits $T$, the coordinator of $T$ must have initiated the commit phase of $T$. With WCF, no conflict can happen in the commit phase and thus $L_2$ can only abort $T$ due to failure, and all the partition leaders agree on an $\mathcal{W}_g \le T.ts$ in the recovery process. However, that is impossible because in that case, $L_1$ should have aborted $T$.
\end{proof}

\begin{claim}
    Primo satisfies the validity property.
\end{claim}
\begin{proof}
Primo trivially satisfies validity because the commit/abort decision of $T$ is either proposed by the coordinator or when the cluster agrees on a global-watermark $\mathcal{W}_g$ during recovery. For the former, the value is explicitly proposed by the coordinator; and for the latter, the value is implicitly proposed depending on whether $T.ts < \mathcal{W}_g$ (i.e., commit) or not (i.e., abort).
\end{proof}

\begin{claim}
    Primo satisfies the termination property.
\end{claim}
\begin{proof}
For a transaction $T$,
since deadlock is resolved by \texttt{WAIT\_DIE} policy, $T$ eventually either aborts in the execution phase (due to potential deadlock) or enters the commit phase. For the former, all the involved partitions would eventually be notified and abort $T$ (the coordinator sends the \texttt{ABORT} messages to the participants, or the participants consult the replicas in the coordinator's partition if the coordinator crashes). For the latter, $T$ waits for $W_g$ to grow beyond its $ts$.
If there is no crash, $W_g$ would eventually be upraised beyond its $ts$ (i.e., commit $T$). That is because, with $T.ts$ being a finite constant and each partition grows $W_p$ at least by one for every $t_m$ interval, $W_g$ would eventually become larger than $T.ts$.
In case of failure, the recovery process agrees on a $\mathcal{W}_g$ which either makes $T$ to commit (if $T.ts < \mathcal{W}_g$) or abort (if $T.ts \ge \mathcal{W}_g$).
\end{proof}
}

%% file: experiments.tex
\section{Evaluation} \label{sec:exp}
In this section, we study the performance of Primo and compare it with state-of-the-art protocols in various aspects.

\subsection{Experimental Setup}
We conduct experiments on a cluster of \texttt{i4p.8xlarge} servers on Alibaba Cloud. Each server has a 2.9GHz Intel Xeon Platinum 8369B CPU (16 cores $\times$ 2 HT), 128GB DRAM, and 126GB SSD. Servers are connected with 16 Gbps Ethernet. Each server runs 16-bit CentOS 7.0 and we compile using GCC 4.8.5 with \texttt{-O3} option enabled.
\subsubsection{Competitors} We include \icde{five}\techrep{six} distributed transaction protocols for comparison.

\begin{itemize}[leftmargin=1.2em]
    \item \textbf{2PL(NW)}: A protocol similar to Spanner \cite{spanner} that combines 2PL with 2PC introduced in Section \ref{sec:2pc}. It adopts \texttt{NO\_WAIT} policy \cite{2pl} to avoid deadlock (i.e., a transaction is aborted if it acquires a lock that is held by another transaction).
    \item \textbf{2PL(WD)}: Another variant of 2PL+2PC but using \texttt{WAIT\_DIE} policy \cite{2pl} for deadlock prevention.
    \item \textbf{Silo}: An optimistic scheme first introduced in \cite{silo}, and we use its state-of-the-art distributed variant introduced in COCO \cite{coco}.
    \item \textbf{Sundial}: A state-of-the-art optimistic scheme based on TicToc \cite{tictoc}. It is augmented with several optimizations for the distributed settings in \cite{sundial}.
    \revision{
    \marginnote{\META{Meta3}\RI{R1O4}\RII{R2O4}}
    \item \textbf{Aria} \cite{aria}: A state-of-the-art deterministic database that does not assume prior knowledge of read-write sets. It executes transactions in batches and saves logging overhead by determinism. It still requires 2PC-like communication for each batch (see Section \ref{sec:background}).}
    \techrep{
    \item \added{\textbf{TAPIR}: A representative of the works that co-optimize 2PC with replication. It removes unnecessary consistency requirements in the replication protocol to improve the performance of distributed transactions \cite{TAPIR}.}}
\end{itemize}

\techrep{
\noindent We also include two logging optimizations as baselines to our watermark-based distributed group commit (WM).
\begin{itemize}[leftmargin=1.2em]
    \item \textbf{COCO}: The state-of-the-art distributed group commit protocol introduced in Section \ref{sec:dgc}. 
    \item \textbf{CLV}: Controlled-Lock-Violation \cite{elr-1} is a fine-grained logging scheme which moves logging out of the critical path like COCO and WM. CLV is similar to Early-Lock-Release (ELR) \cite{elr-2} and we choose CLV as it is reported to perform better than ELR \cite{elr-1}.
\end{itemize}
}

\icde{
\noindent Since Primo employs distributed group commit, we also combine 2PL(NW), 2PL(WD), Silo, and Sundial with the state-of-the-art distributed group commit protocol introduced in \textbf{COCO}.
Aria does not need that because as a deterministic database, it ensures durability by logging transaction commands in a sequencing layer.
}

\added{
The baselines are incorporated with common 2PC optimizations including Presumed-Abort \cite{presumed-opt-1} and Unsolicited-Vote \cite{ingress}.}
Deterministic databases other than Aria \cite{aria,calvin,bohm,pwv,slog,ov2,deter-case,deter-eval,deter-overview,caracal,quecc} 
and access localization protocols \cite{leap,zeus} are not listed for comparison due to their limitations (e.g, read-write sets must be pre-determined, data location seldom changes). Star \cite{star} and Lotus \cite{lotus} also require prior knowledge of transactions (e.g., whether a transaction is local or distributed) to multiplex their execution paths and thus also not included. \icde{We also compared with TAPIR \cite{TAPIR} and showed Primo has 4.07$\times$ to 8.25$\times$ higher throughput over TAPIR in our technical report \cite{techreport}.}

\subsubsection{Workloads}
Our experiments are performed using two popular OLTP workloads.
\begin{itemize}[leftmargin=1.2em]
    \item \textbf{YCSB} \cite{ycsb}. Following \cite{ccbench,aria}, each transaction contains 10 key accesses following the Zipf distribution \cite{zipf}.
    By default, there are 5 reads and 5 read-modify-write operations in each transaction, and the $\texttt{skewness}$ of the Zipf distribution is set to 0.6 (medium contention). We place 1M keys per partition, and 20\% of the transactions are distributed transactions by default.
    The impact of these workload parameters is studied in Section \ref{sec:param}.
    \item \textbf{TPC-C} \cite{tpcc}. A benchmark that models a realistic wholesale business. Following the specification, 10\% of \texttt{NewOrder} and 15\% of \texttt{Payment} transactions access a remote warehouse.
    By default, there are 16 warehouses per partition, and the impact of this number is studied in Section \ref{sec:wh}.
\end{itemize}

\subsubsection{Implementation and configuration}
\icde{
For fair comparisons, all the competitors are implemented using the same code framework, DBx1000 \cite{dbx1000}.} \techrep{For fair comparisons, Primo, 2PL(NW), 2PL(WD), Silo, Sundial, \revision{Aria}, COCO, and CLV are all implemented in the same framework of DBx1000 \cite{dbx1000}.}\techrep{Since Primo employs distributed group commit, we also combine 2PL(NW), 2PL(WD), Silo, and Sundial with the state-of-the-art distributed group commit protocol introduced in COCO.}
Aria, as a deterministic database ensures durability by logging transaction commands in a sequencing layer.
However, that is not implemented in Aria's open-source code \cite{aria-code}. Hence
we adopted a typical sequencing layer provided by another deterministic database Calvin\cite{calvin}. Like in Calvin, the sequencing layer uses 10ms epochs to collect client requests. We confirmed that adding such a sequencing layer does not affect Aria's throughput since it is not the bottleneck.
\techrep{For TAPIR, due to the architectural difference (it uses a special inconsistent storage layer), it is hard to unify it in the same framework as the others. Hence, we directly use its open-source implementation. Due to the same reason, we use a separate Section \ref{sec:tapir} to better interpret the results of the comparison with TAPIR.
}

\techrep{
We optimized DBx1000 to allow each worker-thread to initiate a new transaction when the running transaction is waiting for a lock or a remote response.
We also add the exponential backoff strategy \cite{opportunities} for retrying aborted transactions. Specifically, if a transaction aborts, the initial backoff time is 0.5ms and the duration doubles every time it aborts again.
}

In our experiments, each partition contains three replicas and there are four partitions by default. We scale up to 20 partitions in Section \ref{sec:scale}.
Since the latency is tunable with the distributed group commit (i.e., by tuning the parameters in COCO, and tuning the interval $t_m$ of generating the partition-watermarks in WM), we tune the average latency of them to around 10ms by default for fair comparisons. \revision{The batch size of Aria is tuned to optimal in all experiments.}
All the reported results are the average of five runs.

\techrep{
\begin{figure*}
    \centering
    \begin{minipage}{0.9\linewidth}
    \begin{subfigure}[b]{0.45\textwidth}
    \centering
         \includegraphics[width=\textwidth]{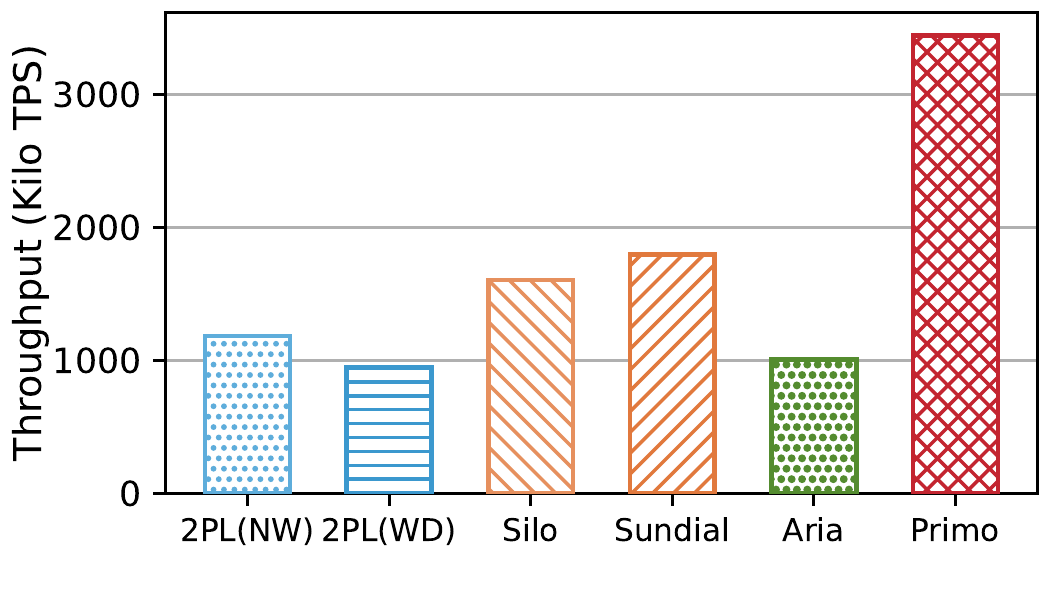}
         \caption{Throughput}
         \label{fig:ycsb_overall}
    \end{subfigure}
    \hfill
    \begin{subfigure}[b]{0.45\textwidth}
    \centering
         \includegraphics[width=\textwidth]{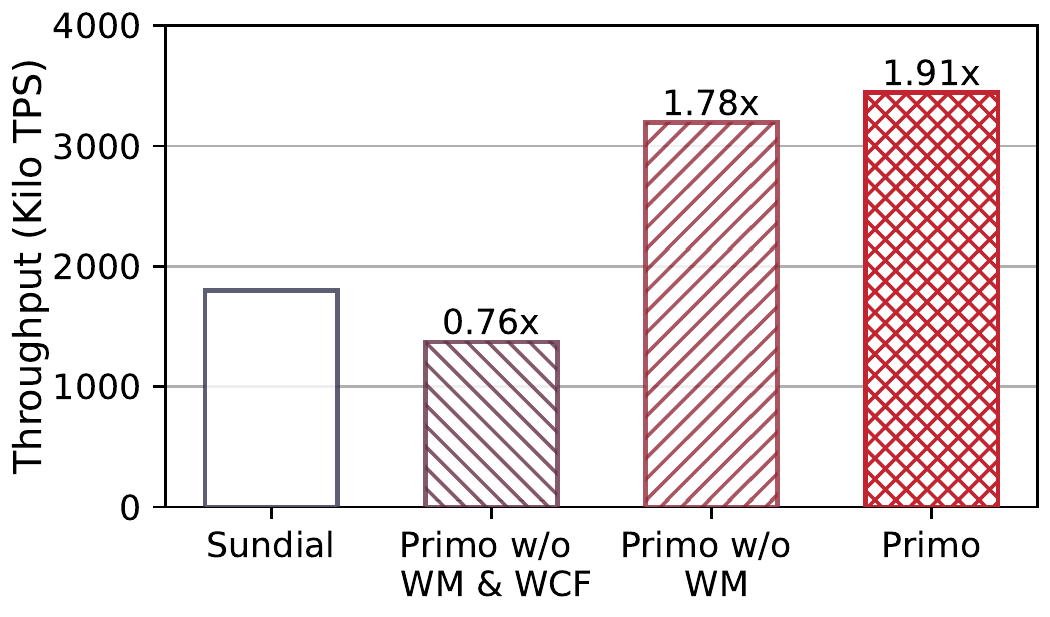}
         \caption{Factor breakdown}
         \label{fig:ycsb_factor}
    \end{subfigure}
    \\
    \begin{subfigure}[b]{0.45\textwidth}
    \centering
         \includegraphics[width=\textwidth]{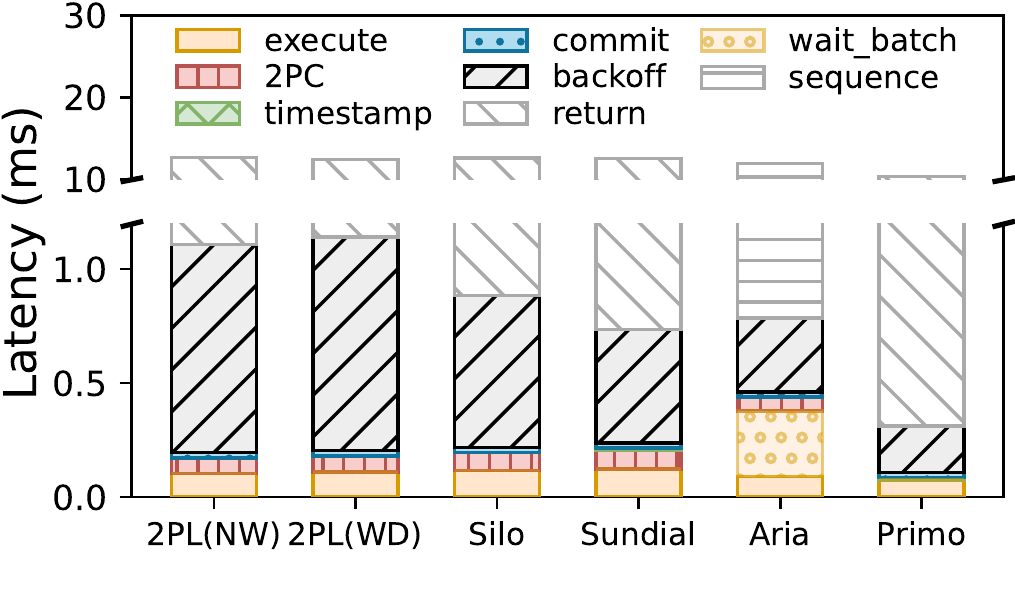}
         \caption{Latency breakdown}
         \label{fig:ycsb_runtime}
    \end{subfigure}
    \hfill
    \begin{subfigure}[b]{0.45\textwidth}
    \centering
         \includegraphics[width=\textwidth]{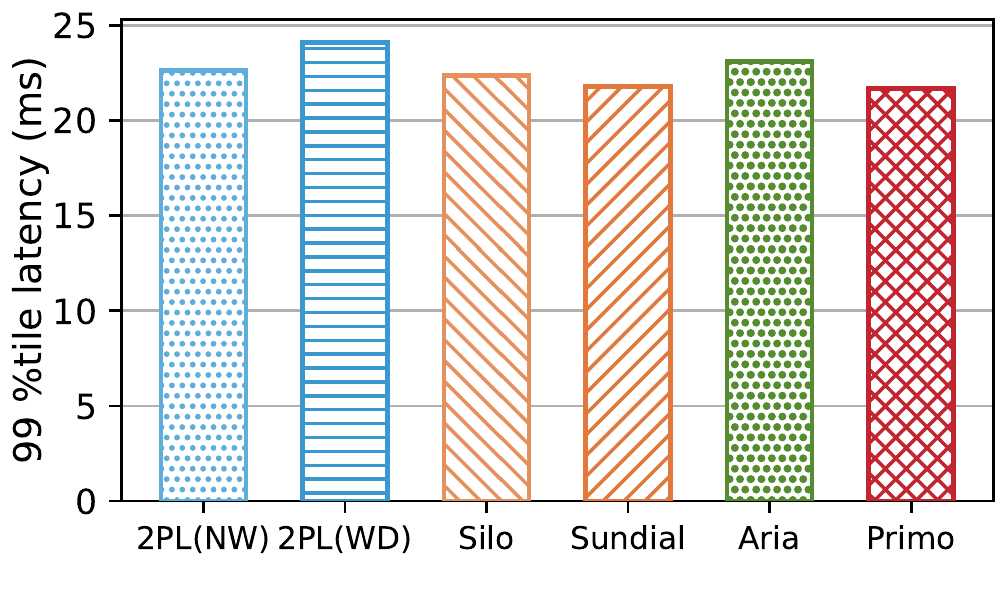}
         \caption{Tail Latency}
         \label{fig:ycsb_tail}
    \end{subfigure}
    \end{minipage}
    \caption{Overall performance and breakdown on YCSB (default setting)}
    \label{fig:ycsb_default}
\end{figure*}

\begin{figure*}
    \centering
    \begin{minipage}{0.9\linewidth}
    \begin{subfigure}[b]{0.45\textwidth}
    \centering
         \includegraphics[width=\textwidth]{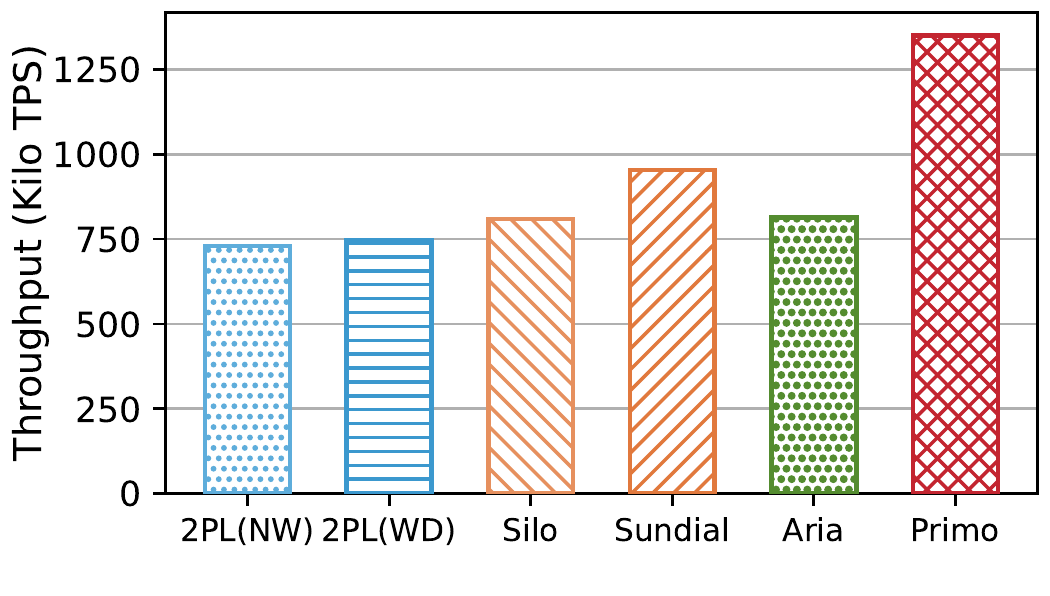}
         \caption{Throughput}
         \label{fig:tpcc_overall}
    \end{subfigure}
    \hfill
    \begin{subfigure}[b]{0.45\textwidth}
    \centering
         \includegraphics[width=\textwidth]{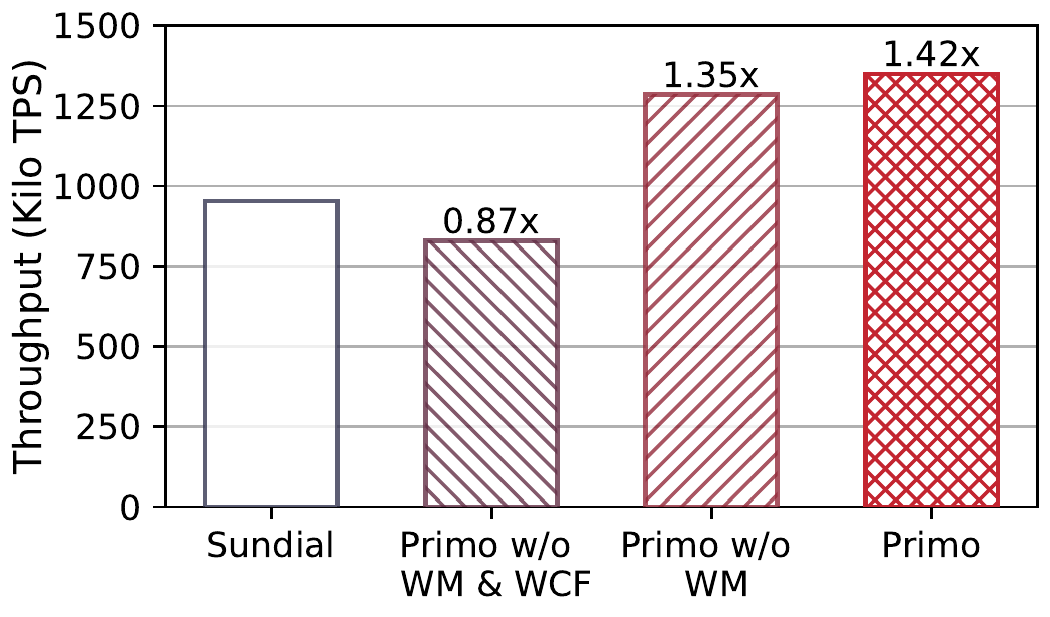}
         \caption{Factor breakdown}
         \label{fig:tpcc_factor}
    \end{subfigure}
    \\
    \begin{subfigure}[b]{0.45\textwidth}
    \centering
         \includegraphics[width=\textwidth]{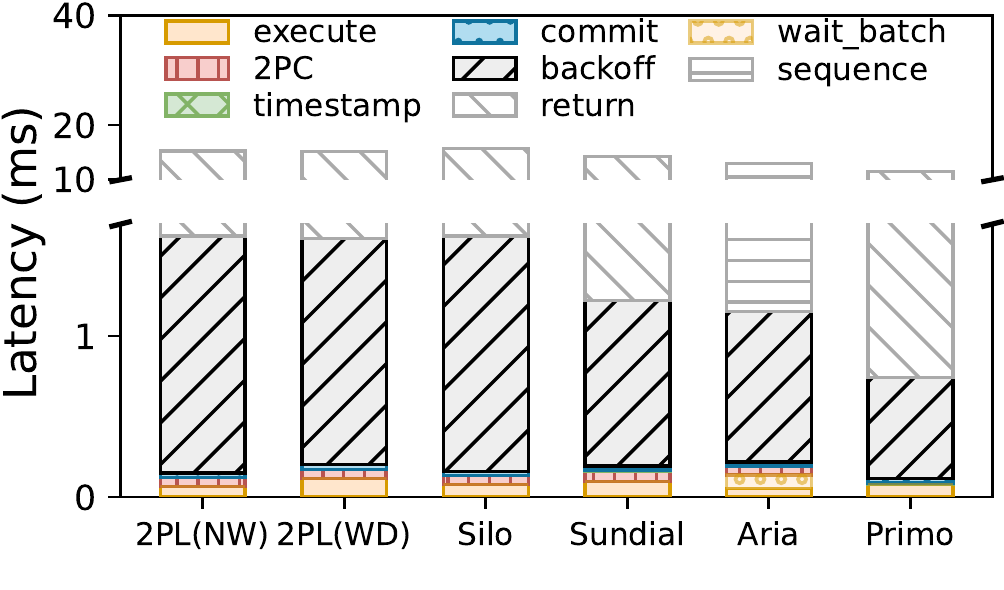}
         \caption{Latency breakdown}
         \label{fig:tpcc_runtime}
    \end{subfigure}
    \hfill
    \begin{subfigure}[b]{0.45\textwidth}
    \centering
         \includegraphics[width=\textwidth]{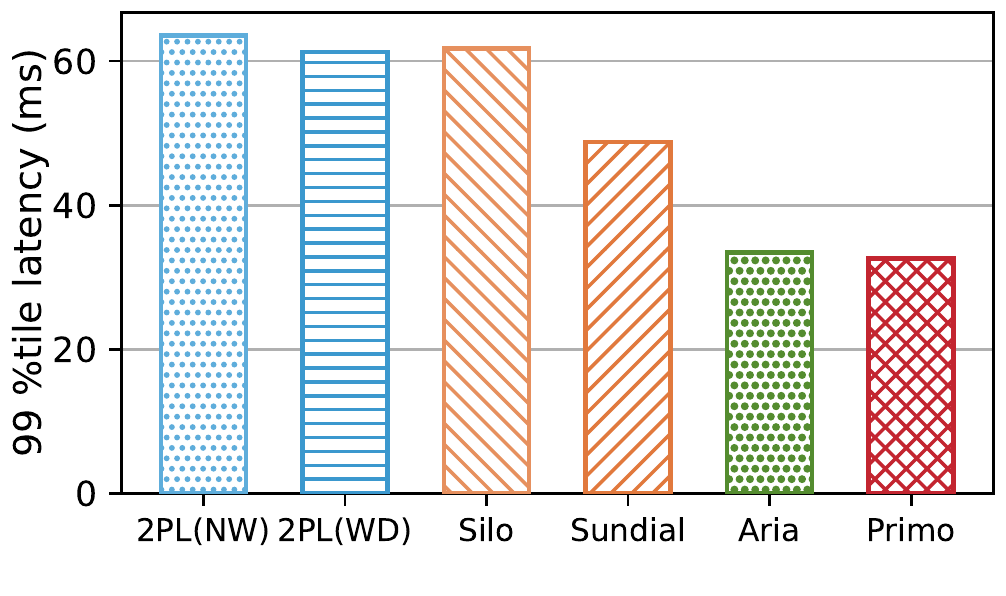}
         \caption{Tail latency}
         \label{fig:tpcc_tail}
    \end{subfigure}
    \end{minipage}
    \caption{Overall performance and breakdown on TPC-C (default setting)}
    \label{fig:tpcc_default}
\end{figure*}
}

\subsection{Overall Performance and Breakdown}\label{sec:exp_overall}

We first evaluate the performance under the default setting. As shown in
Figure \ref{fig:ycsb_overall} and Figure \ref{fig:tpcc_overall},
Primo achieves 1.91$\times$ and 1.42$\times$ higher throughput over the best of five competitors (i.e., Sundial) on YCSB and TPC-C, respectively. 
To understand the factors that contributed to the improvement, we conduct factor breakdowns, where %
we first disable both write-conflict-free concurrency control (WCF) and watermark-based distributed group commit (WM) in Primo, and add them back progressively.
Since WCF is a 2PL variant, when disabling it, we replace it with classic 2PL and add back 2PC for atomicity (but still use TicToc for local transactions).
When WM is disabled, Primo uses COCO for durability.
Figure \ref{fig:ycsb_factor} and Figure \ref{fig:tpcc_factor} show the improvement ratio using Sundial as a reference.
When both WCF and WM are disabled, we show that simply using 2PL+2PC for distributed transactions cannot outperform Sundial, even with TicToc for local transactions.
This demonstrates the effectiveness of WCF as it contributes to 1.78$\times$ and 1.35$\times$ improvement on YCSB and TPC-C, respectively.
\techrep{The improvement of WM is not fully demonstrated with only four partitions in the default setting since its advantage is scalability.
Nonetheless, the WM scheme shines when we scale the number of partitions to 20 in Section \ref{sec:scale}.
}

To explain why Primo outperforms the baselines, we present the latency breakdowns in Figure \ref{fig:ycsb_runtime} and Figure \ref{fig:tpcc_runtime}.
Most component names are self-explanatory. Besides, \texttt{timestamp} means the time for maintaining timestamps (if any). If a transaction aborts, it would backoff for a certain period and then retry, which contributes to the \texttt{backoff} part. 
The \texttt{return} part means the time a transaction waiting distributed group commit for returning the results to the client.
\revision{
\texttt{sequence} and \texttt{wait\_batch} are specific to Aria, which represents the time spent on the sequencing layer and the time waiting for the whole batch to finish execution, respectively.}
Except for the \texttt{backoff}, \texttt{return}, and \texttt{sequence} parts, the other parts all contribute to the contention footprint.
The advantage of Primo is that it eliminates 2PC (no \texttt{2PC} time) such that the contention footprint is significantly reduced.
\revision{
Aria has smaller \texttt{2PC} overhead
\marginnote{\META{Meta3}\RI{R1O4}\RII{R2O4}}
than other 2PC-based protocols, but at the cost of the extra \texttt{wait\_batch} time in the critical path. That explains why Aria cannot outperform Primo.
In fact, Aria's advantage of determinism (save logging and replication overhead) is not evident here because those overheads are also out of the critical path in Primo and other baselines (by using distributed group commit). With the additional overhead of enforcing determinism, Aria does not show a clear advantage over other competitors.} 
The result shows the overhead of maintaining the timestamps in Primo is negligible. 
\revision{All the protocols \marginnote{\RIII{R3O2}} have around 10ms latency because
we unified the epoch size of COCO and the watermark interval of WM to 20ms. 
The competitors have slightly higher latency than Primo because they have higher abort rates (more results on that in Section \ref{sec:exp_contention}), meaning transactions may spend more time retrying.
Moreover,
a transaction in epoch $i$ could possibly retry in epoch $(i+1)$, causing 20ms longer latency. Hence, by having lower abort rates, Primo also has lower tail latency.}

\subsection{Impact of Workload Parameters} \label{sec:param}
We next study how Primo is affected by various workload parameters. Most of the experiments are conducted on YCSB as its parameters are more flexible to configure.

\subsubsection{Impact of contention} \label{sec:exp_contention}
We compare the performance under various contention by tuning the skewness of the Zipf distribution in YCSB, while the other parameters are unchanged and use the default value. %
As shown in Figure \ref{fig:ycsb_skew}, the throughput of Primo is better than the competitors under all levels of contention. That's because WCF shortens the contention footprint by eliminating 2PC.
The benefit of eliminating 2PC is also demonstrated in Figure \ref{fig:ycsb_skew_abort}, which shows Primo always has lower abort rates.
As expected by our analysis in Section \ref{sec:analysis}, Primo gains larger margins under higher contention.
\marginnote{\META{Meta3}\RI{R1O4}\RII{R2O4}}
\revision{
The abort rate of Aria increases quickly under high contention because its batching prolongs the contention footprint.}

\subsubsection{Impact of distributed transactions}
We study the impact of the ratio of the distributed transactions under two contention levels using YCSB.
The skewness is set to 0.0 and 0.9 for low contention and high contention, respectively, with a controlled ratio of distributed transactions.
As shown in Figure \ref{fig:ycsb_remote},
The improvement ratio of Primo increases with a higher ratio of distributed transactions under low contention, which is due to more 2PC runs being saved.
Under high contention, the benefit of eliminating 2PC is more evident, but with more distributed transactions, the cost of the extra exclusive-locks required by WCF also increases.
As a result, Primo is able to win the best of the competitors by 2.46$\times$ but the improvement ratio decreases to 1.96$\times$ when all the transactions are distributed.

\icde{
\begin{figure}[!t]
    \centering
    \begin{minipage}{\linewidth}
    \begin{subfigure}[b]{\linewidth}
        \centering
        \includegraphics[width=0.83\linewidth]{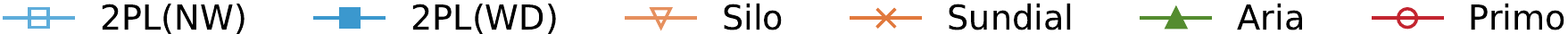}
    \end{subfigure} 
    \end{minipage}
    \begin{minipage}{\linewidth}
    \begin{subfigure}[b]{0.49\linewidth}
        \includegraphics[width=\textwidth]{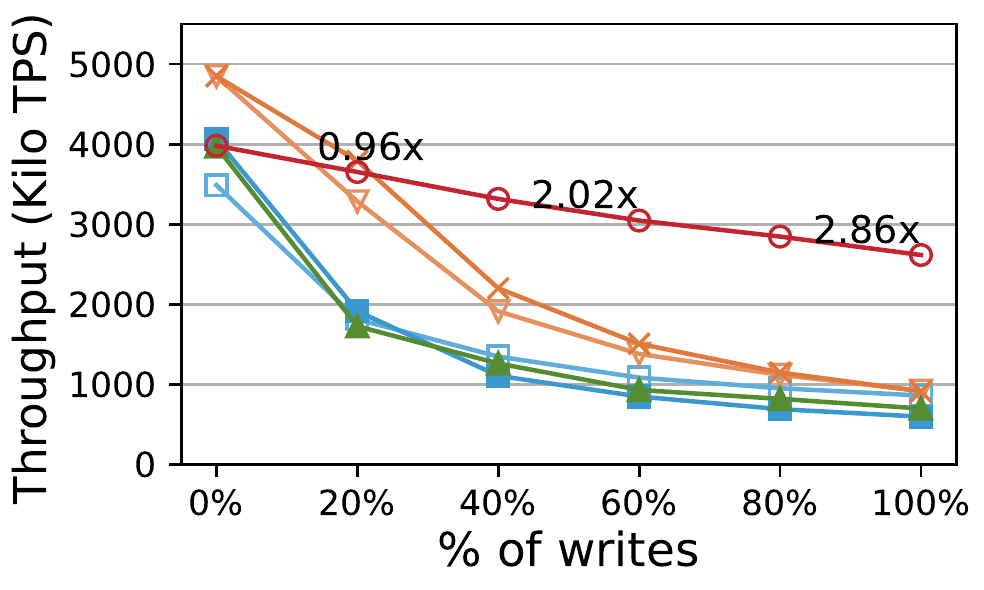}
        \vspace{-0.5em}
        \caption{20\% distributed}
        \label{fig:ycsb_read_2}
    \end{subfigure}
    \hfill
    \begin{subfigure}[b]{0.49\linewidth}
        \includegraphics[width=\textwidth]{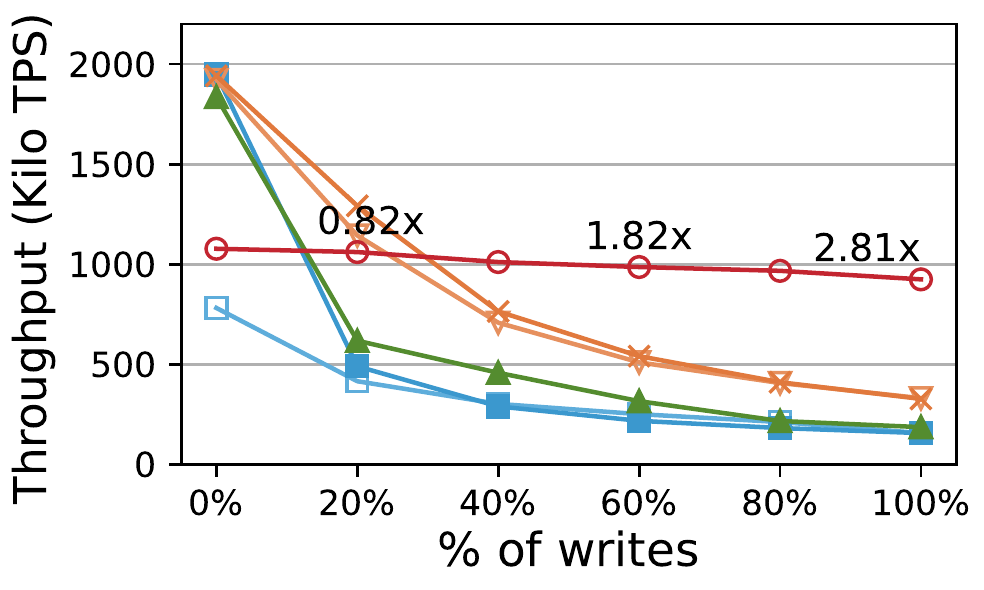}
        \vspace{-0.5em}
        \caption{80\% distributed}
        \label{fig:ycsb_read_8}
    \end{subfigure}  
    \end{minipage}
    \caption{\revision{Impact of the read-write ratio}} \label{fig:ycsb_read_write}
\end{figure}

\begin{figure}[!t]
\centering
\begin{minipage}{.48\linewidth}
  \centering
  \includegraphics[width=\linewidth]{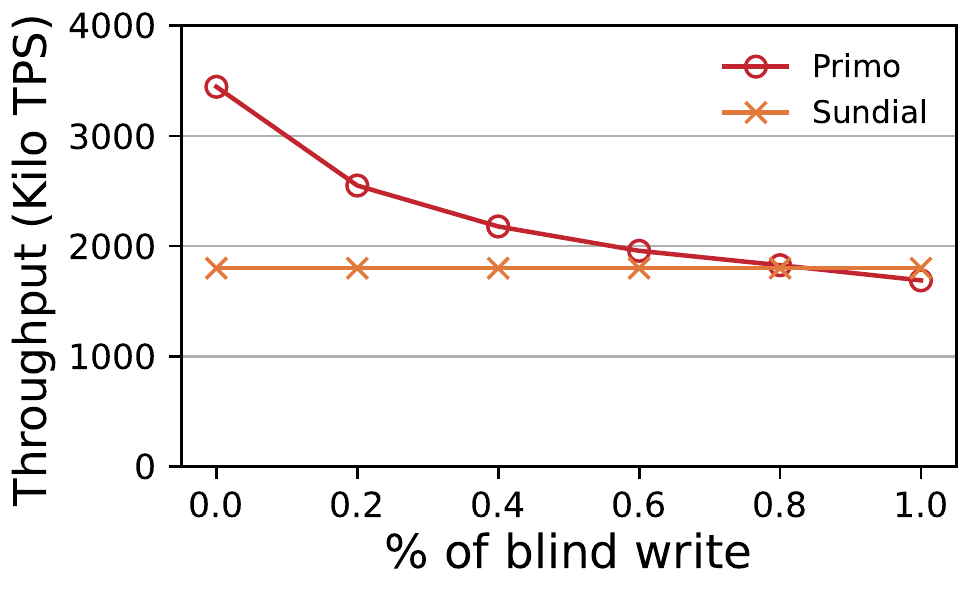}
  \captionof{figure}{Impact of the blind write ratio}

  \label{fig:ycsb_blind_write}
\end{minipage}%
\hfill
\begin{minipage}{.48\linewidth}
  \centering
  \includegraphics[width=\linewidth]{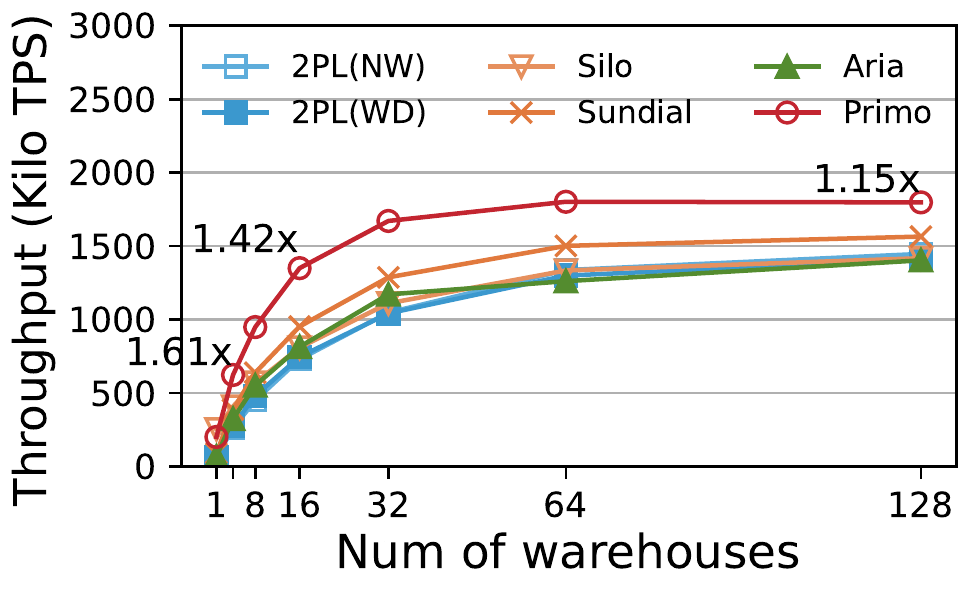}
  \captionof{figure}{\revision{Impact of num of warehourses in TPC-C}}
  \label{fig:tpcc_wh}
\end{minipage}
\end{figure}
}

\techrep{
\begin{figure}[!t]
    \centering
    \begin{minipage}{0.8\linewidth}
    \begin{subfigure}[b]{\linewidth}
        \centering
        \includegraphics[width=0.83\linewidth]{figs/legend.pdf}
    \end{subfigure}
    \end{minipage}
    \begin{minipage}{0.9\linewidth}
    \begin{subfigure}[b]{0.45\linewidth}
        \centering
        \includegraphics[width=\linewidth]{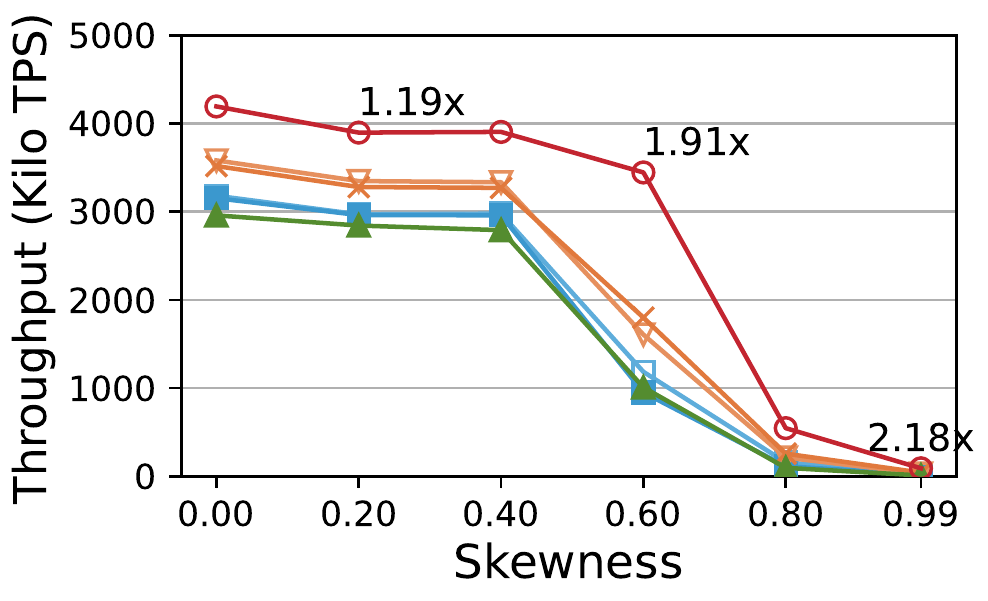}
        \caption{Throughput}
        \label{fig:ycsb_skew}
    \end{subfigure}
    \hfill
    \begin{subfigure}[b]{0.45\linewidth}
        \centering
        \includegraphics[width=\linewidth]{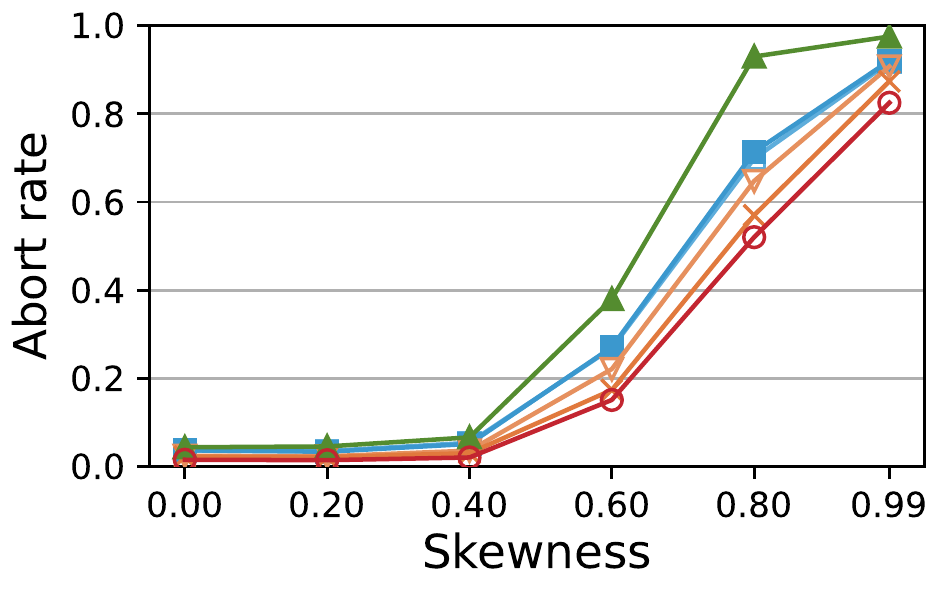}
        \caption{Abort rate}
        \label{fig:ycsb_skew_abort}
    \end{subfigure}
    \end{minipage}
    \caption{Impact of contention}
\end{figure}

\begin{figure}[!t]
    \centering
    \begin{minipage}{0.8\linewidth}
    \begin{subfigure}[b]{\linewidth}
        \centering
        \includegraphics[width=0.85\linewidth]{figs/legend.pdf}
    \end{subfigure}
    \end{minipage}
    \begin{minipage}{0.9\linewidth}
     \begin{subfigure}[b]{0.45\linewidth}
        \centering
        \includegraphics[width=\linewidth]{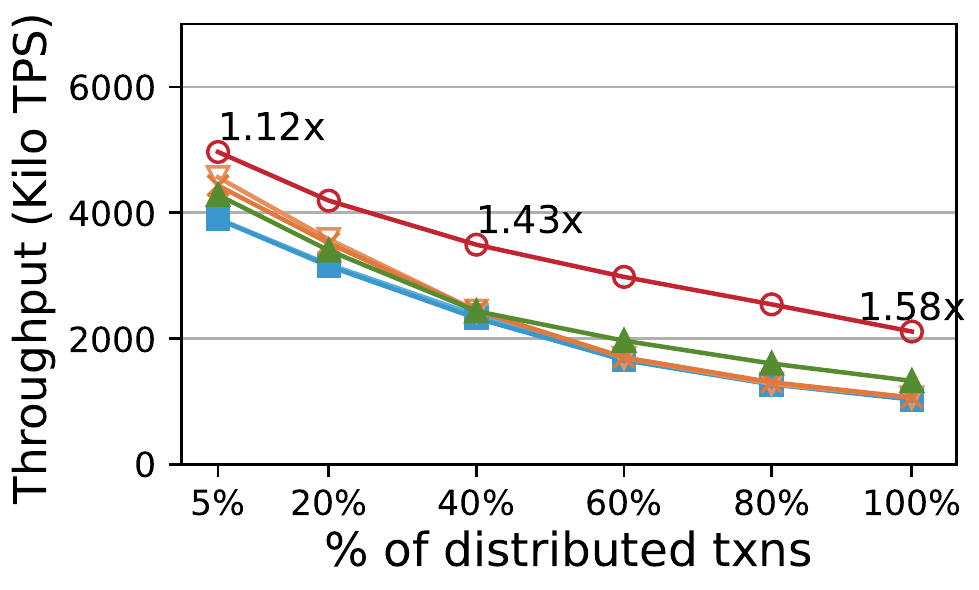}
        \caption{Low contention}
        \label{fig:ycsb_remote_0}
    \end{subfigure}
    \hfill
    \begin{subfigure}[b]{0.45\linewidth}
        \centering
        \includegraphics[width=\linewidth]{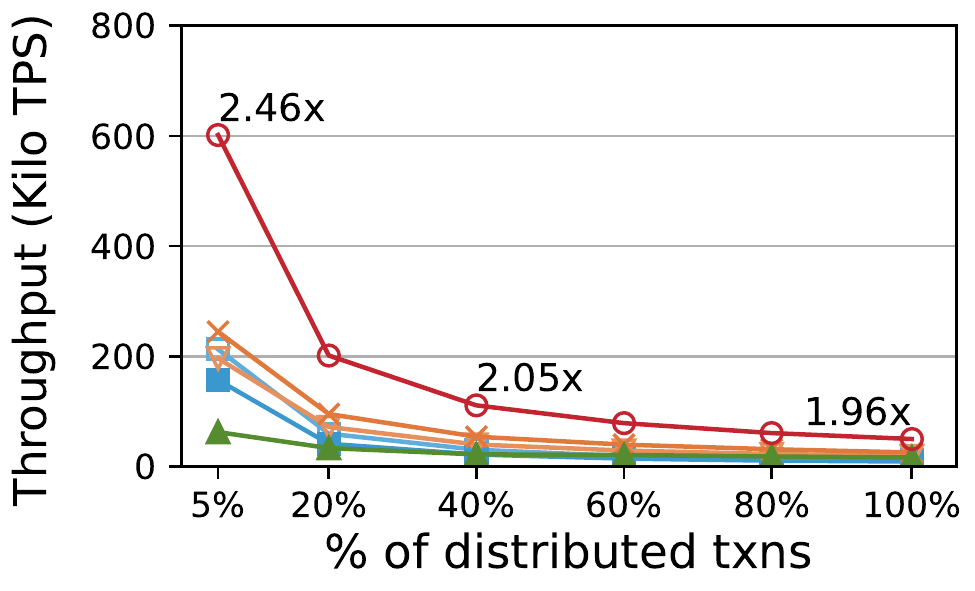}
        \caption{High contention}
        \label{fig:ycsb_remote_9}
    \end{subfigure} 
    \end{minipage}
    \caption{Impact of distributed transactions}
    \label{fig:ycsb_remote}
\end{figure}
}

\subsubsection{Impact of read-write ratio}
We study the effect of the read-write ratio using YCSB under two different percentages of distributed transactions (20\% and 80\%) with a controlled percentage of writes (out of 10 operations in each transaction).
The results in Figure \ref{fig:ycsb_read_2} and Figure \ref{fig:ycsb_read_8} show that Primo offers stable throughput regardless of the percentage of writes while the performances of the competitors all degrade quickly in write-heavy workloads. 
That's because write-heavy workloads exhibit higher contention and Primo is less affected by that as it shortens the contention footprint by eliminating 2PC.
In contrast, since the contention in read-heavy workloads is much lower,
2PC-based protocols readily yield good performances. 
Hence, there is actually no need for removing 2PC in this case.
In the mostly local setting (Figure \ref{fig:ycsb_read_2}), 
the extra overhead of exclusive-locks is small as we only measured less than 2\% of local transactions are aborted due to the extra exclusive locks. Hence, Primo still outperforms competitors even under read-heavy conditions in this case.
The overhead is higher for mostly-distributed read-heavy workloads in Figure \ref{fig:ycsb_read_8}.
As an easy fix, Primo could fall back to 2PC to avoid the overhead in such cases.

\added{
\subsubsection{Impact of blind write}
As discussed in Section \ref{sec:complete_waf}, Primo may require extra roundtrips to handle blind-writes. We study the impact of that in this experiment by a controlled ratio of the blind-writes (e.g., 50\% blind-write means half of the read-modify-write operations in the default YCSB are changed to blind-writes). Figure \ref{fig:ycsb_blind_write} shows the results with Sundial (the best of the competitors) as a reference. 
Primo outperforms Sundial as long as the blind-write ratio is less than 80\%. 
When all writes are blind-writes, Primo cannot win because all distributed transactions in Primo require the same number of roundtrips as 2PC.
Nonetheless, read-modify-writes are more commonly seen in practice than blind-writes (e.g., TPC-C \cite{tpcc}, TATP \cite{tatp}, and Smallbank \cite{smallbank}).
}

\subsubsection{Impact of the number of warehouses in TPC-C} \label{sec:wh}
The number of warehouses affects the table size, the contention, and the ratio of distributed transactions in TPC-C.
Roughly, with more warehouses, the table size is larger, which in turn reduces the contention. Besides that, since there are  more local warehouses per partition, the ratio of the distributed transactions also decreases.
Figure \ref{fig:tpcc_wh} shows the result. %
Primo outperforms all the competitors regardless of the number of warehouses. In fact, it improves more with fewer warehouses because the contention amplifies the advantage of Primo.

\techrep{
\begin{figure}[!t]
    \centering
    \begin{minipage}{0.8\linewidth}
    \begin{subfigure}[b]{\linewidth}
        \centering
        \includegraphics[width=0.85\linewidth]{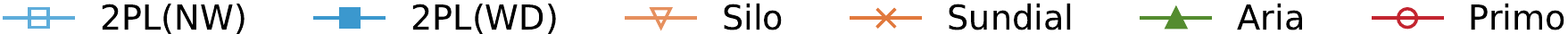}
    \end{subfigure} 
    \end{minipage}
    \begin{minipage}{0.9\linewidth}
    \begin{subfigure}[b]{0.45\linewidth}
        \includegraphics[width=\textwidth]{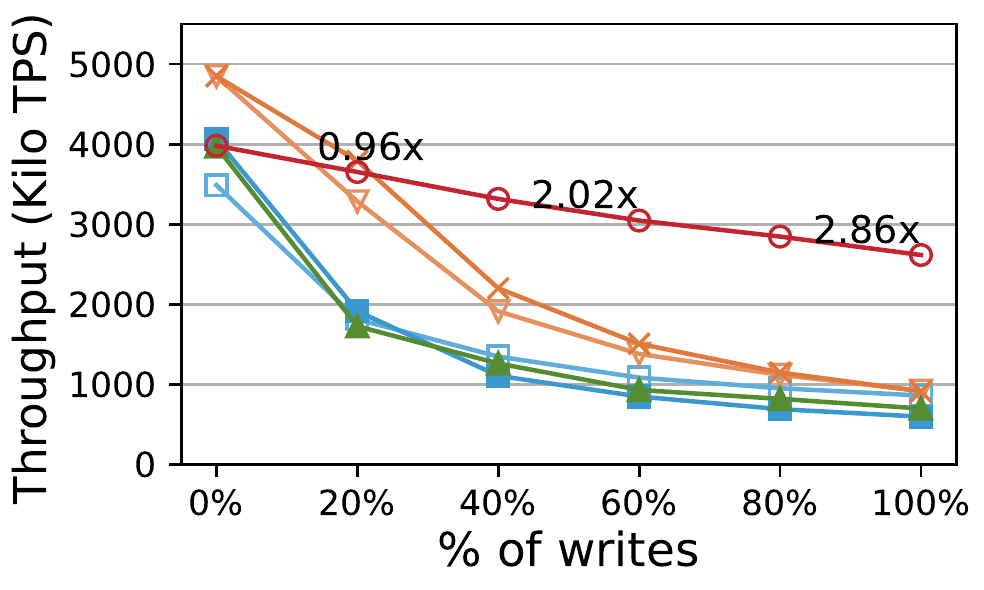}
        \caption{20\% distributed}
        \label{fig:ycsb_read_2}
    \end{subfigure}
    \hfill
    \begin{subfigure}[b]{0.45\linewidth}
        \includegraphics[width=\textwidth]{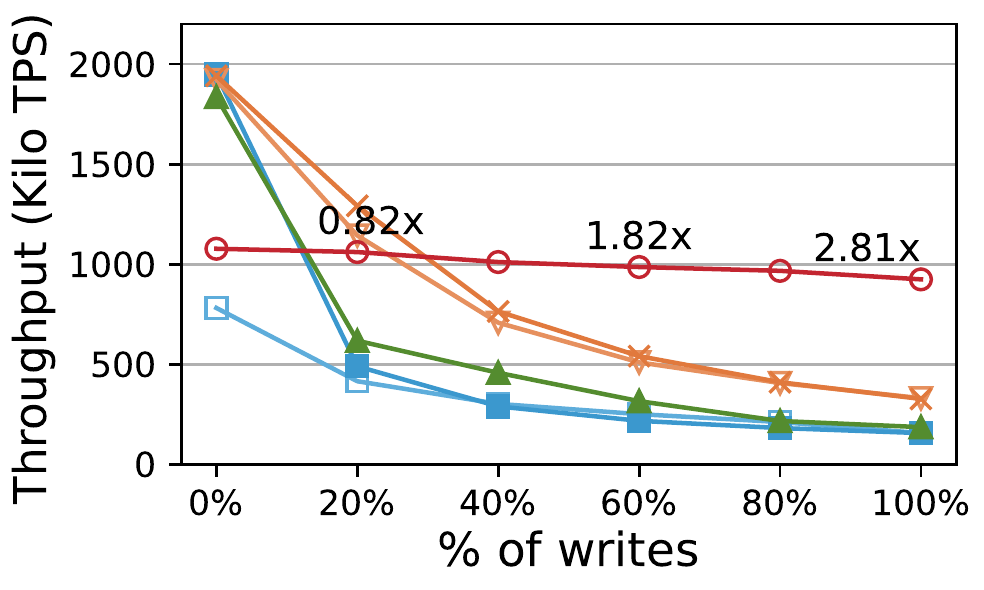}
        \caption{80\% distributed}
        \label{fig:ycsb_read_8}
    \end{subfigure}  
    \end{minipage}
    \caption{Impact of the read-write ratio}
\end{figure}

\begin{figure}[!t]
\centering
\begin{minipage}{0.9\linewidth}
\begin{minipage}{.45\linewidth}
  \centering
  \includegraphics[width=\linewidth]{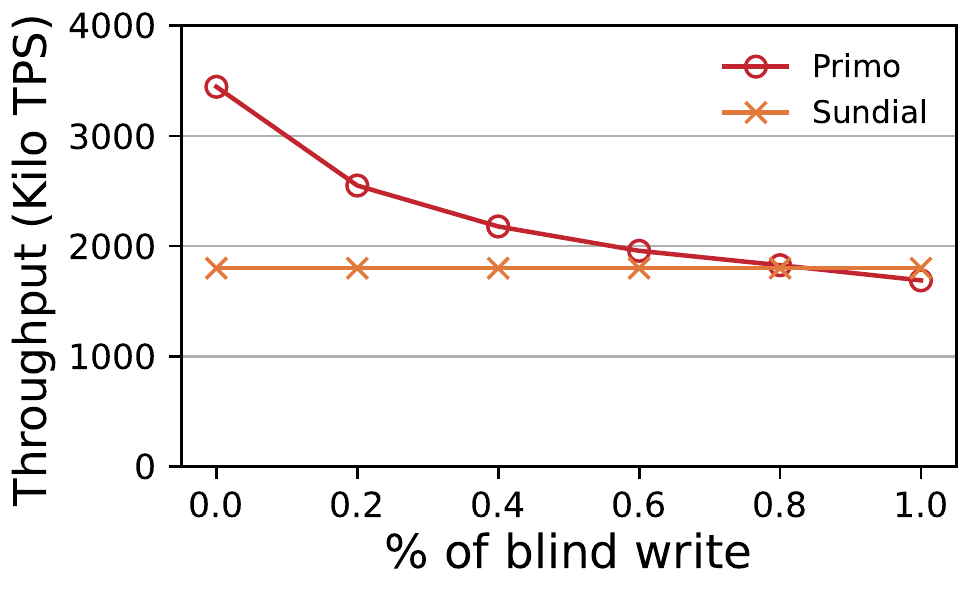}
  \captionof{figure}{Impact of the blind write ratio}
  \label{fig:ycsb_blind_write}
\end{minipage}%
\hfill
\begin{minipage}{.45\linewidth}
  \centering
  \includegraphics[width=\linewidth]{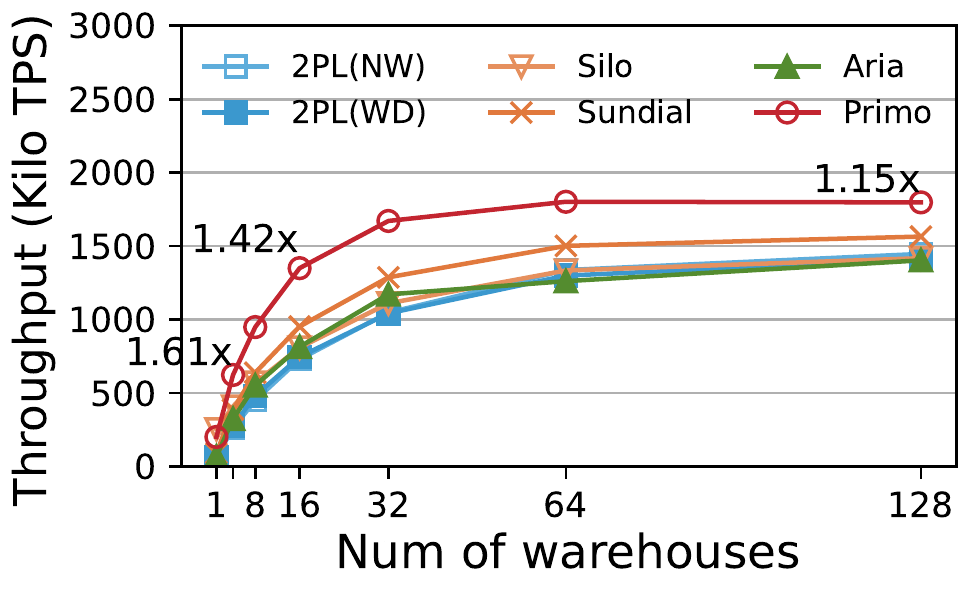}
  \captionof{figure}{Impact of num of warehourses in TPC-C} 
  \label{fig:tpcc_wh}
\end{minipage}
\end{minipage}
\end{figure}
}

\icde{
\deleted{
\begin{figure}[t!]
    \centering
    \begin{minipage}{\linewidth}
    \centering
    \begin{subfigure}[b]{0.5\linewidth}
        \centering
        \includegraphics[width=\linewidth]{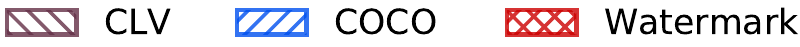}
    \end{subfigure}  
    \end{minipage}
    \begin{minipage}{\linewidth}
    \begin{subfigure}[b]{0.49\linewidth}
        \includegraphics[width=\textwidth]{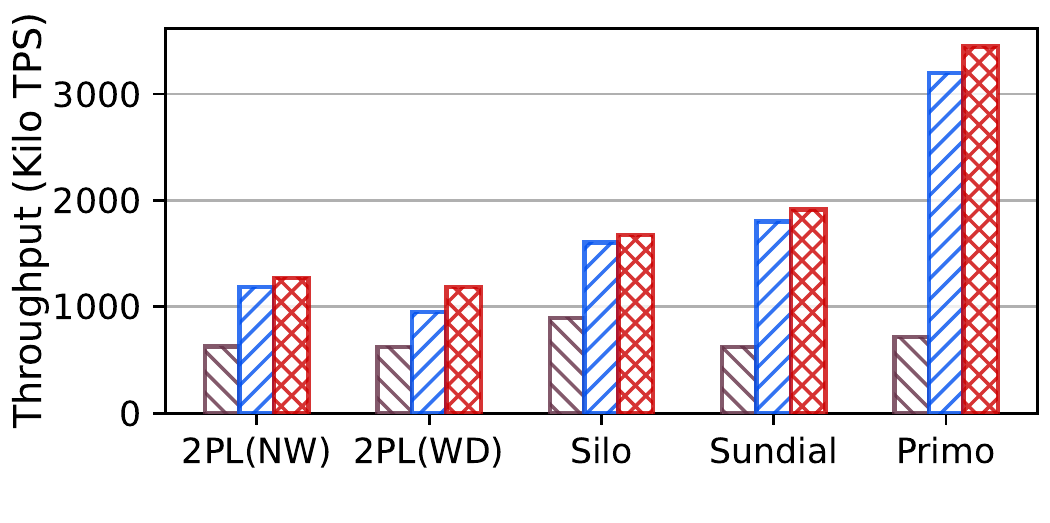}
        \caption{YCSB} \vspace{-0.1cm}
        \label{fig:ycsb_log}
    \end{subfigure}
    \hfill
    \begin{subfigure}[b]{0.49\linewidth}
        \includegraphics[width=\textwidth]{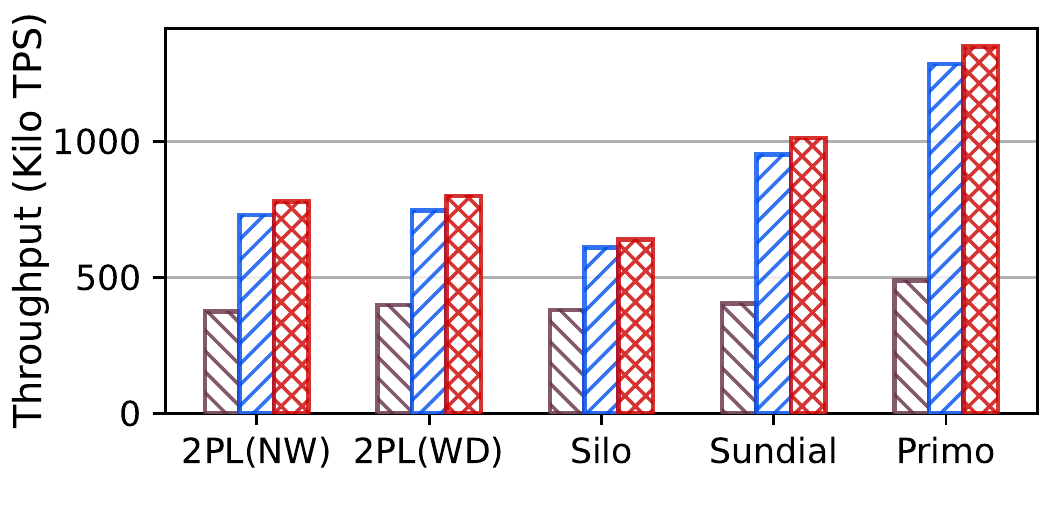}
        \caption{TPC-C} \vspace{-0.1cm}
        \label{fig:tpcc_log}
    \end{subfigure}  
    \end{minipage}
    \caption{Comparison of Logging Optimizations}
    \vspace{-0.2cm}
    \label{fig:log_compare}
\end{figure}
}
}

\techrep{
\begin{figure}[t!]
    \centering
    \begin{minipage}{0.9\linewidth}
    \centering
    \begin{subfigure}[b]{0.5\linewidth}
        \centering
        \includegraphics[width=\linewidth]{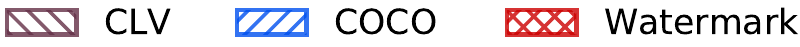}
    \end{subfigure}  
    \end{minipage}
    \begin{minipage}{0.9\linewidth}
    \begin{subfigure}[b]{0.45\linewidth}
        \includegraphics[width=\textwidth]{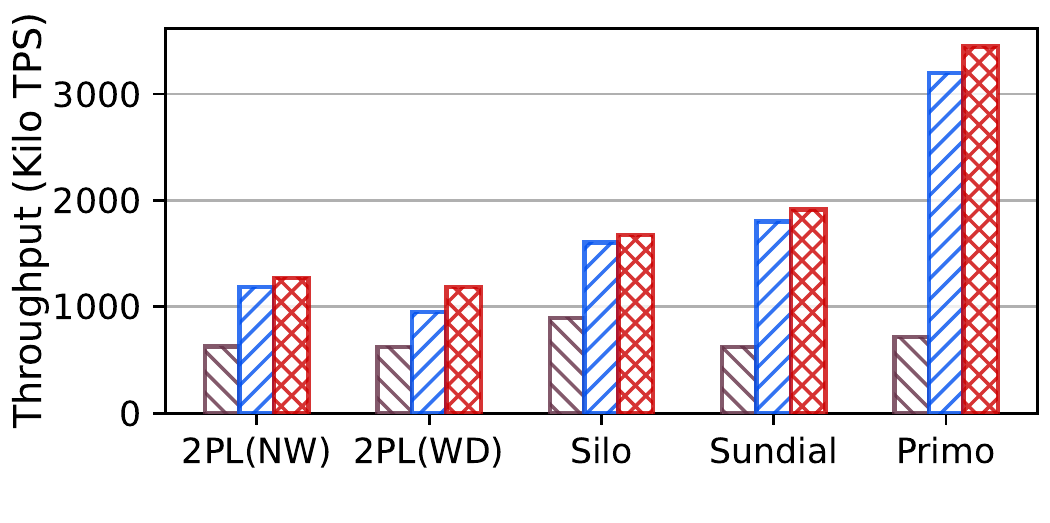}
        \caption{YCSB}
        \label{fig:ycsb_log}
    \end{subfigure}
    \hfill
    \begin{subfigure}[b]{0.45\linewidth}
        \includegraphics[width=\textwidth]{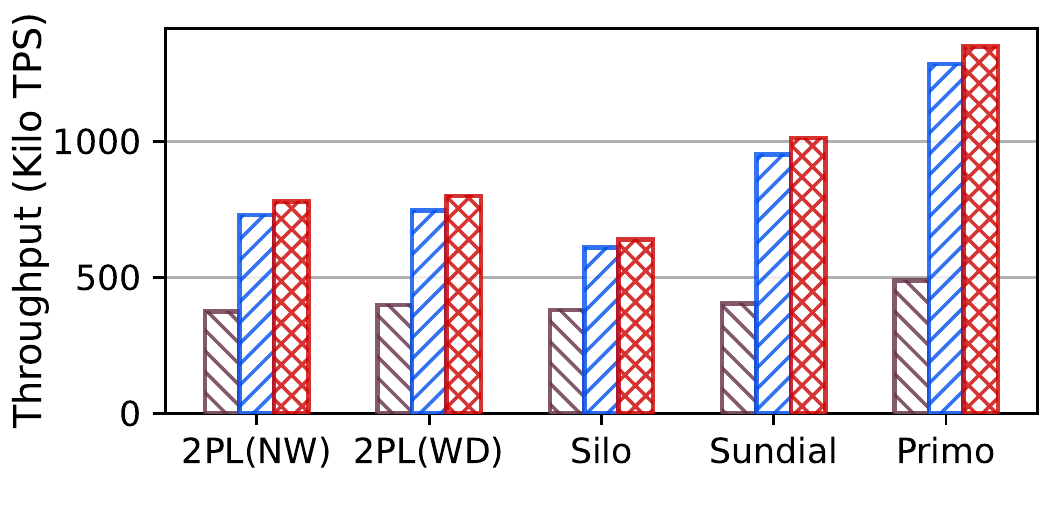}
        \caption{TPC-C}
        \label{fig:tpcc_log}
    \end{subfigure}  
    \end{minipage}
    \caption{Comparison of Logging Optimizations}
    \label{fig:log_compare}
\end{figure}
}

\subsection{Comparison of logging optimizations} \label{sec:log_exp}
\techrep{
In this experiment, we compare our WM scheme with COCO and CLV.
Their performances are compared under various distributed concurrency control schemes in Figure \ref{fig:log_compare}.
\revision{Aria is not included because as a deterministic database, it ensures durability in a special sequencing layer that logs transaction commands only before execution, while WM, COCO, and CLV target non-deterministic databases that log the transactions' write-sets \cite{aries} after execution}.
The results show WM outperforms both COCO and CLV.
The performance of CLV is not as good as group commit protocols (i.e., WM and COCO) due to its overhead of tracking fine-grained dependencies.
}

\techrep{
\begin{figure}[t!]
    \marginnote{\RI{R1O5}\RII{R2O3}}
    \centering
    \begin{minipage}{\linewidth}
    \centering
    \begin{subfigure}[b]{0.34\linewidth}
        \centering
        \includegraphics[width=\linewidth]{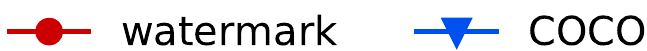}
    \end{subfigure}  
    \end{minipage}
    \begin{minipage}{\linewidth}
    \begin{subfigure}[b]{0.31\linewidth}
        \includegraphics[width=\textwidth]{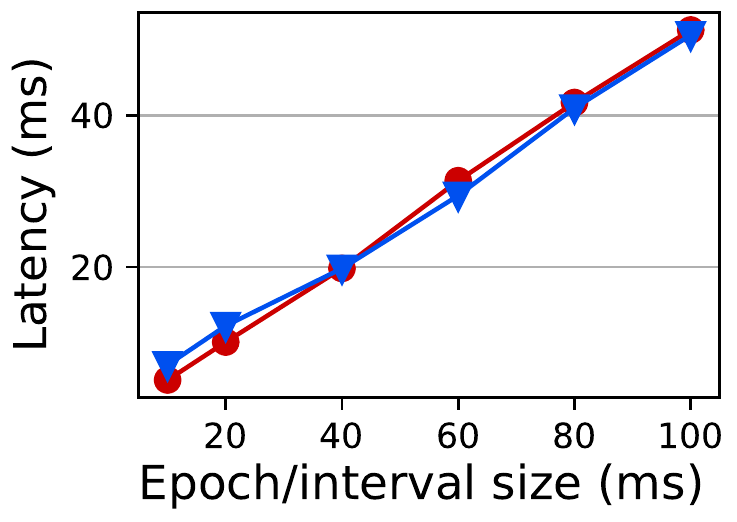}
        \caption{Latency}
        \label{fig:ycsb_epoch_latency}
    \end{subfigure}
    \begin{subfigure}[b]{0.33\linewidth}
        \includegraphics[width=\textwidth]{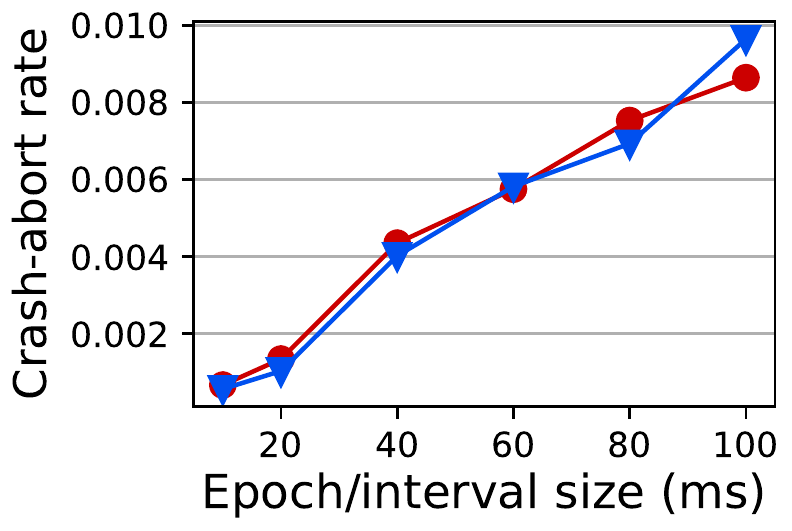}
        \caption{Crash-abort rate}
        \label{fig:ycsb_epoch_abort}
    \end{subfigure}
    \begin{subfigure}[b]{0.32\linewidth}
        \includegraphics[width=\textwidth]{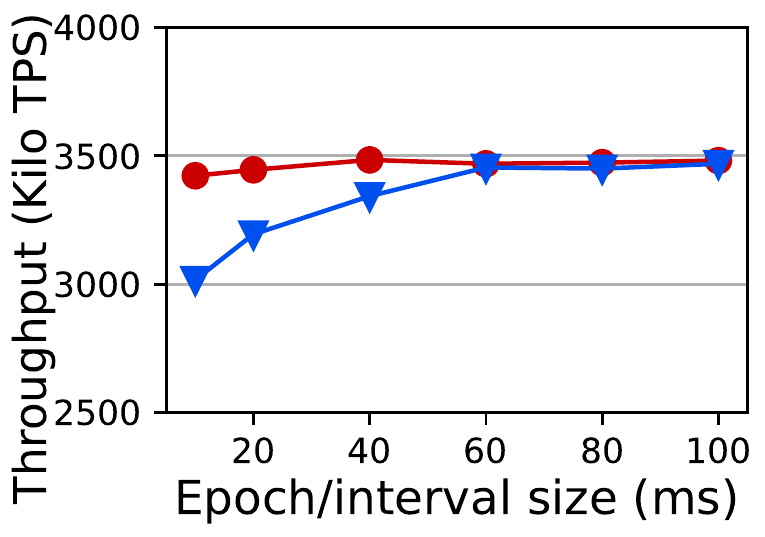}
        \caption{Throughput}
        \label{fig:ycsb_epoch_throughput}
    \end{subfigure}
    \end{minipage}
    \caption{\revision{Impact of Watermark Interval/Epoch Size (YCSB default setting; results on TPC-C shows the same trend)}}
    \label{fig:ycsb_epoch}
\end{figure}

\begin{figure}[t!]
\marginnote{\META{Meta2}\RI{R1O3}}
    \centering
    \begin{minipage}{0.9\linewidth}
    \centering
    \begin{subfigure}[b]{0.7\linewidth}
        \centering
        \includegraphics[width=\linewidth]{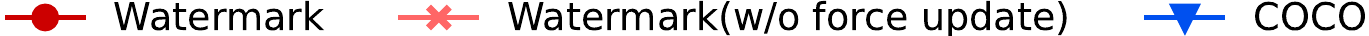}
    \end{subfigure}  
    \end{minipage}
    \begin{minipage}{\linewidth}
    \begin{subfigure}[b]{0.52\linewidth}
        \includegraphics[width=\textwidth]{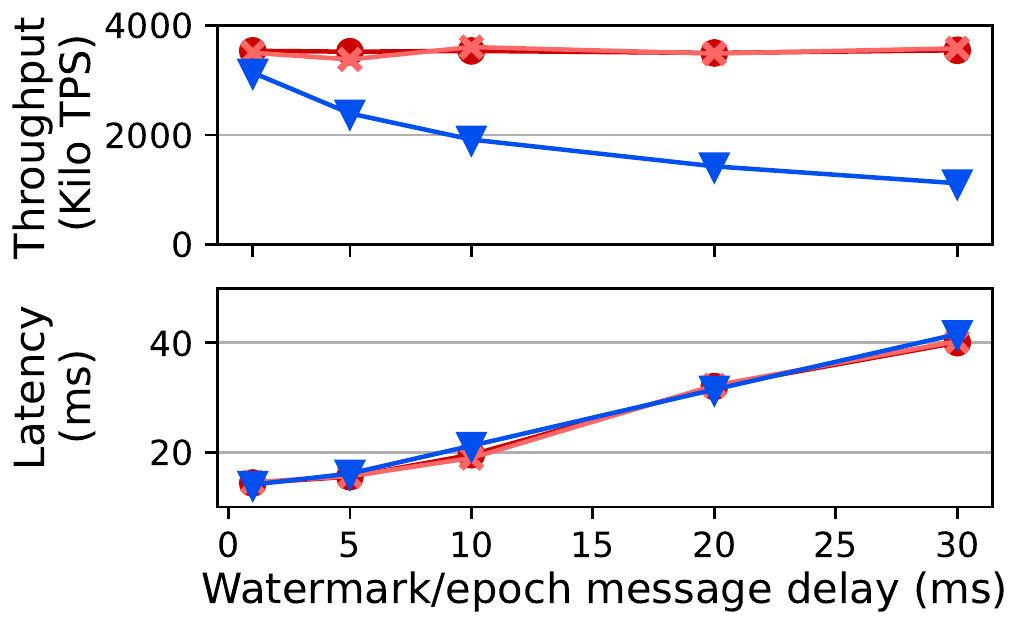}
        \caption{Lagging due to message delay}
        \label{fig:ycsb_add_delay}
    \end{subfigure}
    \hfill
    \begin{subfigure}[b]{0.47\linewidth}
        \includegraphics[width=\textwidth]{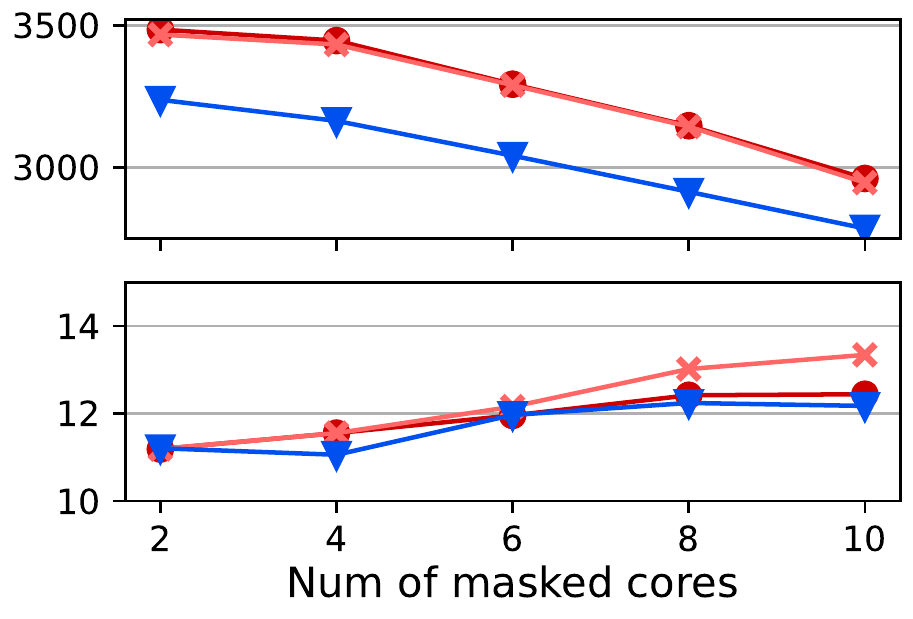}
        \caption{Lagging due to slow partition}
        \label{fig:ycsb_mask_threads}
    \end{subfigure}
    \end{minipage}
    \caption{\revision{Impact of Watermark/Epoch Lagging (YCSB default setting; results on TPC-C shows the same trend)}}
\end{figure}
}

\revision{
\marginnote{\RI{R1O5}\RII{R2O3}}
Our watermark-based group commit (WM) and COCO trade latency and crash-abort rate for throughput. The trade-off is tunable by controlling the watermark interval size in WM or the epoch size in COCO.
Figure \ref{fig:ycsb_epoch} confirms that.
In this experiment, the crash-abort rate (i.e., abort rate due to failure) is measured by killing a partition after running the system for 10s.
For fair comparisons, both WM and COCO use our WCF for distributed concurrency control.
Figures \ref{fig:ycsb_epoch_latency} and \ref{fig:ycsb_epoch_abort} show that the latency and the crash-abort rate are proportional to the epoch/interval size.
Notice that, the crash-abort rate would be negligible in modern servers that rarely crash (e.g., once a month), while the crash-abort rate reported in Figure \ref{fig:ycsb_epoch_abort} simulates an extreme case that the server crashes every 10s.
Figure \ref{fig:ycsb_epoch_throughput} shows that WM offers better throughput than COCO under the same epoch/interval size (i.e., the same latency and crash-abort rate) because it does not need synchronous coordination. Their throughput both saturates with an epoch/interval size beyond 60ms.
}

\revision{
\marginnote{\META{Meta2}\RI{R1O3}}
In distributed group commit protocol like COCO, a lagging partition could block the other partitions. 
Our WM scheme is less affected since it is an asynchronous scheme.
Figure \ref{fig:ycsb_add_delay} confirms that by showing the results when a partition's watermark is lagged due to network delay. 
In this case, 
while the latency inevitably increases in both COCO and WM,
WM shows almost no throughput drop while COCO's throughput drops significantly. That is because, in COCO, a lagging epoch message could detain the execution of the next epoch (stringent epoch synchronization is required for consistency in COCO). In contrast, WM does not need that synchronization but ensures consistency by logical timestamps (Section \ref{sec:watermark}).
Another situation of lagging might be more challenging to WM, where a partition runs slowly and thus grows its watermark slowly due to hardware problems. That seemingly makes WM to have higher latency than COCO who synchronizes the pace of progress in every epoch. 
We simulate this situation by masking some CPU cores of a partition and show the results in Figure \ref{fig:ycsb_mask_threads}.
In this case, both WM and COCO show inevitable throughput drops since we indeed make one partition slow-running.
Figure \ref{fig:ycsb_mask_threads} confirms that our force-updating mechanism (i.e., adaptively add offsets to the watermark of the slow partition) effectively solves the aforementioned problem (with force-updating, WM latency becomes similar to COCO).
}

\icde{
\begin{figure}[t!]
    \marginnote{\META{Meta3}\RI{R1O4}\RII{R2O4}}
    \centering
    \begin{minipage}{\linewidth}
    \begin{subfigure}[b]{\linewidth}
        \centering
        \includegraphics[width=0.95\linewidth]{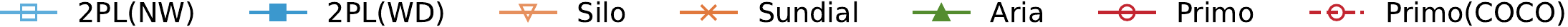}
    \end{subfigure}
    \end{minipage}
    \begin{minipage}{\linewidth}
    \begin{subfigure}[b]{0.5\linewidth}
        \includegraphics[width=\textwidth]{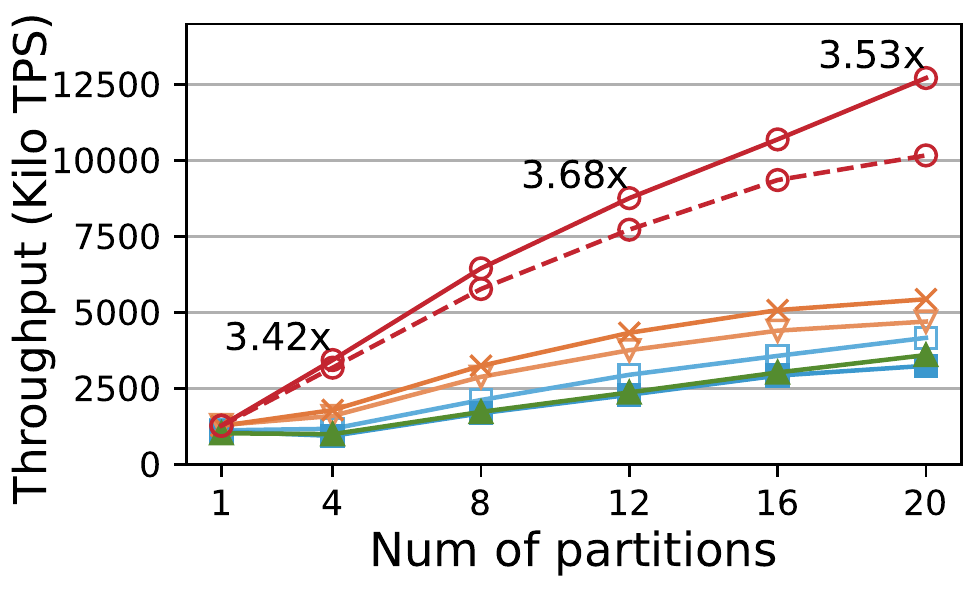}
        \caption{YCSB}

        \label{fig:ycsb_scale}
    \end{subfigure}
    \hfill
    \begin{subfigure}[b]{0.49\linewidth}
        \includegraphics[width=\textwidth]{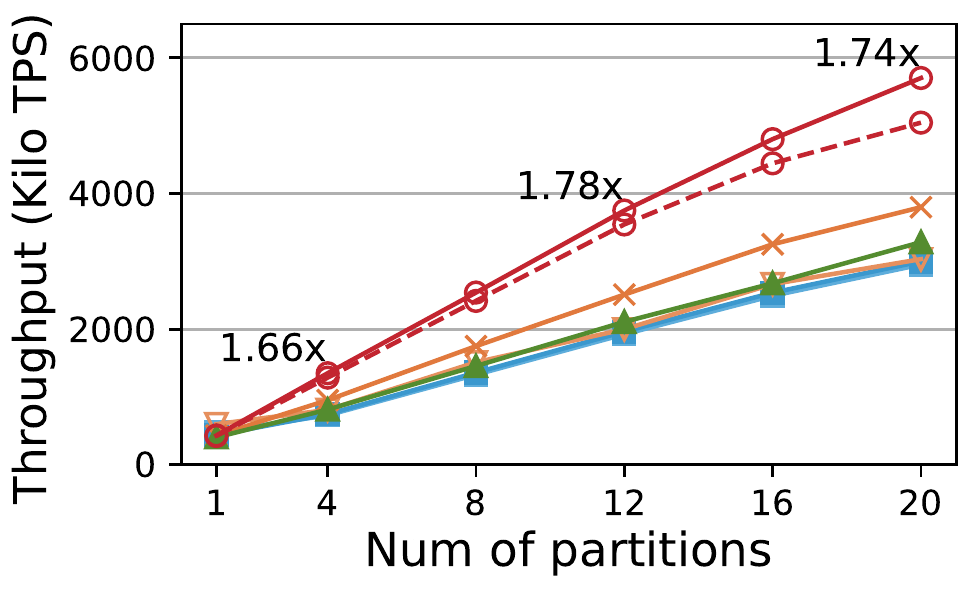}
        \caption{TPC-C}

        \label{fig:tpcc_scale}
    \end{subfigure}
    \end{minipage}
    \caption{\revision{Scalability}}
    \vspace{-1em}
\end{figure}

\deleted{
\begin{figure}[t!]
    \begin{subfigure}[b]{0.5\linewidth}
        \includegraphics[width=\textwidth]{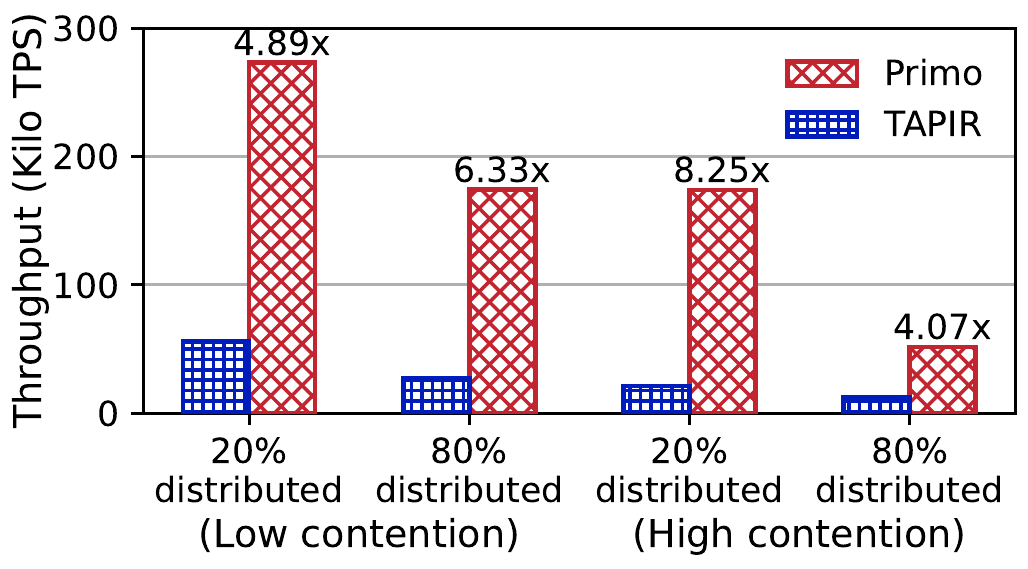}
        \caption{Throughput} \vspace{-0.1cm}
        \label{fig:tapir_throughput}
    \end{subfigure}
    \hfill
    \begin{subfigure}[b]{0.49\linewidth}
        \includegraphics[width=\textwidth]{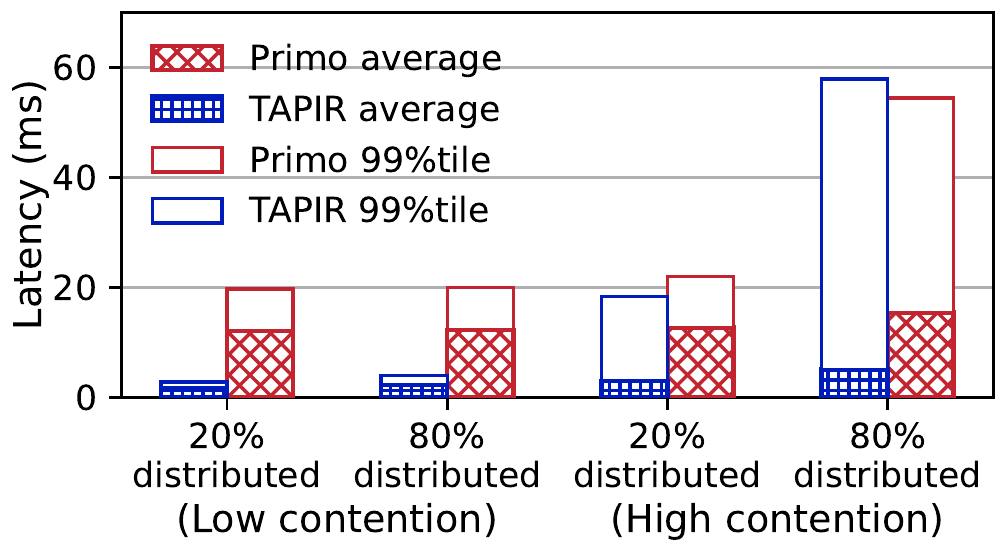}
        \caption{Latency} \vspace{-0.1cm}
        \label{fig:tapir_latency}
    \end{subfigure}
    \caption{Compare with TAPIR}
    \label{fig:tapir}
    \vspace{-0.25cm}
\end{figure}
}
}

\subsection{Scalability} \label{sec:scale}
In this experiment, we scale the number of partitions up to 20 using the default workload parameters.
As shown in Figure \ref{fig:ycsb_scale} and Figure \ref{fig:tpcc_scale}, Primo provides better scalability over the competitors on both YCSB and TPC-C.
That's because Primo eliminates 2PC (2PC harms scalability \cite{calvin}).
To compare the scalability of the distributed group commit schemes (COCO and WM), we include the results of Primo after replacing WM with COCO (i.e., the dashed line).
The results show that COCO suffers from the scalability problem when the number of partitions is greater than 12, while Primo using the WM scheme still scales well.

\techrep{
\begin{figure}[t!]
    \centering
    \begin{minipage}{0.8\linewidth}
    \begin{subfigure}[b]{\linewidth}
        \centering
        \includegraphics[width=\linewidth]{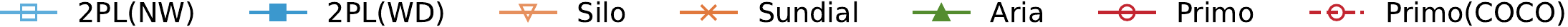}
    \end{subfigure}
    \end{minipage}
    \begin{minipage}{0.9\linewidth}
    \begin{subfigure}[b]{0.45\linewidth}
        \includegraphics[width=\textwidth]{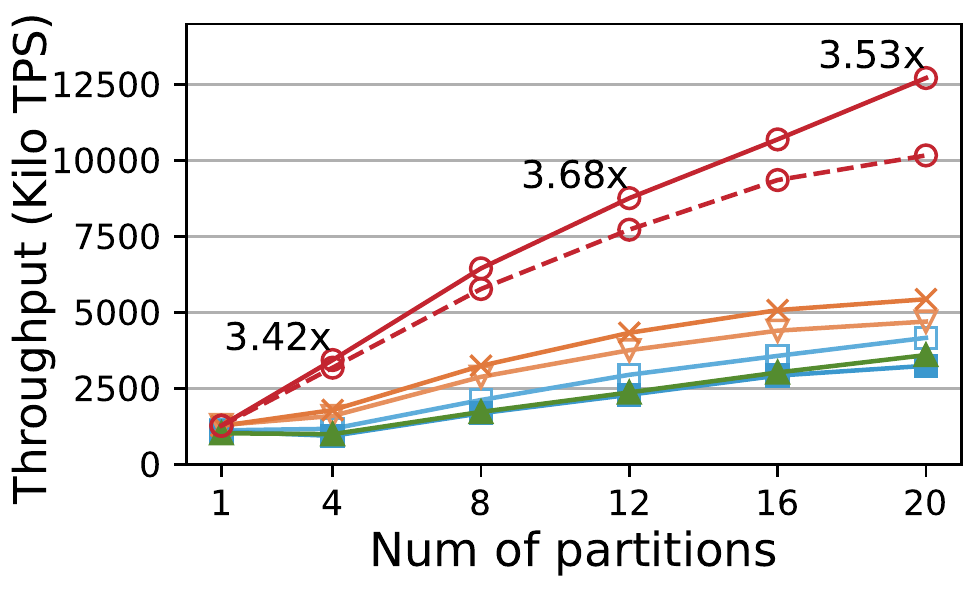}
        \caption{YCSB}
        \label{fig:ycsb_scale}
    \end{subfigure}
    \hfill
    \begin{subfigure}[b]{0.45\linewidth}
        \includegraphics[width=\textwidth]{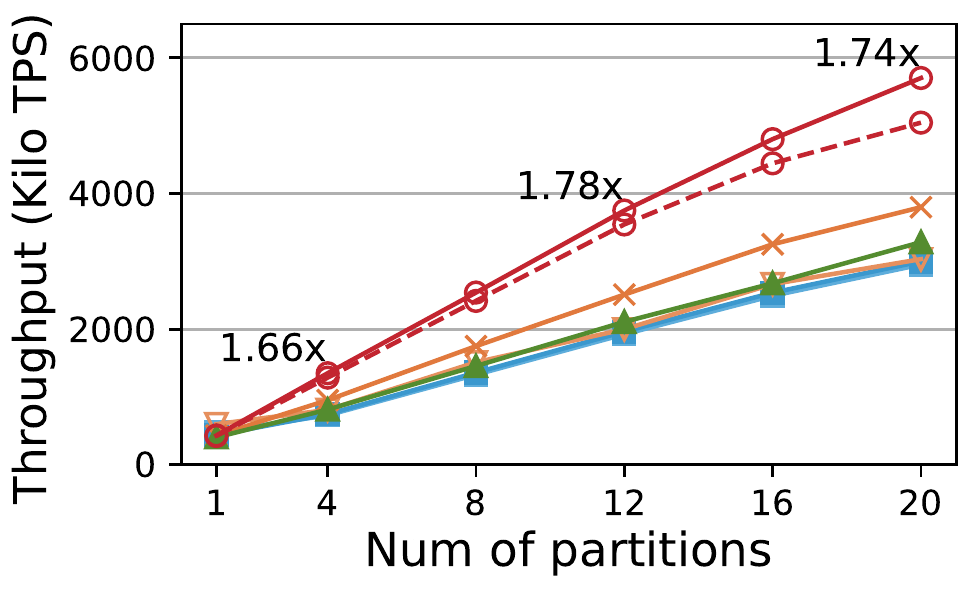}
        \caption{TPC-C}
        \label{fig:tpcc_scale}
    \end{subfigure}
    \end{minipage}
    \caption{Scalability}
\end{figure}

\begin{figure}[t!]
    \begin{minipage}{0.9\linewidth}
    \begin{subfigure}[b]{0.49\linewidth}
        \includegraphics[width=\textwidth]{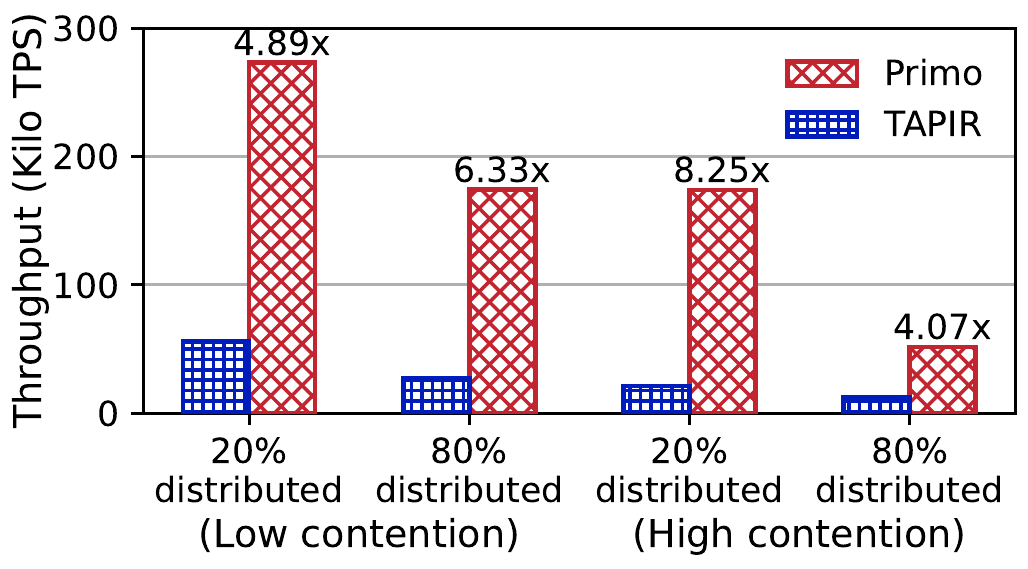}
        \caption{Throughput}
        \label{fig:tapir_throughput}
    \end{subfigure}
    \hfill
    \begin{subfigure}[b]{0.49\linewidth}
        \includegraphics[width=\textwidth]{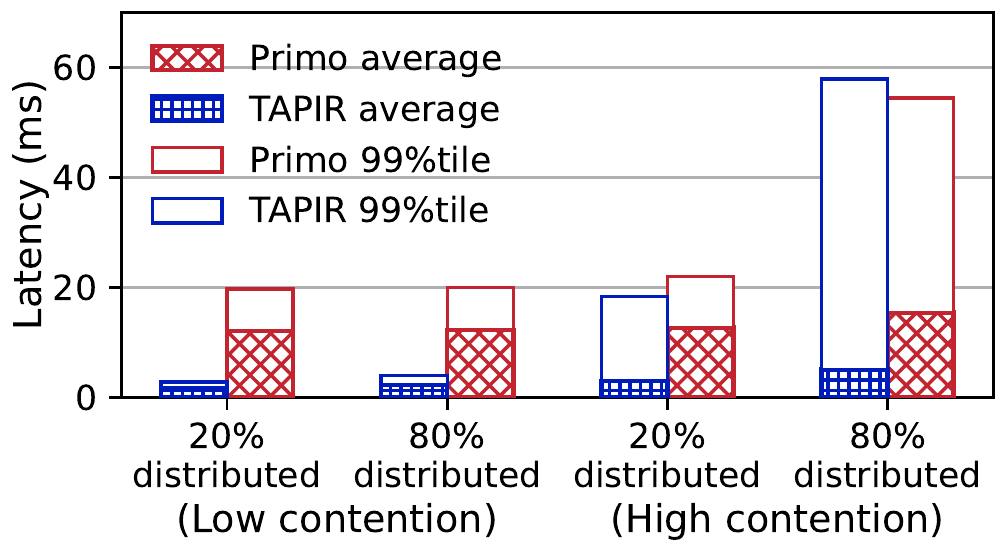}
        \caption{Latency}
        \label{fig:tapir_latency}
    \end{subfigure}
    \caption{Compare with TAPIR}
    \label{fig:tapir}
    \end{minipage}
\end{figure}
}

\techrep{
\subsection{Compare with TAPIR} \label{sec:tapir}
Since the implementation of TAPIR does not support multi-threading, we also restrict Primo to use only one thread per server in this experiment (Although using one thread, both TAPIR and Primo still have concurrent transactions for hiding the network latency).
Figure \ref{fig:tapir} shows the results under various contention and ratios of distributed transactions (we set the skewness to 0.0 and 0.9 in YCSB for low contention and high contention, respectively).
Although it is a cross-system comparison, the orthogonal design goals of Primo and TAPIR are reflected in the results.
Primo offers better throughput by (1) eliminating 2PC and (2) trading latency for throughput (i.e., group commit), while TAPIR features lower latency by having a more lightweight replication protocol.
Interestingly, Primo even has lower tail latency under high contention when 80\% of the transactions are distributed. That's because Primo has lower abort rates due to smaller contention footprint, while transactions in TAPIR retry more times to commit. 
}

%% file: related_work.tex
\section{Related Work} \label{sec:related_work}

\stitle{Distributed Concurrency Control.}
Recently, many distributed concurrency control protocols \cite{spanner,maat,f1,dist-eval,warranties,sundial,centiman,warranties,early-release-1,early-release-2,lotus} are proposed but most of them rely on 2PC. %
Among them, some 2PL variants \cite{early-release-1,early-release-2} allow locks to be released before 2PC at the cost of cascading aborts. Primo effectively achieves the same outcome without that issue.
While some works rely on special hardware \cite{farm,opacity,eris,p4db}, or special transaction models \cite{chains,racoco,carousel}, Primo has no such assumptions.

\stitle{Classic 1PC Protocols.} 
Classic 1PC protocols \cite{coordinator-log-1,coordinator-log-2,IVY,dac,1pc-1,1pc-2,1pc-3,1pc-4,1pc-5,1pc-6} are largely ignored in modern DBMSs due to practicality limitations.
In these protocols, each participant enters the \texttt{PREPARED} state after processing every remote request. 
Thus, the transaction can commit after all the remote operations are successfully processed.
However, to determine whether all the operations have succeeded, each operation must be explicitly acknowledged, which is not the case in modern DBMSs. 
For instance, write operations often do not require explicit acknowledgment because committing a transaction implicitly confirms that all the write operations have been executed successfully. Furthermore, entering the \texttt{PREPARED} state after every operation requires synchronously persisting the state for each operation, resulting in prohibitively expensive overhead, particularly in modern main-memory databases.

\stitle{Deterministic Database.}
Besides the discussions in Section \ref{sec:background}, DDBs have made significant advancements in various aspects.
They provide more efficient deterministic scheduling \cite{pwv, quecc, single-thread-dcc,tpart}, better contention management \cite{caracal}, applications for data migration \cite{mgcrab, migration}, and optimizations to geo-distributed databases \cite{slog, ov2}.
\revision{
QueCC \cite{quecc} also shows that determinism can be optimized to improve core scalability in non-distributed settings.}
However, they still rely on the assumption that transactions' read-write sets can be inferred before execution. This assumption limits them from supporting general distributed transactions.

\stitle{Access Localization.}
Another method of eliminating 2PC is to turn distributed transactions into local transactions. As discussed in Section \ref{sec:intro}, LEAP \cite{leap} pre-localizes data by assuming that the read-write sets of the transactions can be inferred before execution. Thus, it cannot support general distributed transactions.
Zues \cite{zeus} localizes data on-demand, but localizing a record requires shipping the data and updating the ownership directory (on another server) atomically, which effectively brings back 2PC (and even more rounds since a transaction may need to localize multiple records) in the critical path.
Thus they assume a workload where data location seldom changes.
STAR \cite{star} localizes distributed transactions by setting up a centralized replica to hold all data. However, this approach effectively makes the distributed database a monolithic database that cannot scale out.

\stitle{Logging Optimizations.}
In non-distributed setting, while group commit \cite{silo,impementation-tech} is already a common practice, recent research 
advocates logging in parallel \cite{parallel-logging,pacman,taurus,adaptive-logging} and logical logging \cite{logical-logging-1,logical-logging-2}. Primo can incorporate them to further improve logging efficiency.
In distributed setting,  DistDGCC \cite{distdgcc} reduces the time for log construction using dependency information. However, they rely on the prior knowledge of read-write sets, while our WM scheme does not need any assumption.
Early-Lock-Release \cite{elr-2,elr-3} allows releasing locks without waiting for persisting logs, and CLV \cite{elr-1} provides a better implementation of that idea. However, they both require tracking fine-grained dependencies which have high overhead (See experiments in \techrep{Section \ref{sec:exp}}\icde{our technical report \cite{techreport}}).
In contrast, WM does not have that overhead and offers better scalability by removing the global synchronization found in COCO.

\stitle{Replicated Database.}
Recent research co-optimizes 2PC and the replication protocols \cite{mdcc,replicated-commit,min-geo,cockroachdb,TAPIR,consolidating,carousel,logless}.
For example, TAPIR \cite{TAPIR} does not require fully consistent replication to offer serializable distributed transactions. Replicated Commit \cite{replicated-commit} optimizes multi-datacenter replication by overlaying consensus protocol over 2PC instead of the opposite, which reduces the number of inter-datacenter communications.
These techniques are largely orthogonal to Primo, and we plan to co-optimize the WCF protocol with the replication protocol in future work.

\revision{
\stitle{Consensus Problems.} Mutual exclusion \cite{mutex-algorithms,lamport_work,bakery,mutex1,mutex2} and atomic broadcast \cite{dist-book,ab-survey,ab-zookeeper} are consensus problems related to atomic commit. However, they are based on replicated or shared-everything settings while Primo focuses on the shared-nothing setting. 
In general, mutual exclusion focuses on single-item synchronization but Primo is transactional (i.e., multi-item). While a general consensus protocol (e.g., Paxos \cite{consensus-2}) can replace 2PC for atomic commit, Primo eliminates the need for a standalone atomic commit protocol to provide better performance.
}

%% file: conclusion.tex
\section{Conclusion} \label{sec:conclusion}
In this paper, we present Primo, a novel distributed transaction protocol that eliminates 2PC while retaining generality.
It optimizes the transactional throughput by minimizing the contention footprint of the distributed transactions. It allows a transaction to immediately release locks after installing the write-set -- without 2PC and durable log writes in the critical path.
Our experiments on popular OLTP benchmarks show that Primo offers 1.42$\times$ to 8.25$\times$ better throughput than state-of-the-art.

\icde{
\section*{Acknowledgements}
This work is partially supported by 
Alibaba Group (Innovative Research Program and Research Intern Program),
Hong Kong General Research Fund (14200817), Hong Kong AoE/P-404/18, Innovation and Technology Fund (ITS/310/18, ITP/047/19LP) and the Center for Perceptual and Interactive Intelligence (CPII) Ltd under InnoHK supported by the Innovation and Technology Commission (HKSAR Government).
}

%% file: tech_report_main.bbl

%% file: appendix.tex
\appendix
\section{Theoretical Analysis} \label{appendix}
To provide a better understanding of the tradeoff of WCF, we conduct a theoretical analysis to compare the \emph{conflict rates} of WCF and the 2PC-based schemes. 
Formally, the conflict rate of $T$ is the probability that any operation of $T$ conflicts with the operations of other concurrent transactions. 
Given a reasonable implementation of the protocols (e.g., the CPU cycles are well utilized for executing the transactions), a lower conflict rate generally means a better throughput because there are fewer aborts and less lock-waiting.
Since local transactions occupies the majority of common workloads \cite{hstore,slog,granola,voltdb,estore},
we use a local transaction $T_l$ as a representative to study the conflict rate under Primo and 2PC-based schemes, denoted as 
$\mathcal{CR}_{Primo}$ and $\mathcal{CR}_{2PC}$, respectively. 

\stitle{Assumptions and Configurations.}
To simplify the analysis, we make the following assumptions:
\begin{itemize}[leftmargin=2em]
    \item Local transactions are of similar duration \underline{$t_l$}.
    \item Each distributed transaction takes $t_l$ for executing the local logic, and \underline{$t_r$} for a remote access. 
    \item The duration for 2PC is the same as a remote access. Thus, a distributed transaction using 2PC takes time $t_l + 2t_r$, and using WCF reduces time to $t_l + t_r$.
    \item The system is saturated. Specifically, when a worker-thread is waiting for the response of a remote request, it initiates a new transaction to utilize the spared CPU time.
\end{itemize}
We consider a cluster with \underline{$n$} partitions and each server has \underline{$h$} threads.
The ratio of distributed transactions is \underline{$R_d$}, and
every transaction accesses \underline{$m$} different keys where \underline{$R_r$} of them are reads. 
The contention of the workload is captured by \underline{$P_c$}, which is the probability that two randomly selected operations access the same record.
Under this workload, Primo needs to update the \texttt{rts} for \underline{$R_u$} of the read records. 

\stitle{Probability of $T_l$ conflicting with a given transaction.}
With the above setup, we first analyze the probability that $T_l$ conflicts with a given concurrent transaction $T$.
We analyze 2PC and Primo separately.

\sstitle{2PC}: $T_l$ conflicts with $T$ in two cases: (1) $T_l$'s write-set intersects with $T$'s access-set (i.e., read-set $\cup$ write-set); (2) $T_l$'s read-set intersects with $T$'s write-set.
We calculate the probability that $T_l$ conflicts with $T$ by considering its complementary, i.e., \textcolor{blue}{(1)} each pair of keys with one selected from $T_l$'s write-set and another from $T$'s access-set does not match, and \textcolor{red}{(2)} each pair of keys with one selected from $T_l$'s read-set and another from $T$'s read-set does not match:

\begin{equation} \label{eq:c_2pc}
\begin{split}
C_{2PC} &= 1 - \textcolor{blue}{(1 - P_c)^{m(1-R_r) \cdot m}} \textcolor{red}{(1 - P_c)^{mR_r \cdot m(1-R_r)}} \\
        &= 1 - (1 - P_c)^{m^2 (1 - R_r^2)} .
\end{split}
\end{equation}

\sstitle{Primo}: Primo treats local transactions and distributed transactions differently. If $T$ is local, $T_l$ conflicts with $T$ under the same condition as in the 2PC-based protocol. Thus, the probability that $T$ conflicts with $T_l$ is
$C_{Primo\_l} = C_{2PC}$.
If $T$ is a distributed transaction, $T_l$ conflicts with $T$ in three cases:
\textcolor{blue}{(1)} $T_l$'s write-set intersects with $T$'s access-set; \textcolor{red}{(2)} $T_l$'s read records that need to update the \texttt{rts} intersect with $T$'s access-set; and \textcolor{orange}{(3)} the rest of $T_l$'s read records intersects with $T$'s write-set. Therefore, the probability that $T_l$ conflicts with a distributed transaction $T$ is

\begin{equation}
\begin{split}
    C_{Primo\_d} &= 1 - (1 - P_c)^{\textcolor{blue}{m (1-R_r) \cdot m} + \textcolor{red}{m R_r R_u \cdot m} + \textcolor{orange}{m R_r (1-R_u) \cdot m (1-R_r)}} \\
                 &= 1 - (1 - P_c)^{m^2(1-R_r^2+R_r^2R_u)}
\end{split}
\end{equation}

\stitle{Number of concurrent transactions.}
We now analyze the number of transactions that is concurrent to $T_l$. Without loss of generality, we assume $T_l$ starts at time $t$ and finishes at $t + t_l$. We omit the difference between the local transactions in Primo and in 2PC-based schemes, and thus the number of local transactions concurrent to $T_l$ is the same, denoted as $N_l$.

\sstitle{2PC}: A distributed transaction would be concurrent with $T_l$ if it starts within the interval $[t - t_l - 2t_r, t + t_l]$. Since the system would initiate a new transaction when waiting for a remote response, $\frac{2t_l+2t_r}{t_l}$ transactions are initiated in the interval on one worker-thread. Thus, the total number of distributed transactions concurrent to $T_l$ is:

\begin{equation}
    N_{2PC} = R_dnh \frac{2t_l+2t_r}{t_l} = R_dnh(2+2\frac{t_r}{t_l}) \approx 2R_dnh\frac{t_r}{t_l}.
\end{equation}

The approximation is generally valid since $t_r >> t_l$.

\sstitle{Primo}: In Primo, a distributed transaction would be concurrent with $T_l$ if it starts within the interval $[t - t_l - t_r, t+t_l$]. Thus, the number of distributed transactions concurrent to $T_l$ is
\begin{equation}
    N_{Primo} = R_d \frac{2t_r + t_r}{t_l} = R_dnh(2 + \frac{t_r}{t_l}) \approx R_dnh\frac{t_r}{t_l}.
\end{equation}

\stitle{Conflict rate of $T_l$.}
We are now ready to calculate the conflict rate of $T_l$.

\sstitle{2PC}: The conflict rate of $T_l$ can be calculated by considering the complimentary where $T_l$ does not conflict with any concurrent transactions:

\begin{equation} \label{eq:2pc}
\begin{split}
    \mathcal{CR}_{2PC} &= 1 - (1-C_{2PC})^{N_{2PC} + N_{l}} \\
    &= 1 - (1 - P_c)^{m^2R_dnh
    \frac{t_r}{t_l}\mathcolorbox{yellow!50}{(2-2R_r^2)}}(1-C_{2PC})^{N_l}
\end{split}
\end{equation}

\sstitle{Primo}: Similarly in Primo:
\begin{equation} \label{eq:primo}
\begin{split}
    \mathcal{CR}_{Primo} = 1 - (1 - P_c)^{m^2R_dnh\frac{t_r}{t_l}
    \mathcolorbox{yellow!50}{(1-R_r^2+R_r^2R_u)}}  (1-C_{Primo\_l})^{N_l}
\end{split}
\end{equation}

\stitle{Comparisons and Conclusions.}
As we have shown $C_{2PC} = C_{Primo\_l}$,
comparing $\mathcal{CR}_{2PC}$ with $\mathcal{CR}_{Primo}$ only needs to compare the highlighted factors in Equation \ref{eq:2pc} and Equation \ref{eq:primo}, which
depends on the value of $R_r$ and $R_u$ (i.e., the read ratio and the ratio of the read records that need to update \texttt{rts}).
By taking a conservative value of $R_u = 0.6$ (the maximum value seen in our experiments), 
Primo shows a definite advantage when $R_r < 0.8$. 
The improvement is amplified by the contention ($P_c$), the number of accessed records in a transaction ($m$), the number of partitions ($n$), the number of worker-threads per partition ($h$), the network delay ($t_r$) and the ratio of distributed transactions ($R_d$).
In fact, when $R_r > 0.8$, since there are fewer conflicts in such read-mostly workloads, the logical timestamps grow slower and hence there are fewer transactions whose logical timestamps are not in the valid intervals of their read records, which leads to $R_u$ approaching 0. 
In this case, $\mathcal{CR}_{primo} \approx \mathcal{CR}_{2PC}$, meaning that Primo has a similar performance as the 2PC-based schemes when $R_r > 0.8$.
Notice that the above analysis would be less accurate when most of the transactions are distributed transactions because we have been focusing on the conflict rate of a local transaction $T_l$. 
Indeed, in a mostly-distributed read-heavy workload, the performance of Primo would not be as good as 2PC-based protocols due to a large number of extra exclusive-locks.
Nonetheless, since read-heavy workloads have much lower contention, 2PC is less a problem in terms of the contention footprint. Hence, Primo can fallback to use 2PC in this case.

%% file: tech_report_main.bbl
\begin{thebibliography}{110}


\ifx \showCODEN    \undefined \def \showCODEN     #1{\unskip}     \fi
\ifx \showDOI      \undefined \def \showDOI       #1{#1}\fi
\ifx \showISBNx    \undefined \def \showISBNx     #1{\unskip}     \fi
\ifx \showISBNxiii \undefined \def \showISBNxiii  #1{\unskip}     \fi
\ifx \showISSN     \undefined \def \showISSN      #1{\unskip}     \fi
\ifx \showLCCN     \undefined \def \showLCCN      #1{\unskip}     \fi
\ifx \shownote     \undefined \def \shownote      #1{#1}          \fi
\ifx \showarticletitle \undefined \def \showarticletitle #1{#1}   \fi
\ifx \showURL      \undefined \def \showURL       {\relax}        \fi
\providecommand\bibfield[2]{#2}
\providecommand\bibinfo[2]{#2}
\providecommand\natexlab[1]{#1}
\providecommand\showeprint[2][]{arXiv:#2}

\bibitem[tpc(2010)]%
        {tpcc}
 \bibinfo{year}{2010}\natexlab{}.
\newblock \bibinfo{title}{{TPC-C} BENCHMARK Revision 5.11}.
\newblock
\newblock


\bibitem[Abadi and Faleiro(2018)]%
        {deter-overview}
\bibfield{author}{\bibinfo{person}{Daniel~J Abadi} {and}
  \bibinfo{person}{Jose~M Faleiro}.} \bibinfo{year}{2018}\natexlab{}.
\newblock \showarticletitle{An overview of deterministic database systems}.
\newblock \bibinfo{journal}{\emph{Commun. ACM}} \bibinfo{volume}{61},
  \bibinfo{number}{9} (\bibinfo{year}{2018}), \bibinfo{pages}{78--88}.
\newblock


\bibitem[Abdallah(1997)]%
        {1pc-5}
\bibfield{author}{\bibinfo{person}{Maha Abdallah}.}
  \bibinfo{year}{1997}\natexlab{}.
\newblock \showarticletitle{A non-blocking single-phase commit protocol for
  rigorous participants}. In \bibinfo{booktitle}{\emph{In Proceedings of the
  National Conference Bases de Donnes Avances}}. Citeseer.
\newblock


\bibitem[Abdallah et~al\mbox{.}(1998)]%
        {1pc-1}
\bibfield{author}{\bibinfo{person}{Maha Abdallah}, \bibinfo{person}{Rachid
  Guerraoui}, {and} \bibinfo{person}{Philippe Pucheral}.}
  \bibinfo{year}{1998}\natexlab{}.
\newblock \showarticletitle{One-phase commit: does it make sense?}. In
  \bibinfo{booktitle}{\emph{Proceedings 1998 International Conference on
  Parallel and Distributed Systems (Cat. No. 98TB100250)}}. IEEE,
  \bibinfo{pages}{182--192}.
\newblock


\bibitem[Abdallah et~al\mbox{.}(2002)]%
        {dac}
\bibfield{author}{\bibinfo{person}{Maha Abdallah}, \bibinfo{person}{Rachid
  Guerraoui}, {and} \bibinfo{person}{Philippe Pucheral}.}
  \bibinfo{year}{2002}\natexlab{}.
\newblock \showarticletitle{Dictatorial transaction processing: Atomic
  commitment without veto right}.
\newblock \bibinfo{journal}{\emph{Distributed and Parallel Databases}}
  \bibinfo{volume}{11}, \bibinfo{number}{3} (\bibinfo{year}{2002}),
  \bibinfo{pages}{239--268}.
\newblock


\bibitem[Abdallah and Pucheral(1998)]%
        {1pc-3}
\bibfield{author}{\bibinfo{person}{Maha Abdallah} {and}
  \bibinfo{person}{Philippe Pucheral}.} \bibinfo{year}{1998}\natexlab{}.
\newblock \showarticletitle{A single-phase non-blocking atomic commitment
  protocol}. In \bibinfo{booktitle}{\emph{International Conference on Database
  and Expert Systems Applications}}. Springer, \bibinfo{pages}{584--595}.
\newblock


\bibitem[Akidau et~al\mbox{.}(2021)]%
        {flink-1}
\bibfield{author}{\bibinfo{person}{Tyler Akidau}, \bibinfo{person}{Edmon
  Begoli}, \bibinfo{person}{Slava Chernyak}, \bibinfo{person}{Fabian Hueske},
  \bibinfo{person}{Kathryn Knight}, \bibinfo{person}{Kenneth Knowles},
  \bibinfo{person}{Daniel Mills}, {and} \bibinfo{person}{Dan Sotolongo}.}
  \bibinfo{year}{2021}\natexlab{}.
\newblock \bibinfo{booktitle}{\emph{Watermarks in Stream Processing Systems:
  Semantics and Comparative Analysis of {Apache} {Flink} and {Google} {Cloud}
  Dataflow}}.
\newblock \bibinfo{type}{{T}echnical {R}eport}. \bibinfo{institution}{Oak Ridge
  National Lab.(ORNL), Oak Ridge, TN (United States)}.
\newblock


\bibitem[Al-Houmaily and Chrysanthis(1995)]%
        {IVY}
\bibfield{author}{\bibinfo{person}{Y Al-Houmaily} {and} \bibinfo{person}{P
  Chrysanthis}.} \bibinfo{year}{1995}\natexlab{}.
\newblock \showarticletitle{Two-phase commit in gigabit-networked distributed
  databases}. In \bibinfo{booktitle}{\emph{Int. Conf. on Parallel and
  Distributed Computing Systems (PDCS)}}. Citeseer.
\newblock


\bibitem[Al-Houmaily and Chrysanthis(2004a)]%
        {1pc-4}
\bibfield{author}{\bibinfo{person}{Yousef~J Al-Houmaily} {and}
  \bibinfo{person}{Panos~K Chrysanthis}.} \bibinfo{year}{2004}\natexlab{a}.
\newblock \showarticletitle{{1-2PC}: the one-two phase atomic commit protocol}.
  In \bibinfo{booktitle}{\emph{Proceedings of the 2004 ACM symposium on Applied
  computing}}. \bibinfo{pages}{684--691}.
\newblock


\bibitem[Al-Houmaily and Chrysanthis(2004b)]%
        {1pc-6}
\bibfield{author}{\bibinfo{person}{Yousef~J Al-Houmaily} {and}
  \bibinfo{person}{Panos~K Chrysanthis}.} \bibinfo{year}{2004}\natexlab{b}.
\newblock \showarticletitle{{ML-1-2PC}: an adaptive multi-level atomic commit
  protocol}. In \bibinfo{booktitle}{\emph{East European Conference on Advances
  in Databases and Information Systems}}. Springer, \bibinfo{pages}{275--290}.
\newblock


\bibitem[Alomari et~al\mbox{.}(2008)]%
        {smallbank}
\bibfield{author}{\bibinfo{person}{Mohammad Alomari}, \bibinfo{person}{Michael
  Cahill}, \bibinfo{person}{Alan Fekete}, {and} \bibinfo{person}{Uwe Rohm}.}
  \bibinfo{year}{2008}\natexlab{}.
\newblock \showarticletitle{The cost of serializability on platforms that use
  snapshot isolation}. In \bibinfo{booktitle}{\emph{2008 IEEE 24th
  International Conference on Data Engineering}}. IEEE,
  \bibinfo{pages}{576--585}.
\newblock


\bibitem[Androulaki et~al\mbox{.}(2018)]%
        {fabric}
\bibfield{author}{\bibinfo{person}{Elli Androulaki}, \bibinfo{person}{Artem
  Barger}, \bibinfo{person}{Vita Bortnikov}, \bibinfo{person}{Christian
  Cachin}, \bibinfo{person}{Konstantinos Christidis}, \bibinfo{person}{Angelo
  De~Caro}, \bibinfo{person}{David Enyeart}, \bibinfo{person}{Christopher
  Ferris}, \bibinfo{person}{Gennady Laventman}, \bibinfo{person}{Yacov
  Manevich}, {et~al\mbox{.}}} \bibinfo{year}{2018}\natexlab{}.
\newblock \showarticletitle{Hyperledger fabric: a distributed operating system
  for permissioned blockchains}. In \bibinfo{booktitle}{\emph{Proceedings of
  the thirteenth EuroSys conference}}. \bibinfo{pages}{1--15}.
\newblock


\bibitem[Carbone et~al\mbox{.}(2015)]%
        {flink-2}
\bibfield{author}{\bibinfo{person}{Paris Carbone}, \bibinfo{person}{Asterios
  Katsifodimos}, \bibinfo{person}{Stephan Ewen}, \bibinfo{person}{Volker
  Markl}, \bibinfo{person}{Seif Haridi}, {and} \bibinfo{person}{Kostas
  Tzoumas}.} \bibinfo{year}{2015}\natexlab{}.
\newblock \showarticletitle{Apache {Flink}: Stream and batch processing in a
  single engine}.
\newblock \bibinfo{journal}{\emph{Bulletin of the IEEE Computer Society
  Technical Committee on Data Engineering}} \bibinfo{volume}{36},
  \bibinfo{number}{4} (\bibinfo{year}{2015}).
\newblock


\bibitem[Cooper et~al\mbox{.}(2010)]%
        {ycsb}
\bibfield{author}{\bibinfo{person}{Brian~F Cooper}, \bibinfo{person}{Adam
  Silberstein}, \bibinfo{person}{Erwin Tam}, \bibinfo{person}{Raghu
  Ramakrishnan}, {and} \bibinfo{person}{Russell Sears}.}
  \bibinfo{year}{2010}\natexlab{}.
\newblock \showarticletitle{Benchmarking cloud serving systems with {YCSB}}. In
  \bibinfo{booktitle}{\emph{Proceedings of the 1st ACM symposium on Cloud
  computing}}. \bibinfo{pages}{143--154}.
\newblock


\bibitem[Corbett et~al\mbox{.}(2013)]%
        {spanner}
\bibfield{author}{\bibinfo{person}{James~C Corbett}, \bibinfo{person}{Jeffrey
  Dean}, \bibinfo{person}{Michael Epstein}, \bibinfo{person}{Andrew Fikes},
  \bibinfo{person}{Christopher Frost}, \bibinfo{person}{Jeffrey~John Furman},
  \bibinfo{person}{Sanjay Ghemawat}, \bibinfo{person}{Andrey Gubarev},
  \bibinfo{person}{Christopher Heiser}, \bibinfo{person}{Peter Hochschild},
  {et~al\mbox{.}}} \bibinfo{year}{2013}\natexlab{}.
\newblock \showarticletitle{Spanner: Google’s globally distributed database}.
\newblock \bibinfo{journal}{\emph{ACM Transactions on Computer Systems (TOCS)}}
  \bibinfo{volume}{31}, \bibinfo{number}{3} (\bibinfo{year}{2013}),
  \bibinfo{pages}{1--22}.
\newblock


\bibitem[Cowling and Liskov(2012)]%
        {granola}
\bibfield{author}{\bibinfo{person}{James Cowling} {and}
  \bibinfo{person}{Barbara Liskov}.} \bibinfo{year}{2012}\natexlab{}.
\newblock \showarticletitle{Granola: Low-Overhead distributed transaction
  coordination}. In \bibinfo{booktitle}{\emph{2012 USENIX Annual Technical
  Conference (USENIX ATC 12)}}. \bibinfo{pages}{223--235}.
\newblock


\bibitem[D{\'e}fago et~al\mbox{.}(2004)]%
        {ab-survey}
\bibfield{author}{\bibinfo{person}{Xavier D{\'e}fago},
  \bibinfo{person}{Andr{\'e} Schiper}, {and} \bibinfo{person}{P{\'e}ter
  Urb{\'a}n}.} \bibinfo{year}{2004}\natexlab{}.
\newblock \showarticletitle{Total order broadcast and multicast algorithms:
  Taxonomy and survey}.
\newblock \bibinfo{journal}{\emph{ACM Computing Surveys (CSUR)}}
  \bibinfo{volume}{36}, \bibinfo{number}{4} (\bibinfo{year}{2004}),
  \bibinfo{pages}{372--421}.
\newblock


\bibitem[DeWitt et~al\mbox{.}(1984)]%
        {impementation-tech}
\bibfield{author}{\bibinfo{person}{David~J DeWitt}, \bibinfo{person}{Randy~H
  Katz}, \bibinfo{person}{Frank Olken}, \bibinfo{person}{Leonard~D Shapiro},
  \bibinfo{person}{Michael~R Stonebraker}, {and} \bibinfo{person}{David~A
  Wood}.} \bibinfo{year}{1984}\natexlab{}.
\newblock \showarticletitle{Implementation techniques for main memory database
  systems}. In \bibinfo{booktitle}{\emph{Proceedings of the 1984 ACM SIGMOD
  international conference on management of data}}. \bibinfo{pages}{1--8}.
\newblock


\bibitem[Ding et~al\mbox{.}(2015)]%
        {centiman}
\bibfield{author}{\bibinfo{person}{Bailu Ding}, \bibinfo{person}{Lucja Kot},
  \bibinfo{person}{Alan Demers}, {and} \bibinfo{person}{Johannes Gehrke}.}
  \bibinfo{year}{2015}\natexlab{}.
\newblock \showarticletitle{Centiman: elastic, high performance optimistic
  concurrency control by watermarking}. In
  \bibinfo{booktitle}{\emph{Proceedings of the Sixth ACM Symposium on Cloud
  Computing}}. \bibinfo{pages}{262--275}.
\newblock


\bibitem[Dragojevi{\'c} et~al\mbox{.}(2015)]%
        {farm}
\bibfield{author}{\bibinfo{person}{Aleksandar Dragojevi{\'c}},
  \bibinfo{person}{Dushyanth Narayanan}, \bibinfo{person}{Edmund~B
  Nightingale}, \bibinfo{person}{Matthew Renzelmann}, \bibinfo{person}{Alex
  Shamis}, \bibinfo{person}{Anirudh Badam}, {and} \bibinfo{person}{Miguel
  Castro}.} \bibinfo{year}{2015}\natexlab{}.
\newblock \showarticletitle{No compromises: distributed transactions with
  consistency, availability, and performance}. In
  \bibinfo{booktitle}{\emph{Proceedings of the 25th symposium on operating
  systems principles}}. \bibinfo{pages}{54--70}.
\newblock


\bibitem[Elbagir et~al\mbox{.}(2016)]%
        {presumed-opt-1}
\bibfield{author}{\bibinfo{person}{Fadia~A Elbagir}, \bibinfo{person}{Ahmed
  Khalid}, {and} \bibinfo{person}{Khalid Khanfar}.}
  \bibinfo{year}{2016}\natexlab{}.
\newblock \showarticletitle{A survey of commit protocols in distributed real
  time database systems}.
\newblock \bibinfo{journal}{\emph{International Journal of Emerging Trends \&
  Technology in Computer Science}} \bibinfo{volume}{31}, \bibinfo{number}{2}
  (\bibinfo{year}{2016}), \bibinfo{pages}{61--66}.
\newblock


\bibitem[Faleiro and Abadi(2015)]%
        {bohm}
\bibfield{author}{\bibinfo{person}{Jose~M Faleiro} {and}
  \bibinfo{person}{Daniel~J Abadi}.} \bibinfo{year}{2015}\natexlab{}.
\newblock \showarticletitle{Rethinking serializable multiversion concurrency
  control}.
\newblock \bibinfo{journal}{\emph{Proceedings of the VLDB Endowment}}
  \bibinfo{volume}{8}, \bibinfo{number}{11} (\bibinfo{year}{2015}).
\newblock


\bibitem[Faleiro et~al\mbox{.}(2017)]%
        {pwv}
\bibfield{author}{\bibinfo{person}{Jose~M Faleiro}, \bibinfo{person}{Daniel~J
  Abadi}, {and} \bibinfo{person}{Joseph~M Hellerstein}.}
  \bibinfo{year}{2017}\natexlab{}.
\newblock \showarticletitle{High performance transactions via early write
  visibility}.
\newblock \bibinfo{journal}{\emph{Proceedings of the VLDB Endowment}}
  \bibinfo{volume}{10}, \bibinfo{number}{5} (\bibinfo{year}{2017}).
\newblock


\bibitem[Fan and Golab(2019)]%
        {ov1}
\bibfield{author}{\bibinfo{person}{Hua Fan} {and} \bibinfo{person}{Wojciech
  Golab}.} \bibinfo{year}{2019}\natexlab{}.
\newblock \showarticletitle{Ocean vista: gossip-based visibility control for
  speedy geo-distributed transactions}.
\newblock \bibinfo{journal}{\emph{Proceedings of the VLDB Endowment}}
  \bibinfo{volume}{12}, \bibinfo{number}{11} (\bibinfo{year}{2019}),
  \bibinfo{pages}{1471--1484}.
\newblock


\bibitem[Fan and Golab(2021)]%
        {ov2}
\bibfield{author}{\bibinfo{person}{Hua Fan} {and} \bibinfo{person}{Wojciech~M.
  Golab}.} \bibinfo{year}{2021}\natexlab{}.
\newblock \showarticletitle{Gossip-based visibility control for
  high-performance geo-distributed transactions}.
\newblock \bibinfo{journal}{\emph{{VLDB} J.}} \bibinfo{volume}{30},
  \bibinfo{number}{1} (\bibinfo{year}{2021}), \bibinfo{pages}{93--114}.
\newblock


\bibitem[Fekete et~al\mbox{.}(2005)]%
        {ssi-1}
\bibfield{author}{\bibinfo{person}{Alan Fekete}, \bibinfo{person}{Dimitrios
  Liarokapis}, \bibinfo{person}{Elizabeth O'Neil}, \bibinfo{person}{Patrick
  O'Neil}, {and} \bibinfo{person}{Dennis Shasha}.}
  \bibinfo{year}{2005}\natexlab{}.
\newblock \showarticletitle{Making snapshot isolation serializable}.
\newblock \bibinfo{journal}{\emph{ACM Transactions on Database Systems (TODS)}}
  \bibinfo{volume}{30}, \bibinfo{number}{2} (\bibinfo{year}{2005}),
  \bibinfo{pages}{492--528}.
\newblock


\bibitem[Galakatos and Zamanian(2016)]%
        {2pc-overhead}
\bibfield{author}{\bibinfo{person}{Carsten Binnig Andrew Crotty~Alex Galakatos}
  {and} \bibinfo{person}{Tim Kraska~Erfan Zamanian}.}
  \bibinfo{year}{2016}\natexlab{}.
\newblock \showarticletitle{The End of Slow Networks: It’s Time for a
  Redesign}.
\newblock \bibinfo{journal}{\emph{Proceedings of the VLDB Endowment}}
  \bibinfo{volume}{9}, \bibinfo{number}{7} (\bibinfo{year}{2016}).
\newblock


\bibitem[Golab and Hendler(2017)]%
        {mutex1}
\bibfield{author}{\bibinfo{person}{Wojciech Golab} {and} \bibinfo{person}{Danny
  Hendler}.} \bibinfo{year}{2017}\natexlab{}.
\newblock \showarticletitle{Recoverable mutual exclusion in sub-logarithmic
  time}. In \bibinfo{booktitle}{\emph{Proceedings of the ACM Symposium on
  Principles of Distributed Computing}}. \bibinfo{pages}{211--220}.
\newblock


\bibitem[Golab and Hendler(2018)]%
        {mutex2}
\bibfield{author}{\bibinfo{person}{Wojciech Golab} {and} \bibinfo{person}{Danny
  Hendler}.} \bibinfo{year}{2018}\natexlab{}.
\newblock \showarticletitle{Recoverable mutual exclusion under system-wide
  failures}. In \bibinfo{booktitle}{\emph{Proceedings of the 2018 ACM Symposium
  on Principles of Distributed Computing}}. \bibinfo{pages}{17--26}.
\newblock


\bibitem[Graefe et~al\mbox{.}(2013)]%
        {elr-1}
\bibfield{author}{\bibinfo{person}{Goetz Graefe}, \bibinfo{person}{Mark
  Lillibridge}, \bibinfo{person}{Harumi Kuno}, \bibinfo{person}{Joseph Tucek},
  {and} \bibinfo{person}{Alistair Veitch}.} \bibinfo{year}{2013}\natexlab{}.
\newblock \showarticletitle{Controlled lock violation}. In
  \bibinfo{booktitle}{\emph{Proceedings of the 2013 ACM SIGMOD International
  Conference on Management of Data}}. \bibinfo{pages}{85--96}.
\newblock


\bibitem[Gray et~al\mbox{.}(1994)]%
        {zipf}
\bibfield{author}{\bibinfo{person}{Jim Gray}, \bibinfo{person}{Prakash
  Sundaresan}, \bibinfo{person}{Susanne Englert}, \bibinfo{person}{Ken
  Baclawski}, {and} \bibinfo{person}{Peter~J Weinberger}.}
  \bibinfo{year}{1994}\natexlab{}.
\newblock \showarticletitle{Quickly generating billion-record synthetic
  databases}. In \bibinfo{booktitle}{\emph{Proceedings of the 1994 ACM SIGMOD
  international conference on Management of data}}. \bibinfo{pages}{243--252}.
\newblock


\bibitem[Gupta et~al\mbox{.}(1997)]%
        {early-release-1}
\bibfield{author}{\bibinfo{person}{Ramesh Gupta}, \bibinfo{person}{Jayant
  Haritsa}, {and} \bibinfo{person}{Krithi Ramamritham}.}
  \bibinfo{year}{1997}\natexlab{}.
\newblock \showarticletitle{Revisiting commit processing in distributed
  database systems}. In \bibinfo{booktitle}{\emph{Proceedings of the 1997 ACM
  SIGMOD international conference on Management of data}}.
  \bibinfo{pages}{486--497}.
\newblock


\bibitem[Harding et~al\mbox{.}(2017)]%
        {dist-eval}
\bibfield{author}{\bibinfo{person}{Rachael Harding}, \bibinfo{person}{Dana
  Van~Aken}, \bibinfo{person}{Andrew Pavlo}, {and} \bibinfo{person}{Michael
  Stonebraker}.} \bibinfo{year}{2017}\natexlab{}.
\newblock \showarticletitle{An evaluation of distributed concurrency control}.
\newblock \bibinfo{journal}{\emph{Proceedings of the VLDB Endowment}}
  \bibinfo{volume}{10}, \bibinfo{number}{5} (\bibinfo{year}{2017}),
  \bibinfo{pages}{553--564}.
\newblock


\bibitem[Huang et~al\mbox{.}(2017)]%
        {gray-failure}
\bibfield{author}{\bibinfo{person}{Peng Huang}, \bibinfo{person}{Chuanxiong
  Guo}, \bibinfo{person}{Lidong Zhou}, \bibinfo{person}{Jacob~R Lorch},
  \bibinfo{person}{Yingnong Dang}, \bibinfo{person}{Murali Chintalapati}, {and}
  \bibinfo{person}{Randolph Yao}.} \bibinfo{year}{2017}\natexlab{}.
\newblock \showarticletitle{Gray failure: The achilles' heel of cloud-scale
  systems}. In \bibinfo{booktitle}{\emph{Proceedings of the 16th Workshop on
  Hot Topics in Operating Systems}}. \bibinfo{pages}{150--155}.
\newblock


\bibitem[Huang et~al\mbox{.}(2022)]%
        {opportunities}
\bibfield{author}{\bibinfo{person}{Yihe Huang}, \bibinfo{person}{William Qian},
  \bibinfo{person}{Eddie Kohler}, \bibinfo{person}{Barbara Liskov}, {and}
  \bibinfo{person}{Liuba Shrira}.} \bibinfo{year}{2022}\natexlab{}.
\newblock \showarticletitle{Opportunities for optimism in contended main-memory
  multicore transactions}.
\newblock \bibinfo{journal}{\emph{The VLDB Journal}} (\bibinfo{year}{2022}),
  \bibinfo{pages}{1--23}.
\newblock


\bibitem[Jasny et~al\mbox{.}(2022)]%
        {p4db}
\bibfield{author}{\bibinfo{person}{Matthias Jasny}, \bibinfo{person}{Lasse
  Thostrup}, \bibinfo{person}{Tobias Ziegler}, {and} \bibinfo{person}{Carsten
  Binnig}.} \bibinfo{year}{2022}\natexlab{}.
\newblock \showarticletitle{{P4DB}-the case for in-network {OLTP}}. In
  \bibinfo{booktitle}{\emph{Proceedings of the 2022 International Conference on
  Management of Data}}. \bibinfo{pages}{1375--1389}.
\newblock


\bibitem[Jones et~al\mbox{.}(2010)]%
        {early-release-2}
\bibfield{author}{\bibinfo{person}{Evan~PC Jones}, \bibinfo{person}{Daniel~J
  Abadi}, {and} \bibinfo{person}{Samuel Madden}.}
  \bibinfo{year}{2010}\natexlab{}.
\newblock \showarticletitle{Low overhead concurrency control for partitioned
  main memory databases}. In \bibinfo{booktitle}{\emph{Proceedings of the 2010
  ACM SIGMOD International Conference on Management of data}}.
  \bibinfo{pages}{603--614}.
\newblock


\bibitem[Junqueira and Reed(2013)]%
        {zookeeper}
\bibfield{author}{\bibinfo{person}{Flavio Junqueira} {and}
  \bibinfo{person}{Benjamin Reed}.} \bibinfo{year}{2013}\natexlab{}.
\newblock \bibinfo{booktitle}{\emph{ZooKeeper: distributed process
  coordination}}.
\newblock \bibinfo{publisher}{" O'Reilly Media, Inc."}.
\newblock


\bibitem[Kallman et~al\mbox{.}(2008)]%
        {hstore}
\bibfield{author}{\bibinfo{person}{Robert Kallman}, \bibinfo{person}{Hideaki
  Kimura}, \bibinfo{person}{Jonathan Natkins}, \bibinfo{person}{Andrew Pavlo},
  \bibinfo{person}{Alexander Rasin}, \bibinfo{person}{Stanley Zdonik},
  \bibinfo{person}{Evan~PC Jones}, \bibinfo{person}{Samuel Madden},
  \bibinfo{person}{Michael Stonebraker}, \bibinfo{person}{Yang Zhang},
  {et~al\mbox{.}}} \bibinfo{year}{2008}\natexlab{}.
\newblock \showarticletitle{H-store: a high-performance, distributed main
  memory transaction processing system}.
\newblock \bibinfo{journal}{\emph{Proceedings of the VLDB Endowment}}
  \bibinfo{volume}{1}, \bibinfo{number}{2} (\bibinfo{year}{2008}),
  \bibinfo{pages}{1496--1499}.
\newblock


\bibitem[Katsarakis et~al\mbox{.}(2021)]%
        {zeus}
\bibfield{author}{\bibinfo{person}{Antonios Katsarakis}, \bibinfo{person}{Yijun
  Ma}, \bibinfo{person}{Zhaowei Tan}, \bibinfo{person}{Andrew Bainbridge},
  \bibinfo{person}{Matthew Balkwill}, \bibinfo{person}{Aleksandar Dragojevic},
  \bibinfo{person}{Boris Grot}, \bibinfo{person}{Bozidar Radunovic}, {and}
  \bibinfo{person}{Yongguang Zhang}.} \bibinfo{year}{2021}\natexlab{}.
\newblock \showarticletitle{Zeus: Locality-aware distributed transactions}. In
  \bibinfo{booktitle}{\emph{Proceedings of the Sixteenth European Conference on
  Computer Systems}}. \bibinfo{pages}{145--161}.
\newblock


\bibitem[Kimura et~al\mbox{.}(2012)]%
        {elr-2}
\bibfield{author}{\bibinfo{person}{Hideaki Kimura}, \bibinfo{person}{Goetz
  Graefe}, {and} \bibinfo{person}{Harumi~A Kuno}.}
  \bibinfo{year}{2012}\natexlab{}.
\newblock \showarticletitle{Efficient locking techniques for databases on
  modern hardware.}. In \bibinfo{booktitle}{\emph{ADMS@ VLDB}}. Citeseer,
  \bibinfo{pages}{1--12}.
\newblock


\bibitem[Kraska et~al\mbox{.}(2013)]%
        {mdcc}
\bibfield{author}{\bibinfo{person}{Tim Kraska}, \bibinfo{person}{Gene Pang},
  \bibinfo{person}{Michael~J Franklin}, \bibinfo{person}{Samuel Madden}, {and}
  \bibinfo{person}{Alan Fekete}.} \bibinfo{year}{2013}\natexlab{}.
\newblock \showarticletitle{{MDCC}: Multi-data center consistency}. In
  \bibinfo{booktitle}{\emph{Proceedings of the 8th ACM European Conference on
  Computer Systems}}. \bibinfo{pages}{113--126}.
\newblock


\bibitem[Kshemkalyani and Singhal(2011)]%
        {dist-book}
\bibfield{author}{\bibinfo{person}{Ajay~D Kshemkalyani} {and}
  \bibinfo{person}{Mukesh Singhal}.} \bibinfo{year}{2011}\natexlab{}.
\newblock \bibinfo{booktitle}{\emph{Distributed computing: principles,
  algorithms, and systems}}.
\newblock \bibinfo{publisher}{Cambridge University Press}.
\newblock


\bibitem[Lamport(2001)]%
        {consensus-2}
\bibfield{author}{\bibinfo{person}{Leslie Lamport}.}
  \bibinfo{year}{2001}\natexlab{}.
\newblock \showarticletitle{Paxos made simple}.
\newblock \bibinfo{journal}{\emph{ACM SIGACT News (Distributed Computing
  Column) 32, 4 (Whole Number 121, December 2001)}} (\bibinfo{year}{2001}),
  \bibinfo{pages}{51--58}.
\newblock


\bibitem[Lamport(2019)]%
        {bakery}
\bibfield{author}{\bibinfo{person}{Leslie Lamport}.}
  \bibinfo{year}{2019}\natexlab{}.
\newblock \showarticletitle{A new solution of Dijkstra's concurrent programming
  problem}.
\newblock In \bibinfo{booktitle}{\emph{Concurrency: the works of leslie
  lamport}}. \bibinfo{pages}{171--178}.
\newblock


\bibitem[Lampson and Lomet(1993)]%
        {NPrC}
\bibfield{author}{\bibinfo{person}{Butler Lampson} {and} \bibinfo{person}{David
  Lomet}.} \bibinfo{year}{1993}\natexlab{}.
\newblock \showarticletitle{A new presumed commit optimization for two phase
  commit}. In \bibinfo{booktitle}{\emph{19th International Conference on Very
  Large Data Bases (VLDB'93)}}. \bibinfo{pages}{630--640}.
\newblock


\bibitem[Larson et~al\mbox{.}(2011)]%
        {mvcc-1}
\bibfield{author}{\bibinfo{person}{Per-{\AA}ke Larson}, \bibinfo{person}{Spyros
  Blanas}, \bibinfo{person}{Cristian Diaconu}, \bibinfo{person}{Craig
  Freedman}, \bibinfo{person}{Jignesh~M Patel}, {and} \bibinfo{person}{Mike
  Zwilling}.} \bibinfo{year}{2011}\natexlab{}.
\newblock \showarticletitle{High-performance concurrency control mechanisms for
  main-memory databases}.
\newblock \bibinfo{journal}{\emph{Proceedings of the VLDB Endowment}}
  \bibinfo{volume}{5}, \bibinfo{number}{4} (\bibinfo{year}{2011}),
  \bibinfo{pages}{298--309}.
\newblock


\bibitem[Lee and Yeom(2002)]%
        {1pc-2}
\bibfield{author}{\bibinfo{person}{Inseon Lee} {and}
  \bibinfo{person}{Heon~Young Yeom}.} \bibinfo{year}{2002}\natexlab{}.
\newblock \showarticletitle{A single phase distributed commit protocol for main
  memory database systems}. In \bibinfo{booktitle}{\emph{Proceedings 16th
  International Parallel and Distributed Processing Symposium}}. IEEE,
  \bibinfo{pages}{8--pp}.
\newblock


\bibitem[Li et~al\mbox{.}(2017)]%
        {eris}
\bibfield{author}{\bibinfo{person}{Jialin Li}, \bibinfo{person}{Ellis Michael},
  {and} \bibinfo{person}{Dan~RK Ports}.} \bibinfo{year}{2017}\natexlab{}.
\newblock \showarticletitle{Eris: Coordination-free consistent transactions
  using in-network concurrency control}. In
  \bibinfo{booktitle}{\emph{Proceedings of the 26th Symposium on Operating
  Systems Principles}}. \bibinfo{pages}{104--120}.
\newblock


\bibitem[Lim et~al\mbox{.}(2017)]%
        {cicada}
\bibfield{author}{\bibinfo{person}{Hyeontaek Lim}, \bibinfo{person}{Michael
  Kaminsky}, {and} \bibinfo{person}{David~G Andersen}.}
  \bibinfo{year}{2017}\natexlab{}.
\newblock \showarticletitle{Cicada: Dependably fast multi-core in-memory
  transactions}. In \bibinfo{booktitle}{\emph{Proceedings of the 2017 ACM
  International Conference on Management of Data}}. \bibinfo{pages}{21--35}.
\newblock


\bibitem[Lin et~al\mbox{.}(2016)]%
        {leap}
\bibfield{author}{\bibinfo{person}{Qian Lin}, \bibinfo{person}{Pengfei Chang},
  \bibinfo{person}{Gang Chen}, \bibinfo{person}{Beng~Chin Ooi},
  \bibinfo{person}{Kian-Lee Tan}, {and} \bibinfo{person}{Zhengkui Wang}.}
  \bibinfo{year}{2016}\natexlab{}.
\newblock \showarticletitle{Towards a Non-{2PC} Transaction Management in
  Distributed Database Systems}. In \bibinfo{booktitle}{\emph{Proceedings of
  the 2016 International Conference on Management of Data}}.
  \bibinfo{pages}{1659--1674}.
\newblock


\bibitem[Lin et~al\mbox{.}(2019)]%
        {mgcrab}
\bibfield{author}{\bibinfo{person}{Yu-Shan Lin}, \bibinfo{person}{Shao-Kan Pi},
  \bibinfo{person}{Meng-Kai Liao}, \bibinfo{person}{Ching Tsai},
  \bibinfo{person}{Aaron Elmore}, {and} \bibinfo{person}{Shan-Hung Wu}.}
  \bibinfo{year}{2019}\natexlab{}.
\newblock \showarticletitle{MgCrab: transaction crabbing for live migration in
  deterministic database systems}.
\newblock \bibinfo{journal}{\emph{Proceedings of the VLDB Endowment}}
  \bibinfo{volume}{12}, \bibinfo{number}{5} (\bibinfo{year}{2019}),
  \bibinfo{pages}{597--610}.
\newblock


\bibitem[Lin et~al\mbox{.}(2021)]%
        {migration}
\bibfield{author}{\bibinfo{person}{Yu-Shan Lin}, \bibinfo{person}{Ching Tsai},
  \bibinfo{person}{Tz-Yu Lin}, \bibinfo{person}{Yun-Sheng Chang}, {and}
  \bibinfo{person}{Shan-Hung Wu}.} \bibinfo{year}{2021}\natexlab{}.
\newblock \showarticletitle{Don't Look Back, Look into the Future: Prescient
  Data Partitioning and Migration for Deterministic Database Systems}. In
  \bibinfo{booktitle}{\emph{Proceedings of the 2021 International Conference on
  Management of Data}}. \bibinfo{pages}{1156--1168}.
\newblock


\bibitem[Liu et~al\mbox{.}(2014)]%
        {warranties}
\bibfield{author}{\bibinfo{person}{Jed Liu}, \bibinfo{person}{Tom Magrino},
  \bibinfo{person}{Owen Arden}, \bibinfo{person}{Michael~D George}, {and}
  \bibinfo{person}{Andrew~C Myers}.} \bibinfo{year}{2014}\natexlab{}.
\newblock \showarticletitle{Warranties for faster strong consistency}. In
  \bibinfo{booktitle}{\emph{11th USENIX Symposium on Networked Systems Design
  and Implementation (NSDI 14)}}. \bibinfo{pages}{503--517}.
\newblock


\bibitem[Lomet et~al\mbox{.}(2011)]%
        {logical-logging-2}
\bibfield{author}{\bibinfo{person}{David Lomet}, \bibinfo{person}{Kostas
  Tzoumas}, {and} \bibinfo{person}{Michael Zwilling}.}
  \bibinfo{year}{2011}\natexlab{}.
\newblock \showarticletitle{Implementing Performance Competitive Logical
  Recovery}.
\newblock \bibinfo{journal}{\emph{Proceedings of the VLDB Endowment}}
  \bibinfo{volume}{4}, \bibinfo{number}{7} (\bibinfo{year}{2011}).
\newblock


\bibitem[Lu(2020)]%
        {aria-code}
\bibfield{author}{\bibinfo{person}{Yi Lu}.} \bibinfo{year}{2020}\natexlab{}.
\newblock \bibinfo{title}{Aria open-source code}.
\newblock \bibinfo{howpublished}{\url{https://github.com/luyi0619/aria}}.
\newblock
\newblock
\shownote{Accessed: 2023-1-6}.


\bibitem[Lu et~al\mbox{.}(2020)]%
        {aria}
\bibfield{author}{\bibinfo{person}{Yi Lu}, \bibinfo{person}{Xiangyao Yu},
  \bibinfo{person}{Lei Cao}, {and} \bibinfo{person}{Samuel Madden}.}
  \bibinfo{year}{2020}\natexlab{}.
\newblock \showarticletitle{Aria: a fast and practical deterministic {OLTP}
  database}.
\newblock \bibinfo{journal}{\emph{Proceedings of the VLDB Endowment}}
  \bibinfo{volume}{13}, \bibinfo{number}{12} (\bibinfo{year}{2020}),
  \bibinfo{pages}{2047--2060}.
\newblock


\bibitem[Lu et~al\mbox{.}(2021)]%
        {coco}
\bibfield{author}{\bibinfo{person}{Yi Lu}, \bibinfo{person}{Xiangyao Yu},
  \bibinfo{person}{Lei Cao}, {and} \bibinfo{person}{Samuel Madden}.}
  \bibinfo{year}{2021}\natexlab{}.
\newblock \showarticletitle{Epoch-based commit and replication in distributed
  {OLTP} databases}.
\newblock \bibinfo{journal}{\emph{Proceedings of the VLDB Endowment}}
  \bibinfo{volume}{14}, \bibinfo{number}{5} (\bibinfo{year}{2021}),
  \bibinfo{pages}{743--756}.
\newblock


\bibitem[Lu et~al\mbox{.}(2019)]%
        {star}
\bibfield{author}{\bibinfo{person}{Yi Lu}, \bibinfo{person}{Xiangyao Yu}, {and}
  \bibinfo{person}{Samuel Madden}.} \bibinfo{year}{2019}\natexlab{}.
\newblock \showarticletitle{{STAR}: Scaling Transactions through Asymmetric
  Replication}.
\newblock \bibinfo{journal}{\emph{Proceedings of the VLDB Endowment}}
  \bibinfo{volume}{12}, \bibinfo{number}{11} (\bibinfo{year}{2019}).
\newblock


\bibitem[Mahmoud et~al\mbox{.}(2013)]%
        {replicated-commit}
\bibfield{author}{\bibinfo{person}{Hatem Mahmoud}, \bibinfo{person}{Faisal
  Nawab}, \bibinfo{person}{Alexander Pucher}, \bibinfo{person}{Divyakant
  Agrawal}, {and} \bibinfo{person}{Amr El~Abbadi}.}
  \bibinfo{year}{2013}\natexlab{}.
\newblock \showarticletitle{Low-latency multi-datacenter databases using
  replicated commit}.
\newblock \bibinfo{journal}{\emph{Proceedings of the VLDB Endowment}}
  \bibinfo{volume}{6}, \bibinfo{number}{9} (\bibinfo{year}{2013}),
  \bibinfo{pages}{661--672}.
\newblock


\bibitem[Mahmoud et~al\mbox{.}(2014)]%
        {maat}
\bibfield{author}{\bibinfo{person}{Hatem~A Mahmoud}, \bibinfo{person}{Vaibhav
  Arora}, \bibinfo{person}{Faisal Nawab}, \bibinfo{person}{Divyakant Agrawal},
  {and} \bibinfo{person}{Amr El~Abbadi}.} \bibinfo{year}{2014}\natexlab{}.
\newblock \showarticletitle{Maat: Effective and scalable coordination of
  distributed transactions in the cloud}.
\newblock \bibinfo{journal}{\emph{Proceedings of the VLDB Endowment}}
  \bibinfo{volume}{7}, \bibinfo{number}{5} (\bibinfo{year}{2014}),
  \bibinfo{pages}{329--340}.
\newblock


\bibitem[Malkhi(2019)]%
        {lamport_work}
\bibfield{author}{\bibinfo{person}{Dahlia Malkhi}.}
  \bibinfo{year}{2019}\natexlab{}.
\newblock \bibinfo{booktitle}{\emph{Concurrency: The Works of Leslie Lamport}}.
\newblock \bibinfo{publisher}{ACM}.
\newblock


\bibitem[Malviya et~al\mbox{.}(2014)]%
        {logical-logging-1}
\bibfield{author}{\bibinfo{person}{Nirmesh Malviya}, \bibinfo{person}{Ariel
  Weisberg}, \bibinfo{person}{Samuel Madden}, {and} \bibinfo{person}{Michael
  Stonebraker}.} \bibinfo{year}{2014}\natexlab{}.
\newblock \showarticletitle{Rethinking main memory {OLTP} recovery}. In
  \bibinfo{booktitle}{\emph{2014 IEEE 30th International Conference on Data
  Engineering}}. IEEE, \bibinfo{pages}{604--615}.
\newblock


\bibitem[Medeiros(2012)]%
        {ab-zookeeper}
\bibfield{author}{\bibinfo{person}{Andr{\'e} Medeiros}.}
  \bibinfo{year}{2012}\natexlab{}.
\newblock \showarticletitle{ZooKeeper’s atomic broadcast protocol: Theory and
  practice}.
\newblock \bibinfo{journal}{\emph{Aalto University School of Science}}
  \bibinfo{volume}{20} (\bibinfo{year}{2012}).
\newblock


\bibitem[Mohan et~al\mbox{.}(1992)]%
        {aries}
\bibfield{author}{\bibinfo{person}{Chandrasekaran Mohan}, \bibinfo{person}{Don
  Haderle}, \bibinfo{person}{Bruce Lindsay}, \bibinfo{person}{Hamid Pirahesh},
  {and} \bibinfo{person}{Peter Schwarz}.} \bibinfo{year}{1992}\natexlab{}.
\newblock \showarticletitle{{ARIES}: A transaction recovery method supporting
  fine-granularity locking and partial rollbacks using write-ahead logging}.
\newblock \bibinfo{journal}{\emph{ACM Transactions on Database Systems (TODS)}}
  \bibinfo{volume}{17}, \bibinfo{number}{1} (\bibinfo{year}{1992}),
  \bibinfo{pages}{94--162}.
\newblock


\bibitem[Moniz et~al\mbox{.}(2017)]%
        {blotter}
\bibfield{author}{\bibinfo{person}{Henrique Moniz}, \bibinfo{person}{Jo{\~a}o
  Leit{\~a}o}, \bibinfo{person}{Ricardo~J Dias}, \bibinfo{person}{Johannes
  Gehrke}, \bibinfo{person}{Nuno Pregui{\c{c}}a}, {and}
  \bibinfo{person}{Rodrigo Rodrigues}.} \bibinfo{year}{2017}\natexlab{}.
\newblock \showarticletitle{Blotter: Low latency transactions for
  geo-replicated storage}. In \bibinfo{booktitle}{\emph{Proceedings of the 26th
  International Conference on World Wide Web}}. \bibinfo{pages}{263--272}.
\newblock


\bibitem[Mu et~al\mbox{.}(2014)]%
        {racoco}
\bibfield{author}{\bibinfo{person}{Shuai Mu}, \bibinfo{person}{Yang Cui},
  \bibinfo{person}{Yang Zhang}, \bibinfo{person}{Wyatt Lloyd}, {and}
  \bibinfo{person}{Jinyang Li}.} \bibinfo{year}{2014}\natexlab{}.
\newblock \showarticletitle{Extracting more concurrency from distributed
  transactions}. In \bibinfo{booktitle}{\emph{11th USENIX Symposium on
  Operating Systems Design and Implementation (OSDI 14)}}.
  \bibinfo{pages}{479--494}.
\newblock


\bibitem[Mu et~al\mbox{.}(2016)]%
        {consolidating}
\bibfield{author}{\bibinfo{person}{Shuai Mu}, \bibinfo{person}{Lamont Nelson},
  \bibinfo{person}{Wyatt Lloyd}, {and} \bibinfo{person}{Jinyang Li}.}
  \bibinfo{year}{2016}\natexlab{}.
\newblock \showarticletitle{Consolidating concurrency control and consensus for
  commits under conflicts}. In \bibinfo{booktitle}{\emph{12th USENIX Symposium
  on Operating Systems Design and Implementation (OSDI 16)}}.
  \bibinfo{pages}{517--532}.
\newblock


\bibitem[Nakamura et~al\mbox{.}(2019)]%
        {parallel-logging}
\bibfield{author}{\bibinfo{person}{Yasuhiro Nakamura},
  \bibinfo{person}{Hideyuki Kawashima}, {and} \bibinfo{person}{Osamu Tatebe}.}
  \bibinfo{year}{2019}\natexlab{}.
\newblock \showarticletitle{Integration of {TicToc} Concurrency Control
  Protocol with Parallel Write Ahead Logging Protocol}.
\newblock \bibinfo{journal}{\emph{International Journal of Networking and
  Computing}} \bibinfo{volume}{9}, \bibinfo{number}{2} (\bibinfo{year}{2019}),
  \bibinfo{pages}{339--353}.
\newblock


\bibitem[Nathan et~al\mbox{.}(2019)]%
        {bcr}
\bibfield{author}{\bibinfo{person}{Senthil Nathan}, \bibinfo{person}{Chander
  Govindarajan}, \bibinfo{person}{Adarsh Saraf}, \bibinfo{person}{Manish
  Sethi}, {and} \bibinfo{person}{Praveen Jayachandran}.}
  \bibinfo{year}{2019}\natexlab{}.
\newblock \showarticletitle{Blockchain meets database: design and
  implementation of a blockchain relational database}.
\newblock \bibinfo{journal}{\emph{Proceedings of the VLDB Endowment}}
  \bibinfo{volume}{12}, \bibinfo{number}{11} (\bibinfo{year}{2019}),
  \bibinfo{pages}{1539--1552}.
\newblock


\bibitem[Nawab et~al\mbox{.}(2015)]%
        {min-geo}
\bibfield{author}{\bibinfo{person}{Faisal Nawab}, \bibinfo{person}{Vaibhav
  Arora}, \bibinfo{person}{Divyakant Agrawal}, {and} \bibinfo{person}{Amr
  El~Abbadi}.} \bibinfo{year}{2015}\natexlab{}.
\newblock \showarticletitle{Minimizing commit latency of transactions in
  geo-replicated data stores}. In \bibinfo{booktitle}{\emph{Proceedings of the
  2015 ACM SIGMOD International Conference on Management of Data}}.
  \bibinfo{pages}{1279--1294}.
\newblock


\bibitem[Ongaro and Ousterhout(2014)]%
        {raft}
\bibfield{author}{\bibinfo{person}{Diego Ongaro} {and} \bibinfo{person}{John
  Ousterhout}.} \bibinfo{year}{2014}\natexlab{}.
\newblock \showarticletitle{In search of an understandable consensus
  algorithm}. In \bibinfo{booktitle}{\emph{2014 USENIX Annual Technical
  Conference (Usenix ATC 14)}}. \bibinfo{pages}{305--319}.
\newblock


\bibitem[Ports and Grittner(2012)]%
        {ssi-2}
\bibfield{author}{\bibinfo{person}{Dan~RK Ports} {and} \bibinfo{person}{Kevin
  Grittner}.} \bibinfo{year}{2012}\natexlab{}.
\newblock \showarticletitle{Serializable Snapshot Isolation in {PostgreSQL}}.
\newblock \bibinfo{journal}{\emph{Proceedings of the VLDB Endowment}}
  \bibinfo{volume}{5}, \bibinfo{number}{12} (\bibinfo{year}{2012}).
\newblock


\bibitem[Qadah and Sadoghi(2018)]%
        {quecc}
\bibfield{author}{\bibinfo{person}{Thamir~M Qadah} {and}
  \bibinfo{person}{Mohammad Sadoghi}.} \bibinfo{year}{2018}\natexlab{}.
\newblock \showarticletitle{Quecc: A queue-oriented, control-free concurrency
  architecture}. In \bibinfo{booktitle}{\emph{Proceedings of the 19th
  International Middleware Conference}}. \bibinfo{pages}{13--25}.
\newblock


\bibitem[Qin et~al\mbox{.}(2021)]%
        {caracal}
\bibfield{author}{\bibinfo{person}{Dai Qin}, \bibinfo{person}{Angela~Demke
  Brown}, {and} \bibinfo{person}{Ashvin Goel}.}
  \bibinfo{year}{2021}\natexlab{}.
\newblock \showarticletitle{Caracal: Contention Management with Deterministic
  Concurrency Control}. In \bibinfo{booktitle}{\emph{Proceedings of the ACM
  SIGOPS 28th Symposium on Operating Systems Principles}}.
  \bibinfo{pages}{180--194}.
\newblock


\bibitem[Raynal and Beeson(1986)]%
        {mutex-algorithms}
\bibfield{author}{\bibinfo{person}{Michel Raynal} {and} \bibinfo{person}{D
  Beeson}.} \bibinfo{year}{1986}\natexlab{}.
\newblock \bibinfo{booktitle}{\emph{Algorithms for mutual exclusion}}.
\newblock \bibinfo{publisher}{MIT press}.
\newblock


\bibitem[Ren et~al\mbox{.}(2019)]%
        {slog}
\bibfield{author}{\bibinfo{person}{Kun Ren}, \bibinfo{person}{Dennis Li}, {and}
  \bibinfo{person}{Daniel~J Abadi}.} \bibinfo{year}{2019}\natexlab{}.
\newblock \showarticletitle{{SLOG}: Serializable, Low-latency, Geo-replicated
  Transactions}.
\newblock \bibinfo{journal}{\emph{Proceedings of the VLDB Endowment}}
  \bibinfo{volume}{12}, \bibinfo{number}{11} (\bibinfo{year}{2019}).
\newblock


\bibitem[Ren et~al\mbox{.}(2014)]%
        {deter-eval}
\bibfield{author}{\bibinfo{person}{Kun Ren}, \bibinfo{person}{Alexander
  Thomson}, {and} \bibinfo{person}{Daniel~J Abadi}.}
  \bibinfo{year}{2014}\natexlab{}.
\newblock \showarticletitle{An evaluation of the advantages and disadvantages
  of deterministic database systems}.
\newblock \bibinfo{journal}{\emph{Proceedings of the VLDB Endowment}}
  \bibinfo{volume}{7}, \bibinfo{number}{10} (\bibinfo{year}{2014}),
  \bibinfo{pages}{821--832}.
\newblock


\bibitem[Sciascia et~al\mbox{.}(2012)]%
        {update-replication}
\bibfield{author}{\bibinfo{person}{Daniele Sciascia}, \bibinfo{person}{Fernando
  Pedone}, {and} \bibinfo{person}{Flavio Junqueira}.}
  \bibinfo{year}{2012}\natexlab{}.
\newblock \showarticletitle{Scalable deferred update replication}. In
  \bibinfo{booktitle}{\emph{IEEE/IFIP International Conference on Dependable
  Systems and Networks (DSN 2012)}}. IEEE, \bibinfo{pages}{1--12}.
\newblock


\bibitem[Shamis et~al\mbox{.}(2019)]%
        {opacity}
\bibfield{author}{\bibinfo{person}{Alex Shamis}, \bibinfo{person}{Matthew
  Renzelmann}, \bibinfo{person}{Stanko Novakovic}, \bibinfo{person}{Georgios
  Chatzopoulos}, \bibinfo{person}{Aleksandar Dragojevi{\'c}},
  \bibinfo{person}{Dushyanth Narayanan}, {and} \bibinfo{person}{Miguel
  Castro}.} \bibinfo{year}{2019}\natexlab{}.
\newblock \showarticletitle{Fast general distributed transactions with
  opacity}. In \bibinfo{booktitle}{\emph{Proceedings of the 2019 International
  Conference on Management of Data}}. \bibinfo{pages}{433--448}.
\newblock


\bibitem[Shute et~al\mbox{.}(2013)]%
        {f1}
\bibfield{author}{\bibinfo{person}{Jeff Shute}, \bibinfo{person}{Radek
  Vingralek}, \bibinfo{person}{Bart Samwel}, \bibinfo{person}{Ben Handy},
  \bibinfo{person}{Chad Whipkey}, \bibinfo{person}{Eric Rollins},
  \bibinfo{person}{Mircea Oancea~Kyle Littlefield}, \bibinfo{person}{David
  Menestrina}, \bibinfo{person}{Stephan Ellner~John Cieslewicz},
  \bibinfo{person}{Ian Rae}, {et~al\mbox{.}}} \bibinfo{year}{2013}\natexlab{}.
\newblock \showarticletitle{F1: A Distributed {SQL} Database That Scales}.
\newblock \bibinfo{journal}{\emph{Proceedings of the VLDB Endowment}}
  \bibinfo{volume}{6}, \bibinfo{number}{11} (\bibinfo{year}{2013}).
\newblock


\bibitem[Soisalon-Soininen and Yl{\"o}nen(1995)]%
        {elr-3}
\bibfield{author}{\bibinfo{person}{Eljas Soisalon-Soininen} {and}
  \bibinfo{person}{Tatu Yl{\"o}nen}.} \bibinfo{year}{1995}\natexlab{}.
\newblock \showarticletitle{Partial strictness in two-phase locking}. In
  \bibinfo{booktitle}{\emph{International Conference on Database Theory}}.
  Springer, \bibinfo{pages}{139--147}.
\newblock


\bibitem[Stamos and Cristian(1990)]%
        {coordinator-log-1}
\bibfield{author}{\bibinfo{person}{James~W Stamos} {and}
  \bibinfo{person}{Flaviu Cristian}.} \bibinfo{year}{1990}\natexlab{}.
\newblock \showarticletitle{A low-cost atomic commit protocol}. In
  \bibinfo{booktitle}{\emph{Proceedings Ninth Symposium on Reliable Distributed
  Systems}}. IEEE, \bibinfo{pages}{66--75}.
\newblock


\bibitem[Stamos and Cristian(1993)]%
        {coordinator-log-2}
\bibfield{author}{\bibinfo{person}{James~W Stamos} {and}
  \bibinfo{person}{Flaviu Cristian}.} \bibinfo{year}{1993}\natexlab{}.
\newblock \showarticletitle{Coordinator log transaction execution protocol}.
\newblock \bibinfo{journal}{\emph{Distributed and Parallel Databases}}
  \bibinfo{volume}{1} (\bibinfo{year}{1993}), \bibinfo{pages}{383--408}.
\newblock


\bibitem[Stonebraker(1979)]%
        {ingress}
\bibfield{author}{\bibinfo{person}{Michael Stonebraker}.}
  \bibinfo{year}{1979}\natexlab{}.
\newblock \showarticletitle{Concurrency control and consistency of multiple
  copies of data in distributed INGRES}.
\newblock \bibinfo{journal}{\emph{IEEE Transactions on software Engineering}}
  \bibinfo{number}{3} (\bibinfo{year}{1979}), \bibinfo{pages}{188--194}.
\newblock


\bibitem[Stonebraker and Weisberg(2013)]%
        {voltdb}
\bibfield{author}{\bibinfo{person}{Michael Stonebraker} {and}
  \bibinfo{person}{Ariel Weisberg}.} \bibinfo{year}{2013}\natexlab{}.
\newblock \showarticletitle{The {VoltDB} Main Memory {DBMS}.}
\newblock \bibinfo{journal}{\emph{IEEE Data Eng. Bull.}} \bibinfo{volume}{36},
  \bibinfo{number}{2} (\bibinfo{year}{2013}), \bibinfo{pages}{21--27}.
\newblock


\bibitem[Taft et~al\mbox{.}(2014)]%
        {estore}
\bibfield{author}{\bibinfo{person}{Rebecca Taft}, \bibinfo{person}{Essam
  Mansour}, \bibinfo{person}{Marco Serafini}, \bibinfo{person}{Jennie Duggan},
  \bibinfo{person}{Aaron~J Elmore}, \bibinfo{person}{Ashraf Aboulnaga},
  \bibinfo{person}{Andrew Pavlo}, {and} \bibinfo{person}{Michael Stonebraker}.}
  \bibinfo{year}{2014}\natexlab{}.
\newblock \showarticletitle{E-store: Fine-grained elastic partitioning for
  distributed transaction processing systems}.
\newblock \bibinfo{journal}{\emph{Proceedings of the VLDB Endowment}}
  \bibinfo{volume}{8}, \bibinfo{number}{3} (\bibinfo{year}{2014}),
  \bibinfo{pages}{245--256}.
\newblock


\bibitem[Taft et~al\mbox{.}(2020)]%
        {cockroachdb}
\bibfield{author}{\bibinfo{person}{Rebecca Taft}, \bibinfo{person}{Irfan
  Sharif}, \bibinfo{person}{Andrei Matei}, \bibinfo{person}{Nathan
  VanBenschoten}, \bibinfo{person}{Jordan Lewis}, \bibinfo{person}{Tobias
  Grieger}, \bibinfo{person}{Kai Niemi}, \bibinfo{person}{Andy Woods},
  \bibinfo{person}{Anne Birzin}, \bibinfo{person}{Raphael Poss},
  {et~al\mbox{.}}} \bibinfo{year}{2020}\natexlab{}.
\newblock \showarticletitle{{CockroachDB}: The Resilient Geo-Distributed {SQL}
  Database}. In \bibinfo{booktitle}{\emph{Proceedings of the 2020 ACM SIGMOD
  International Conference on Management of Data}}.
  \bibinfo{pages}{1493--1509}.
\newblock


\bibitem[Tanabe et~al\mbox{.}(2020)]%
        {ccbench}
\bibfield{author}{\bibinfo{person}{Takayuki Tanabe}, \bibinfo{person}{Takashi
  Hoshino}, \bibinfo{person}{Hideyuki Kawashima}, {and} \bibinfo{person}{Osamu
  Tatebe}.} \bibinfo{year}{2020}\natexlab{}.
\newblock \showarticletitle{An analysis of concurrency control protocols for
  in-memory databases with {CCBench}}.
\newblock \bibinfo{journal}{\emph{Proceedings of the VLDB Endowment}}
  \bibinfo{volume}{13}, \bibinfo{number}{13} (\bibinfo{year}{2020}),
  \bibinfo{pages}{3531--3544}.
\newblock


\bibitem[Thomson and Abadi(2010)]%
        {deter-case}
\bibfield{author}{\bibinfo{person}{Alexander Thomson} {and}
  \bibinfo{person}{Daniel~J Abadi}.} \bibinfo{year}{2010}\natexlab{}.
\newblock \showarticletitle{The case for determinism in database systems}.
\newblock \bibinfo{journal}{\emph{Proceedings of the VLDB Endowment}}
  \bibinfo{volume}{3}, \bibinfo{number}{1-2} (\bibinfo{year}{2010}),
  \bibinfo{pages}{70--80}.
\newblock


\bibitem[Thomson et~al\mbox{.}(2012)]%
        {calvin}
\bibfield{author}{\bibinfo{person}{Alexander Thomson},
  \bibinfo{person}{Thaddeus Diamond}, \bibinfo{person}{Shu-Chun Weng},
  \bibinfo{person}{Kun Ren}, \bibinfo{person}{Philip Shao}, {and}
  \bibinfo{person}{Daniel~J Abadi}.} \bibinfo{year}{2012}\natexlab{}.
\newblock \showarticletitle{Calvin: fast distributed transactions for
  partitioned database systems}. In \bibinfo{booktitle}{\emph{Proceedings of
  the 2012 ACM SIGMOD international conference on management of data}}.
  \bibinfo{pages}{1--12}.
\newblock


\bibitem[Tu et~al\mbox{.}(2013)]%
        {silo}
\bibfield{author}{\bibinfo{person}{Stephen Tu}, \bibinfo{person}{Wenting
  Zheng}, \bibinfo{person}{Eddie Kohler}, \bibinfo{person}{Barbara Liskov},
  {and} \bibinfo{person}{Samuel Madden}.} \bibinfo{year}{2013}\natexlab{}.
\newblock \showarticletitle{Speedy transactions in multicore in-memory
  databases}. In \bibinfo{booktitle}{\emph{Proceedings of the 24th ACM
  Symposium on Operating Systems Principles}}.
\newblock


\bibitem[Wang et~al\mbox{.}(2022)]%
        {sensitivity}
\bibfield{author}{\bibinfo{person}{Yang Wang}, \bibinfo{person}{Miao Yu},
  \bibinfo{person}{Yujie Hui}, \bibinfo{person}{Fang Zhou},
  \bibinfo{person}{Yuyang Huang}, \bibinfo{person}{Rui Zhu},
  \bibinfo{person}{Xueyuan Ren}, \bibinfo{person}{Tianxi Li}, {and}
  \bibinfo{person}{Xiaoyi Lu}.} \bibinfo{year}{2022}\natexlab{}.
\newblock \showarticletitle{A Study of Database Performance Sensitivity to
  Experiment Settings.}
\newblock \bibinfo{journal}{\emph{Proceedings of the VLDB Endowment}}
  \bibinfo{volume}{15}, \bibinfo{number}{7} (\bibinfo{year}{2022}).
\newblock


\bibitem[Weikum and Vossen(2001)]%
        {2pl}
\bibfield{author}{\bibinfo{person}{Gerhard Weikum} {and}
  \bibinfo{person}{Gottfried Vossen}.} \bibinfo{year}{2001}\natexlab{}.
\newblock \bibinfo{booktitle}{\emph{Transactional information systems: theory,
  algorithms, and the practice of concurrency control and recovery}}.
\newblock \bibinfo{publisher}{Elsevier}.
\newblock


\bibitem[Wolski(2009)]%
        {tatp}
\bibfield{author}{\bibinfo{person}{A Wolski}.} \bibinfo{year}{2009}\natexlab{}.
\newblock \bibinfo{title}{{TATP} benchmark description (version 1.0)}.
\newblock
\newblock


\bibitem[Wu et~al\mbox{.}(2016)]%
        {tpart}
\bibfield{author}{\bibinfo{person}{Shan-Hung Wu}, \bibinfo{person}{Tsai-Yu
  Feng}, \bibinfo{person}{Meng-Kai Liao}, \bibinfo{person}{Shao-Kan Pi}, {and}
  \bibinfo{person}{Yu-Shan Lin}.} \bibinfo{year}{2016}\natexlab{}.
\newblock \showarticletitle{T-part: Partitioning of transactions for
  forward-pushing in deterministic database systems}. In
  \bibinfo{booktitle}{\emph{Proceedings of the 2016 International Conference on
  Management of Data}}. \bibinfo{pages}{1553--1565}.
\newblock


\bibitem[Wu et~al\mbox{.}(2017)]%
        {pacman}
\bibfield{author}{\bibinfo{person}{Yingjun Wu}, \bibinfo{person}{Wentian Guo},
  \bibinfo{person}{Chee-Yong Chan}, {and} \bibinfo{person}{Kian-Lee Tan}.}
  \bibinfo{year}{2017}\natexlab{}.
\newblock \showarticletitle{Fast failure recovery for main-memory {DBMSs} on
  multicores}. In \bibinfo{booktitle}{\emph{Proceedings of the 2017 ACM
  International Conference on Management of Data}}. \bibinfo{pages}{267--281}.
\newblock


\bibitem[Xia et~al\mbox{.}(2020)]%
        {taurus}
\bibfield{author}{\bibinfo{person}{Yu Xia}, \bibinfo{person}{Xiangyao Yu},
  \bibinfo{person}{Andrew Pavlo}, {and} \bibinfo{person}{Srinivas Devadas}.}
  \bibinfo{year}{2020}\natexlab{}.
\newblock \showarticletitle{Taurus: Lightweight parallel logging for in-memory
  database management systems}.
\newblock \bibinfo{journal}{\emph{Proceedings of the VLDB Endowment}}
  \bibinfo{volume}{14}, \bibinfo{number}{2} (\bibinfo{year}{2020}),
  \bibinfo{pages}{189--201}.
\newblock


\bibitem[Yan et~al\mbox{.}(2018)]%
        {carousel}
\bibfield{author}{\bibinfo{person}{Xinan Yan}, \bibinfo{person}{Linguan Yang},
  \bibinfo{person}{Hongbo Zhang}, \bibinfo{person}{Xiayue~Charles Lin},
  \bibinfo{person}{Bernard Wong}, \bibinfo{person}{Kenneth Salem}, {and}
  \bibinfo{person}{Tim Brecht}.} \bibinfo{year}{2018}\natexlab{}.
\newblock \showarticletitle{Carousel: Low-latency transaction processing for
  globally-distributed data}. In \bibinfo{booktitle}{\emph{Proceedings of the
  2018 International Conference on Management of Data}}.
  \bibinfo{pages}{231--243}.
\newblock


\bibitem[Yang et~al\mbox{.}(2022)]%
        {oceanbase}
\bibfield{author}{\bibinfo{person}{Zhenkun Yang}, \bibinfo{person}{Chuanhui
  Yang}, \bibinfo{person}{Fusheng Han}, \bibinfo{person}{Mingqiang Zhuang},
  \bibinfo{person}{Bing Yang}, \bibinfo{person}{Zhifeng Yang},
  \bibinfo{person}{Xiaojun Cheng}, \bibinfo{person}{Yuzhong Zhao},
  \bibinfo{person}{Wenhui Shi}, \bibinfo{person}{Huafeng Xi}, {et~al\mbox{.}}}
  \bibinfo{year}{2022}\natexlab{}.
\newblock \showarticletitle{{OceanBase}: a 707 million {tpmC} distributed
  relational database system}.
\newblock \bibinfo{journal}{\emph{Proceedings of the VLDB Endowment}}
  \bibinfo{volume}{15}, \bibinfo{number}{12} (\bibinfo{year}{2022}),
  \bibinfo{pages}{3385--3397}.
\newblock


\bibitem[Yao et~al\mbox{.}(2016a)]%
        {single-thread-dcc}
\bibfield{author}{\bibinfo{person}{Chang Yao}, \bibinfo{person}{Divyakant
  Agrawal}, \bibinfo{person}{Gang Chen}, \bibinfo{person}{Qian Lin},
  \bibinfo{person}{Beng~Chin Ooi}, \bibinfo{person}{Weng-Fai Wong}, {and}
  \bibinfo{person}{Meihui Zhang}.} \bibinfo{year}{2016}\natexlab{a}.
\newblock \showarticletitle{Exploiting single-threaded model in multi-core
  in-memory systems}.
\newblock \bibinfo{journal}{\emph{IEEE Transactions on Knowledge and Data
  Engineering}} \bibinfo{volume}{28}, \bibinfo{number}{10}
  (\bibinfo{year}{2016}), \bibinfo{pages}{2635--2650}.
\newblock


\bibitem[Yao et~al\mbox{.}(2016b)]%
        {adaptive-logging}
\bibfield{author}{\bibinfo{person}{Chang Yao}, \bibinfo{person}{Divyakant
  Agrawal}, \bibinfo{person}{Gang Chen}, \bibinfo{person}{Beng~Chin Ooi}, {and}
  \bibinfo{person}{Sai Wu}.} \bibinfo{year}{2016}\natexlab{b}.
\newblock \showarticletitle{Adaptive logging: Optimizing logging and recovery
  costs in distributed in-memory databases}. In
  \bibinfo{booktitle}{\emph{Proceedings of the 2016 International Conference on
  Management of Data}}. \bibinfo{pages}{1119--1134}.
\newblock


\bibitem[Yao et~al\mbox{.}(2018)]%
        {distdgcc}
\bibfield{author}{\bibinfo{person}{Chang Yao}, \bibinfo{person}{Meihui Zhang},
  \bibinfo{person}{Qian Lin}, \bibinfo{person}{Beng~Chin Ooi}, {and}
  \bibinfo{person}{Jiatao Xu}.} \bibinfo{year}{2018}\natexlab{}.
\newblock \showarticletitle{Scaling distributed transaction processing and
  recovery based on dependency logging}.
\newblock \bibinfo{journal}{\emph{The VLDB Journal}} \bibinfo{volume}{27},
  \bibinfo{number}{3} (\bibinfo{year}{2018}), \bibinfo{pages}{347--368}.
\newblock


\bibitem[Yu et~al\mbox{.}(2014)]%
        {dbx1000}
\bibfield{author}{\bibinfo{person}{Xiangyao Yu}, \bibinfo{person}{George
  Bezerra}, \bibinfo{person}{Andrew Pavlo}, \bibinfo{person}{Srinivas Devadas},
  {and} \bibinfo{person}{Michael Stonebraker}.}
  \bibinfo{year}{2014}\natexlab{}.
\newblock \showarticletitle{Staring into the abyss: An evaluation of
  concurrency control with one thousand cores}.
\newblock  (\bibinfo{year}{2014}).
\newblock


\bibitem[Yu et~al\mbox{.}(2016)]%
        {tictoc}
\bibfield{author}{\bibinfo{person}{Xiangyao Yu}, \bibinfo{person}{Andrew
  Pavlo}, \bibinfo{person}{Daniel Sanchez}, {and} \bibinfo{person}{Srinivas
  Devadas}.} \bibinfo{year}{2016}\natexlab{}.
\newblock \showarticletitle{{TicToc}: Time traveling optimistic concurrency
  control}. In \bibinfo{booktitle}{\emph{Proceedings of the 2016 International
  Conference on Management of Data}}. \bibinfo{pages}{1629--1642}.
\newblock


\bibitem[Yu et~al\mbox{.}(2018)]%
        {sundial}
\bibfield{author}{\bibinfo{person}{Xiangyao Yu}, \bibinfo{person}{Yu Xia},
  \bibinfo{person}{Andrew Pavlo}, \bibinfo{person}{Daniel Sanchez},
  \bibinfo{person}{Larry Rudolph}, {and} \bibinfo{person}{Srinivas Devadas}.}
  \bibinfo{year}{2018}\natexlab{}.
\newblock \showarticletitle{Sundial: harmonizing concurrency control and
  caching in a distributed {OLTP} database management system}.
\newblock \bibinfo{journal}{\emph{Proceedings of the VLDB Endowment}}
  \bibinfo{volume}{11}, \bibinfo{number}{10} (\bibinfo{year}{2018}),
  \bibinfo{pages}{1289--1302}.
\newblock


\bibitem[Zhang et~al\mbox{.}(2018)]%
        {TAPIR}
\bibfield{author}{\bibinfo{person}{Irene Zhang}, \bibinfo{person}{Naveen~Kr
  Sharma}, \bibinfo{person}{Adriana Szekeres}, \bibinfo{person}{Arvind
  Krishnamurthy}, {and} \bibinfo{person}{Dan~RK Ports}.}
  \bibinfo{year}{2018}\natexlab{}.
\newblock \showarticletitle{Building consistent transactions with inconsistent
  replication}.
\newblock \bibinfo{journal}{\emph{ACM Transactions on Computer Systems (TOCS)}}
  \bibinfo{volume}{35}, \bibinfo{number}{4} (\bibinfo{year}{2018}),
  \bibinfo{pages}{1--37}.
\newblock


\bibitem[Zhang et~al\mbox{.}(2013)]%
        {chains}
\bibfield{author}{\bibinfo{person}{Yang Zhang}, \bibinfo{person}{Russell
  Power}, \bibinfo{person}{Siyuan Zhou}, \bibinfo{person}{Yair Sovran},
  \bibinfo{person}{Marcos~K Aguilera}, {and} \bibinfo{person}{Jinyang Li}.}
  \bibinfo{year}{2013}\natexlab{}.
\newblock \showarticletitle{Transaction chains: achieving serializability with
  low latency in geo-distributed storage systems}. In
  \bibinfo{booktitle}{\emph{Proceedings of the Twenty-Fourth ACM Symposium on
  Operating Systems Principles}}. \bibinfo{pages}{276--291}.
\newblock


\bibitem[Zhou et~al\mbox{.}(2022)]%
        {lotus}
\bibfield{author}{\bibinfo{person}{Xinjing Zhou}, \bibinfo{person}{Xiangyao
  Yu}, \bibinfo{person}{Goetz Graefe}, {and} \bibinfo{person}{Michael
  Stonebraker}.} \bibinfo{year}{2022}\natexlab{}.
\newblock \showarticletitle{Lotus: scalable multi-partition transactions on
  single-threaded partitioned databases}.
\newblock \bibinfo{journal}{\emph{Proceedings of the VLDB Endowment}}
  \bibinfo{volume}{15}, \bibinfo{number}{11} (\bibinfo{year}{2022}),
  \bibinfo{pages}{2939--2952}.
\newblock


\bibitem[Zhu et~al\mbox{.}(2020)]%
        {logless}
\bibfield{author}{\bibinfo{person}{Yuqing Zhu}, \bibinfo{person}{Philip~S Yu},
  \bibinfo{person}{Guolei Yi}, \bibinfo{person}{Mengying Guo},
  \bibinfo{person}{Wenlong Ma}, \bibinfo{person}{Jianxun Liu}, {and}
  \bibinfo{person}{Yungang Bao}.} \bibinfo{year}{2020}\natexlab{}.
\newblock \showarticletitle{Logless one-phase commit made possible for
  highly-available datastores}.
\newblock \bibinfo{journal}{\emph{Distributed and Parallel Databases}}
  \bibinfo{volume}{38} (\bibinfo{year}{2020}), \bibinfo{pages}{101--126}.
\newblock


\end{thebibliography}
